\documentclass[hyper,12pt,letterpaper]{JHEP3}
\usepackage{amsmath,amssymb,amsfonts,bm}
\usepackage{graphicx}
\usepackage{multirow}
\usepackage{verbatim}
\usepackage{appendix}

\renewcommand{\circ}{{\phi}}
\newcommand{\Lag}{{M}}
\newcommand{\mb}{\mathbf}

\newcommand{\ie}{{\emph{i.e.}}}
\newcommand{\eg}{{\emph{e.g.}}}
\newcommand{\cf}{{\emph{cf.}}}

\def\Z{\relax\ifmmode\mathchoice
{\hbox{\cmss Z\kern-.4em Z}}{\hbox{\cmss Z\kern-.4em Z}} {\lower.9pt\hbox{\cmsss Z\kern-.4em Z}}
{\lower1.2pt\hbox{\cmsss Z\kern-.4em Z}}\else{\cmss Z\kern-.4em Z}\fi}
\def\IZ{\mathbb{Z}}

\def\S{{\bf S}}

\def\R{{\bf R}}
\def\C{{\bf C}}

\def\cp{{\mathbb{C}}{\bf P}}

\newcommand{\CF}{{\mathcal F}}

\def\CA {{\cal A}}

\def\CF {{\cal F}}

\def\CL {{\cal L}}
\def\CM {{\cal M}}
\def\CN {{\cal N}}
\def\CO {{\cal O}}
\def\CP {{\cal P}}

\def\CV {{\cal V}}
\def\CW {{\cal W}}

\def\CZ {{\cal Z}}

\def\be{\begin{equation}}
\def\ee{\end{equation}}
\def\a{\alpha}

\def\e{\epsilon}

\def\m{\mu}

\def\la{\lambda}

\def\bar{\overline}

\def\Tr{{\rm Tr}}

\def\Fl{{\CF {\kern -1.2pt \ell} }}

\def\leadsto{\rightsquigarrow}

\def\example#1{\bgroup\narrower\footnotefont\baselineskip\footskip\bigbreak
\hrule\medskip\nobreak\noindent {\bf Example}. {\it #1\/}\par\nobreak}
\def\endexample{\medskip\nobreak\hrule\bigbreak\egroup}

\def\btimes{~{{{\lower1pt\hbox{$\square$}} \kern-7.6pt \times}}~}
\def\TT{{\Bbb{T}}}
\def\LL{{\Bbb{L}}}
\def\Weyl{{\cal W}}

\def\C{{\Bbb{C}}}
\def\Z{{\Bbb{Z}}}

\def\m{{\mathfrak m}}

\def\dual#1{{^L\negthinspace #1}}

\def\LG{\dual{G}}

\title{Vortex Counting and Lagrangian 3-manifolds}

\author{Tudor Dimofte,$^1$ Sergei Gukov,$^{1,2}$ Lotte Hollands,$^{1,2}$
\\
\\
$^1$ California Institute of Technology, Pasadena, CA 91125, USA \\
$^2$ University of California, Santa Barbara, CA 93106, USA}

\preprint{~}

\abstract{
To every 3-manifold $M$ one can associate a two-dimensional $\CN=(2,2)$ supersymmetric
field theory by compactifying five-dimensional $\CN=2$ super-Yang-Mills theory on $M$.
This system naturally appears in the study of half-BPS surface operators in
four-dimensional $\CN=2$ gauge theories on one hand,
and in the geometric approach to knot homologies, on the other.
We study the relation between vortex counting in such
two-dimensional $\CN=(2,2)$ supersymmetric field theories
and the refined BPS invariants of the dual geometries.
In certain cases, this counting can be also mapped to the computation
of degenerate conformal blocks in two-dimensional~CFT's.
Degenerate limits of vertex operators in CFT receive a simple interpretation
via geometric transitions in BPS counting.
\\
\\
\\
{\tt ~} }

\begin{document}

\addtolength{\parskip}{.7mm}
\addtolength{\baselineskip}{.2mm}
\addtolength{\abovedisplayskip}{.8mm}
\addtolength{\belowdisplayskip}{.8mm}


\section{Motivation}
\label{sec:intro}

One motivation for the present paper comes from a rather surprising correspondence
between two seemingly different systems, each described by its own partition function.
One is a three-dimensional theory --- such as Chern-Simons gauge theory with complex
gauge group $G_{\C}$ --- whose partition function is a quantum invariant
of a 3-manifold $M$ (possibly with boundary).
Much as in a two-dimensional CFT, this partition function consists of products
of holomorphic and anti-holomorphic pieces, each of which is related to the analytic
continuation of Chern-Simons theory with compact gauge group~$G$ \cite{Witten-AC},
and takes the following general form
\be
Z (M;\hbar) \; = \; \exp\left( \frac{1}{\hbar} S_0 
 \,+\,\sum_{n=0}^\infty S_{n+1} \, \hbar^{n} \right)\,.
\label{Zcspert}
\ee
Here $\hbar \in \C$ is the perturbative expansion parameter (the coupling constant)
and for simplicity we are suppressing the dependence on other parameters
of the theory as well as the geometry of $M$.
For example, if $M$ is a hyperbolic 3-manifold,
the partition function \eqref{Zcspert}
depends on the so-called shape parameters of $M$
and, in the simplest case of $SL(2,\C)$ Chern-Simons theory,
can be expressed as a multiple contour integral~\cite{DGLZ}:
\be
Z (M;\hbar) \; = \; \oint \prod_{i} \frac{d \varphi_i}{\sqrt{4\pi\hbar}} \;
\prod_{j} \Phi_{\hbar} \big( \Delta_j \big)^{\pm1}\,,
\ee
where $\Phi_{\hbar} (\Delta_j)$ is the quantum dilogarithm function,
associated to the $j$-th tetrahedron $\Delta_j$ in a triangulation of $M$,
and depending on its shape parameter $\varphi$.\\

The second system is a two-dimensional (gauge) theory with $\CN=(2,2)$ supersymmetry.
Any such theory can be subject to the $\Omega$-deformation defined by the action
of the rotation symmetry group $SO(2)_E \simeq U(1)_E$ on the two-dimensional (Euclidean) space-time $\R^2$.
As we explain in more detail in section \ref{sec:gauge}, the partition function of
the $\Omega$-deformed $\CN=(2,2)$ gauge theory ``counts'' finite-energy supersymmetric
field configurations on $\R^2$, {\it i.e.} vortices.
For this reason, the resulting partition function will be called
the {\it vortex partition function} and denoted $Z^{{\rm vortex}}$.
Essentially by definition,
the partition function $Z^{{\rm vortex}}$ can be expressed as a perturbative series
\be
Z^{{\rm vortex}} (\hbar) = \exp \Bigl(\, \frac{1}{\hbar} \CW (\hbar) \, \Bigr)
= \exp\left( \frac{1}{\hbar}S_0 
+S_1 + \hbar\,S_2 + \ldots\right)\,,
\label{zvortpert}
\ee
where $\hbar$ is the generator of the equivariant cohomology
$H^*_{U(1)_E} ({\rm pt}) \cong \C [\hbar]$ of a point.
Indeed, as in four dimensions \cite{Nekrasov},
the $\Omega$-deformation can be thought of as a way to regularize
a two-dimensional gauge theory on $\R^2$, such that
$$
{\rm Vol} (\R^2) = \int_{\R^2} 1 = \frac{1}{\hbar} \,.
$$
As a result, the path integral of a two-dimensional
gauge theory in the $\Omega$-background has the form \eqref{zvortpert},
where the twisted superpotential $\CW (\hbar)$ is a regular function of~$\hbar$.
(Notice that, just like in \eqref{Zcspert}, $\hbar$ is a formal complex parameter.)
Besides its dependence on $\hbar$, the vortex partition function \eqref{zvortpert}
also depends on various couplings of the two-dimensional $\CN=(2,2)$ gauge theory
that we suppress in our notations.

By comparing \eqref{Zcspert} and \eqref{zvortpert},
it is clear that vortex partition functions of $\CN=(2,2)$ gauge theories
in two dimensions have exactly the same form as perturbative quantum invariants of 3-manifolds.
Therefore, starting with this observation it is natural to seek a direct correspondence,
where a 3-manifold $M$ (possibly with boundary) defines a two-dimensional
$\CN=(2,2)$ ``effective'' field theory:
\be
\boxed{\phantom{\int} \begin{array}{c@{\qquad}c@{\qquad}c}
\text{3-manifold $M$} & \leadsto & \text{two-dimensional $\CN=(2,2)$ theory} \,
\label{conjduality}
\end{array}\phantom{\int}}
\ee
such that the physics of the resulting two-dimensional theory reflects
the geometric structures on $M$ and
\be
Z (M;\hbar) \; = \; Z^{{\rm vortex}} (\hbar) \,.
\label{zmzvortex}
\ee
This correspondence 
should be viewed as a 3-dimensional analog of the AGT correspondence \cite{AGT}
that, in a similar way, associates to a Riemann surface $C$ (possibly with punctures)
a four-dimensional $\CN=2$ ``effective'' gauge theory:
\be \begin{array}{c@{\qquad}c@{\qquad}c}
\text{2-manifold $C$} & \leadsto & \text{four-dimensional $\CN=2$ theory} \,,
\end{array}
\ee
such that
\be
Z^{{\rm CFT}} (C; \epsilon_1, \epsilon_2) \; = \; Z^{{\rm inst}} (\epsilon_1, \epsilon_2) \,.
\ee
Note that in both cases the partition function of a {\it non-supersymmetric}
quantum field theory is expressed via instanton counting in a different {\it supersymmetric}
gauge theory in the $\Omega$-background.

As in the AGT correspondence, one can approach the proposed duality
\eqref{conjduality} - \eqref{zmzvortex} by starting with a higher-dimensional
supersymmetric theory on the space-time manifold $\R^2_{\hbar} \times M$,
such that the dimensional reduction along $M$ gives the ``effective''
two-dimensional $\CN=(2,2)$ theory \eqref{conjduality}, while the reduction along $\R^2_{\hbar}$
gives a quantum theory of $M$.
In the present case, the appropriate five-dimensional theory is
the maximally supersymmetric ($\CN=2$) Yang-Mills theory
with gauge group $G$ (we assume $G$ to be compact and simple):
$$
\begin{array}{ccccc}
\; & \; & \text{5d $\CN=2$ super-Yang-Mills} & \; & \; \\
\; & \; & \text{on $\R^2_{\hbar} \times M$} & \; & \; \\
\; & \swarrow & \; & \searrow & \; \\
\text{quantum $G_{\C}$ invariant} & \; & \; & \; & \text{2d $\CN=(2,2)$ theory} \\
\text{of $M$} & \; & \; & \; & \text{on $\R^2_{\hbar}$}
\end{array}
$$
Since $M$ can be arbitrary,
the five-dimensional super-Yang-Mills theory must be partially twisted (along $M$)
in order to preserve $\CN=(2,2)$ supersymmetry in two dimensions.
The topological twist along a three-dimensional part of the space-time
is essentially equivalent to that of $\CN=4$ supersymmetric gauge theory in three dimensions,
and there are two natural choices (see {\it e.g.} \cite{BTbranes,BTeucl}):
one is a dimensional reduction of the twist in the Donaldson-Witten theory \cite{WittenDonaldson},
while the other leads to a topological theory whose supersymmetric field configurations are
flat $G_{\C}$ connections on $M$.
Anticipating a relation with $G_{\C}$ quantum invariants, the reader may have correctly guessed
that here we shall need the latter.

In order to describe the partial topological twist in more detail,
we recall that the maximally supersymmetric Yang-Mills theory in five
dimensions can be obtained by dimensional reduction from
super-Yang-Mills in ten dimensions. Under this reduction,
the $SO(10)_E$ symmetry of the Euclidean ten-dimensional theory
is broken to the symmetry group
\be
SO(5)_E \times SO(5)_R
\label{sofivefive}
\ee
where $SO(5)_R$ is the R-symmetry.
The bosonic fields of the five-dimensional $\CN=2$ super-Yang-Mills include
a gauge field and five Higgs scalars, which transform under the symmetry \eqref{sofivefive}
as $({\bf 5}, {\bf 1})$ and $({\bf 1}, {\bf 5})$, respectively.
The fermions transform as $({\bf 4}, {\bf 4})$ under the symmetry group \eqref{sofivefive}.

The partial topological twist on $\R^2 \times M$ breaks the Euclidean rotation symmetry
group $SO(5)_E$ to a subgroup $SO(3)_E \times U(1)_E$ (where $SO(2)_E \simeq U(1)_E$
is the rotation symmetry of $\R^2$) and, similarly, the R-symmetry group $SO(5)_R$
to a subgroup $SO(3)_R \times U(1)_R$.
Then, the new rotation symmetry group of the twisted Euclidean theory, $SO(3)_E'$,
is defined to be a diagonal subgroup of $SO(3)_E \times SO(3)_R$,
so that the full symmetry group of the partially twisted theory is $SO(3)_E' \times U(1)_E \times U(1)_R$.
It is easy to see that under this new symmetry group
the fields of the five-dimensional $\CN=2$ super-Yang-Mills transform as
\be \begin{array}{r@{\qquad}l}
\text{bosons}: & ({\bf 5}, {\bf 1} ) \oplus ({\bf 1}, {\bf 5}) \to 2 \times {\bf 3}^{(0,0)} \oplus {\bf 1}^{(\pm 2,0)} \oplus {\bf 1}^{(0, \pm 2)} \\
\text{fermions}: & ({\bf 4}, {\bf 4} ) \to {\bf 3}^{(\pm 1,\pm 1)} \oplus {\bf 1}^{(\pm 1,\pm 1)} \,
\end{array}
\label{twistedspectrum}
\ee
where all sign combinations have to be considered.
In particular, it is clear that supersymmetry charges (which transform in the same way as fermions)
contain four singlets with respect to $SO(3)_E'$; these are the unbroken supercharges
of the $\CN=(2,2)$ supersymmetric theory on $\R^2$.
It is also clear from \eqref{twistedspectrum} that after the topological twist
a triplet of the Higgs scalars, $\phi$, becomes a 1-form on $M$.
It can be naturally combined with the components of the original gauge field $A$
into a complexified gauge connection $\CA = A + i \phi$.
Moreover, the supersymmetry equations in the twisted theory become
the flatness condition for the $G_{\C}$ gauge connection $\CA$:
\be
\CF = d \CA + \CA \wedge \CA = 0 \,,
\ee
so that the classical vacua of the ``effective'' two-dimensional
$\CN=(2,2)$ supersymmetric field theory \eqref{conjduality}
are precisely the flat $G_{\C}$ connections on~$M$.

It is very well known that (partially) twisted topological gauge theories
can be realized on the world-volume of D-branes (partially) wrapped on
supersymmetric cycles in special holonomy manifolds \cite{BSV}.
In particular, a partial twist of the five-dimensional $\CN=2$ super-Yang-Mills theory
relevant to our discussion here is realized on the world-volume of the D4-branes
supported on a special Lagrangian 3-cycle $M$ in a Calabi-Yau 3-fold $X$:
\be
\begin{matrix}
{\mbox{\rm space-time:}} & \qquad & \R^4 & \times & X \\
& \qquad & \cup &  & \cup \\
{\mbox{\rm D4-brane:}} & \qquad & \R^2 & \times & M
\end{matrix}
\label{sureng}
\ee
Locally, in the vicinity of $M$, the geometry of $X$ always looks like $T^* M$,
where the D4-branes are supported on the zero section $M \subset T^* M$
and the normal bundle is parametrized by the Higgs fields~$\phi$.\\

Our original motivation for considering a (partially twisted) topological field theory
on the D4-branes was to understand the proposed duality \eqref{conjduality}
between quantum invariants of 3-manifolds and $\CN=(2,2)$ supersymmetric
gauge theories in two dimensions.
Now, let us look at the system \eqref{sureng} from the vantage point
of the effective four-dimensional theory on $\R^4$, obtained via
compactification of type II string theory on a Calabi-Yau 3-fold $X$.
Clearly, this four-dimensional theory has $\CN=2$ supersymmetry.
In fact, many familiar $\CN=2$ gauge theories in four dimensions
can be {\it geometrically engineered} in this way,
via compactification on non-compact toric 3-folds~\cite{engineering}.

In this setup, D4-branes supported on a supersymmetric 3-cycle $M$
in a Calabi-Yau manifold $X$ represent non-local operators in
the four-dimensional $\CN=2$ gauge theory. To be more precise,
these operators are localized on a two-dimensional surface $\R^2 \subset \R^4$
and preserve half of the supersymmetry.
In other words, they are half-BPS surface operators of the
four-dimensional $\CN=2$ gauge theory \cite{Ramified}.
Moreover, in this framework
the ``effective'' $\CN=(2,2)$ supersymmetric field theory \eqref{conjduality}
is simply a two-dimensional theory on the surface operator.

Surface operators in $\CN=2$ supersymmetric gauge theories
play a key role in the gauge theory approach to homological knot invariants \cite{RTN}.
More recently, they have been studied in the context of the AGT correspondence \cite{AGGTV},
where it has been argued that a certain class of half-BPS surface operators
corresponds to degenerate vertex operators in the Liouville CFT.

\begin{figure}[t] \centering \includegraphics[width=5.0in]{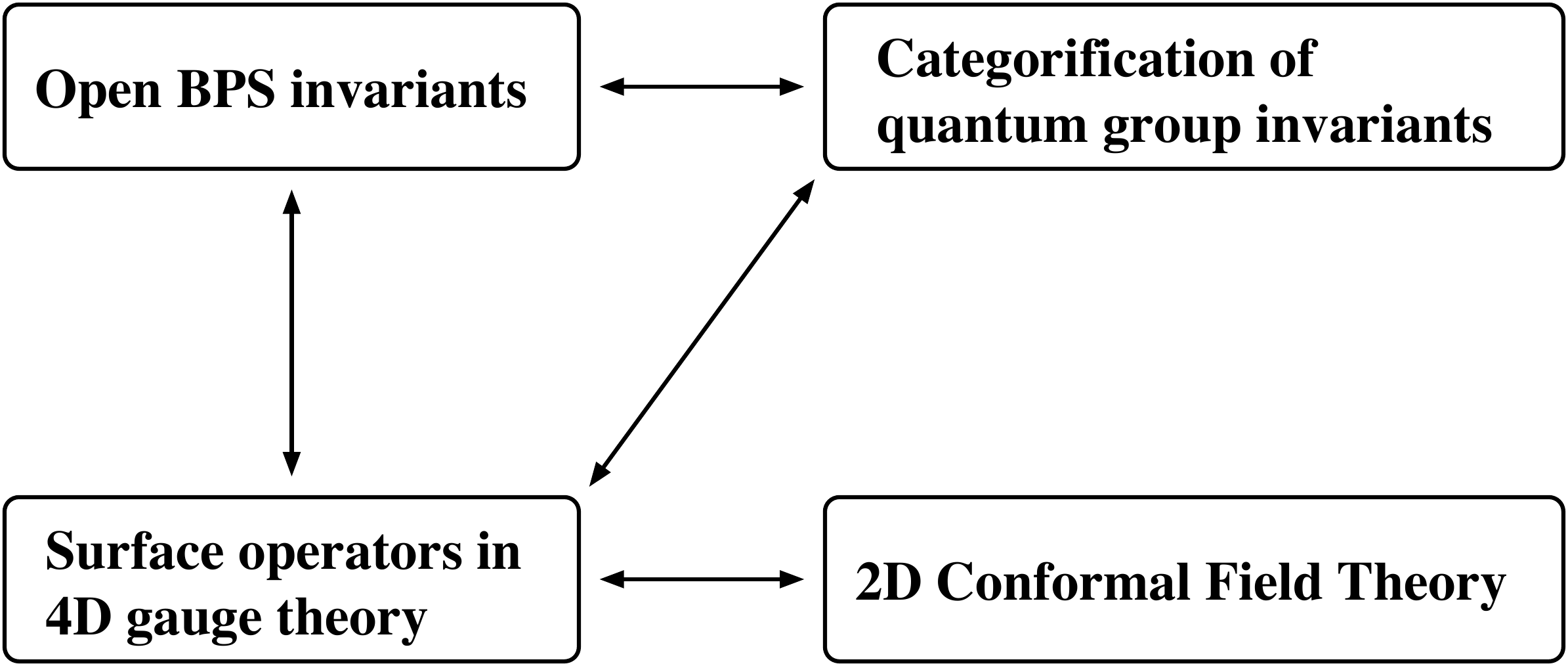}
\caption{The duality web.\label{bigpicture}}
\end{figure}

As a first step towards the duality \eqref{conjduality}, in this paper
we focus on a simple class of (non-compact) Lagrangian 3-manifolds,
which enjoy the toric symmetry of $X$.
Following the dualities in Figure \ref{bigpicture},
we identify the corresponding half-BPS surface operators
in $\CN=2$ four-dimensional gauge theories and
the dual vertex operators in two-dimensional conformal field theories.
Further aspects of the correspondence \eqref{conjduality} will be discussed elsewhere~\cite{inprogress}.


\section{Surface Operators and Vortex Counting}
\label{sec:gauge}

In this section, we study non-perturbative effects in $\CN=2$ supersymmetric
gauge theories in the presence of surface operators,
which --- somewhat like Wilson and 't Hooft operators --- are non-local operators supported
on submanifolds in the four-dimensional space-time, namely on two-dimensional surfaces \cite{Ramified}.
Whereas line operators in general are labeled by discrete data (such as weights and coweights),
surface operators are typically labeled by both discrete and continuous parameters.
The choice of discrete parameters is somewhat analogous to that of line operators,
while continuous labels are a new feature of surface operators:
as explained in \cite{Ramified}, much interesting physics associated with
surface operators can be understood through these continuous parameters.

In four dimensions, surface operators are rather special since their support
is exactly mid-dimensional. In other words, along the support $D$ of a surface
operator the tangent bundle of the space-time 4-manifold $W$ splits as $\R^2 \oplus \R^2$.
In particular, both tangent and normal bundles of $D$ have dimension 2,
which agrees with the degree of the curvature 2-form $F$,
suggesting that four-dimensional gauge theory should be a perfect home for surface operators.
Indeed, some of the continuous parameters of surface operators
come from integrating the curvature 2-form $F$ along $D$,
while some come from utilizing the components of $F$ in the directions of the normal bundle.
These continuous parameters are usually denoted by $\eta$ and $\alpha$, respectively.

The middle-dimensional nature of surface operators also makes them ideal observables
for the problem of equivariant instanton counting in four-dimensional gauge theories
with $\CN=2$ supersymmetry \cite{Nekrasov}.
Indeed, half-BPS surface operators supported on $D = \R^2$ preserve the same
symmetries and supersymmetries as the $\Omega$-deformation
of $\CN=2$ supersymmetric gauge theory on $W = \R^4$.
We recall that the $\Omega$-deformation is defined by the action of
the rotation symmetry group ${\bf T}^2_E = SO(2)_1 \times SO(2)_2 \subset SO(4)$
on $\R^4 = \R^2 \times \R^2$.
This symmetry is preserved by a surface operator supported on one of the $\R^2$'s.
Therefore, as in \cite{Nekrasov}, one can consider equivariant integrals
\be
\label{zkmdef}
Z^{{\rm inst}}_{k,\m} (\epsilon_1, \epsilon_2) = \oint_{\CM_{k,\m}} 1
\ee
on the moduli space, $\CM_{k,\m}$, of ``ramified instantons'' on $W \setminus D$
labeled by the ordinary instanton number $k := c_2 (E)$ 
and the monopole number
\begin{equation}
\label{mnumber}
\m = \frac{1}{2\pi} \int_{D} F \qquad\qquad {\rm (``monopole~number'')}
\end{equation}
that measures the magnetic charge of the gauge bundle $E$ restricted to $D$.
Relegating further details to the rest of this section, here we only mention that
the equivariant integrals \eqref{zkmdef} are rational functions of $\epsilon_1$ and $\epsilon_2$,
the generators of the equivariant cohomology
$H^*_{{\bf T}^2_E} ({\rm pt}) \cong \C [\epsilon_1, \epsilon_2]$ of a point.
We assemble these integrals in a generating function $Z^{{\rm inst}} (\epsilon_1, \epsilon_2, \ldots)$
that, besides $\epsilon_1$ and $\epsilon_2$, depends on all other parameters
of the gauge theory and the surface operator.

One of the main goals of the present paper is to compute the instanton partition
function $Z^{{\rm inst}} (\epsilon_1, \epsilon_2, \ldots)$
in $\CN=2$ gauge theories in the presence of surface operators.
One can do it either directly or by using various dualities that relate
$\CN=2$ gauge theories to other systems, such as type II strings on Calabi-Yau 3-folds
and conformal field theories in two dimensions, as in Figure \ref{bigpicture}.
Therefore, as a necessary prerequisite to such computations, we need to extend
these dualities to surface operators and identify the corresponding objects
in the dual systems. This will be done in sections \ref{sec:bps} and \ref{sec:cft}.


\subsection{Half-BPS Surface Operators in $\CN=2$ Gauge Theories}

As we explained above,
in general surface operators are labeled by a set of discrete and continuous parameters.
In the pure $\CN=2$ super Yang-Mills theory with gauge group $G$ --- which will be our main example ---
there exists a simple class of half-BPS surface operators for which
the discrete parameter corresponds to a choice of the Levi subgroup $\LL \subseteq G$,
while the continuous parameters form a $\Weyl_{\LL}$-invariant pair
\be
\label{aeta}
(\alpha,\eta) \in \TT \times \dual{\TT} \,,
\ee
where $\TT$ is a maximal torus of $G$,
$\dual{\TT}$ is a maximal torus of the dual group $\LG$,
and $\Weyl_{\LL}$ is the Weyl group of $\LL$.
The Levi subgroup $\LL$ can be interpreted as part of the gauge symmetry group
preserved by the surface operator.
More precisely, in the presence of a surface operator supported on $D$
the gauge theory path integral is defined by allowing
$\LL$-valued gauge transformations along $D$.
The extreme choices --- which will be referred to as the maximal and minimal --- are $\LL = G$ and $\LL = \TT$.

The continuous parameter $\alpha$ defines a singularity for the gauge field:
\be
\label{asing}
A = \alpha d \theta + \ldots \,,
\ee
where $x^2 + i x^3 = r e^{i \theta}$ is a local complex coordinate,
normal to the surface $D \subset W$, and the dots stand for less singular terms.
In order to obey the supersymmetry equations, the parameter $\alpha$
must take values in the $\LL$-invariant part of $\frak t$,
the Lie algebra of the maximal torus $\TT$ of $G$.
Moreover, gauge transformations shift values of $\alpha$
by elements of the cocharacter lattice, $\Lambda_{{\rm cochar}}$.
Hence, $\alpha$ takes values in $\TT = {\frak t} / \Lambda_{{\rm cochar}}$.

In addition to the ``magnetic'' parameter $\alpha$ surface operators of this type
are also labeled by the ``electric'' parameter $\eta$,
which enters the path integral through the phase factor
\be
\label{etaphase}
\exp \left( i \eta \cdot \m \right)
\ee
The monopole number $\m$ takes values in the $\Weyl_{\LL}$-invariant part of the cocharacter lattice,
$\Lambda_{{\rm cochar}}$, which we denote as $\Lambda_{\LL}$.
The lattice $\Lambda_{\LL}$ is isomorphic to the second cohomology group
of the flag manifold $G/\LL$, a fact that will be useful to us later.
Therefore,
\begin{equation}
\label{mlattice}
\m ~\in~ \Lambda_{\LL} \cong H_2 (G/\LL ; \Z)
\end{equation}
and the character $\eta$ of the abelian magnetic charges $\m$
takes values in ${\rm Hom} (\Lambda_{\LL},U(1))$, which is
precisely the $\Weyl_{\LL}$-invariant part of $\dual{\TT}$.

To keep things simple, in what follows we mostly focus on
$\CN=2$ super Yang-Mills theory with gauge group $G=SU(N)$ (or, a closely related theory with $G = U(N)$)
and consider half-BPS surface operators with the next-to-maximal Levi subgroup, $\LL = SU(N-1) \times U(1)$.
Then, $G/ \LL \cong \cp^{N-1}$ and the lattice $\Lambda_{\LL}$ is one-dimensional,
so that $\m \in \Z$.
Of course, other choices of $G$ and $\LL$ are also interesting.\\

Now, let us consider the effect of the $\Omega$ deformation \cite{Nekrasov,NW}.
The path integral of the $\Omega$-deformed $\CN=2$ gauge theory
in four dimensions localizes on solutions to the BPS equations \cite{Nekrasov}.
Without surface operators, these are the familiar instanton equations, $F^+ = 0$,
so that the resulting partition function\footnote{Note,
we do not include the perturbative part in the definition of $Z^{{\rm inst}}$.}
is a power series expansion in $\Lambda^N$:
\be
\label{zinst}
Z^{{\rm inst}} (\epsilon_1, \epsilon_2)
= \sum_{k=0}^{\infty} \Lambda^{2Nk} Z^{{\rm inst}}_k (\epsilon_1, \epsilon_2) \,,
\ee
with coefficients given by equivariant integrals on instanton moduli spaces.
(See~\cite{NYlectures} for an excellent set of lectures on this subject.)
In the presence of a half-BPS surface operator in $\CN=2$ gauge theory,
the BPS equations are the modified instanton equations~\cite{RTN}:
\be
F^+ = 2 \pi \alpha (\delta_D)^+
\label{surfselfdual}
\ee
where $\delta_D$ is a two-form delta function that is Poincar\'e dual to $D$.
The moduli space, $\CM_{k,\m}$, of solutions to the BPS equations \eqref{surfselfdual}
has been extensively studied in the mathematical literature (see {\it e.g.} \cite{KM1,KM2,Braverman}).
For example, if $W$ is a closed 4-manifold and $G = SU(2)$, we have
\be
\dim~ \CM_{k,\m} = 8k + 4 \m - 3 (b^+ (W) - b^1 (W) + 1) - (2g(D) - 2) \,.
\ee
Closer to the subject of the present paper is the case\footnote{More generally,
one can take $W$ to be a complex surface with a divisor $D \subset W$
invariant under~${\bf T}^2_E$.} where $W = \R^4$ and $D = \R^2$,
which enjoys the action of the rotation symmetry group ${\bf T}^2_E = U(1)_1 \times U(1)_2$.
Then, the path integral of the $\Omega$-deformed $\CN=2$ gauge theory
has the following general form, {\it cf.} \eqref{zinst},
\begin{equation}
\label{zinstkm}
\boxed{~~
Z^{{\rm inst}} (a,\Lambda,\epsilon;\LL,t)
= \sum_{k=0}^{\infty} ~\sum_{\m \in \Lambda_{\LL}}~
\Lambda^{2Nk} e^{i t \cdot \m} ~Z^{{\rm inst}}_{k,\m} (a,\epsilon) ~~}
\end{equation}
where the coefficients $Z^{{\rm inst}}_{k,\m}$ are precisely the integrals \eqref{zkmdef}.
As was first pointed out by Braverman \cite{Braverman}, the double sum
over $k$ and $\m$ can be naturally combined into a sum over the elements
of the {\it affine} lattice $\Lambda_{\LL}^{{\rm aff}} \equiv \Z \times \Lambda_{\LL}$.

This observation can be regarded as a first hint that the physical quantities of
$\CN=2$ gauge theories in the presence of surface operators
are closely related to representation theory of affine Lie algebras.
Indeed, it was shown in \cite{Braverman} that in the presence of a surface operator
the instanton partition function \eqref{zinstkm} is an eigenfunction of
a deformation of the quadratic affine Toda Hamiltonian (see also \cite{AT}).
In what follows, we will discuss various generalizations of these results,
in particular, the so-called K-theoretic version of the partition function \eqref{zinstkm},
where $\epsilon_1$ and $\epsilon_2$ are replaced by $q_1$ and $q_2$.
This latter generalization is especially important since, as we explain below,
it is the K-theoretic version of the partition function that is directly
related to the geometric setup of section \ref{sec:bps} and to homological knot invariants.

In the classical limit,
\be
\Lambda \to 0 \,,
\ee
only the terms with instanton number $k=0$ contribute to the partition function \eqref{zinstkm},
which then counts only finite-energy supersymmetric field configurations localized
near the surface $D$. For this reason, the resulting partition function will be called
the {\it vortex partition function}:
\be
\label{zvortex-limit}
Z^{{\rm vortex}} 
= \sum_{\m \in \Lambda_{\LL}}~ e^{i t \cdot \m} ~Z^{{\rm inst}}_{0,\m} (a,\epsilon) \,.
\ee\\

In our previous discussion surface operators are defined somewhat
like 't Hooft operators, by removing the surface $D$ from the space-time
4-manifold and prescribing certain boundary conditions for the gauge field
(and, possibly, other fields) around~$D$.
Alternatively, one can try to define surface operators by introducing
additional degrees of freedom supported on~$D$.
In this definition, reminiscent of how one defines Wilson lines,
we introduce a two-dimensional quantum field theory on $D$ with a symmetry group~$G$
(which then can be gauged in coupling to the four-dimensional gauge theory on $\R^4$).
Therefore, depending on whether this two-dimensional field theory is a gauge theory
or a sigma-model, we obtain at least three general constructions of surface operators \cite{Ramified}:

$i)$ as a singularity for the four-dimensional gauge field;

$ii)$ via coupling to a two-dimensional gauge theory on $D$;

$iii)$ via coupling to a sigma-model on $D$.

\noindent
In what follows, we will show how to use each of these constructions to compute
the instanton partition function, $Z^{{\rm inst}}$, in the presence of a surface operator.
However, as emphasized in \cite{Ramified}, the last two methods have a limited range of validity.
For example, while the periodicity of the continuous parameters $\alpha$
is manifest in the first approach, it is not obvious in the last two methods.
Nevertheless, all three approaches agree for small values of $\alpha$,
and this will be sufficient for our analysis.

In addition, there exist various string constructions of surface operators.
The one relevant to our discussion here is based on the brane realization of
$\CN=2$ gauge theory in type IIA string theory \cite{Witten},
where basic surface operators (with next-to-maximal $\LL$)
can be described by introducing semi-infinite D2-branes \cite{AGGTV}:
\begin{align}\label{iiabranes}
\hbox{NS5} &: \quad 012345 \cr
\hbox{D4}  &: \quad 0123~~~6 \cr
\hbox{D2}  &: \quad 01~~~~~~~~7
\end{align}
Lifting this configuration to M-theory, we obtain a M5-brane
with world-volume $\R^4 \times \Sigma$ and a M2-brane (ending on the M5-brane)
with world-volume $\R^2 \times \R_+$.
Here, $D = \R^2$ is the support of the surface operator in
the four-dimensional space-time $W = \R^4$, and $\Sigma$ is the Seiberg-Witten curve
of the $\CN=2$ gauge theory.

\begin{figure}[t] \centering \includegraphics[width=5.0in]{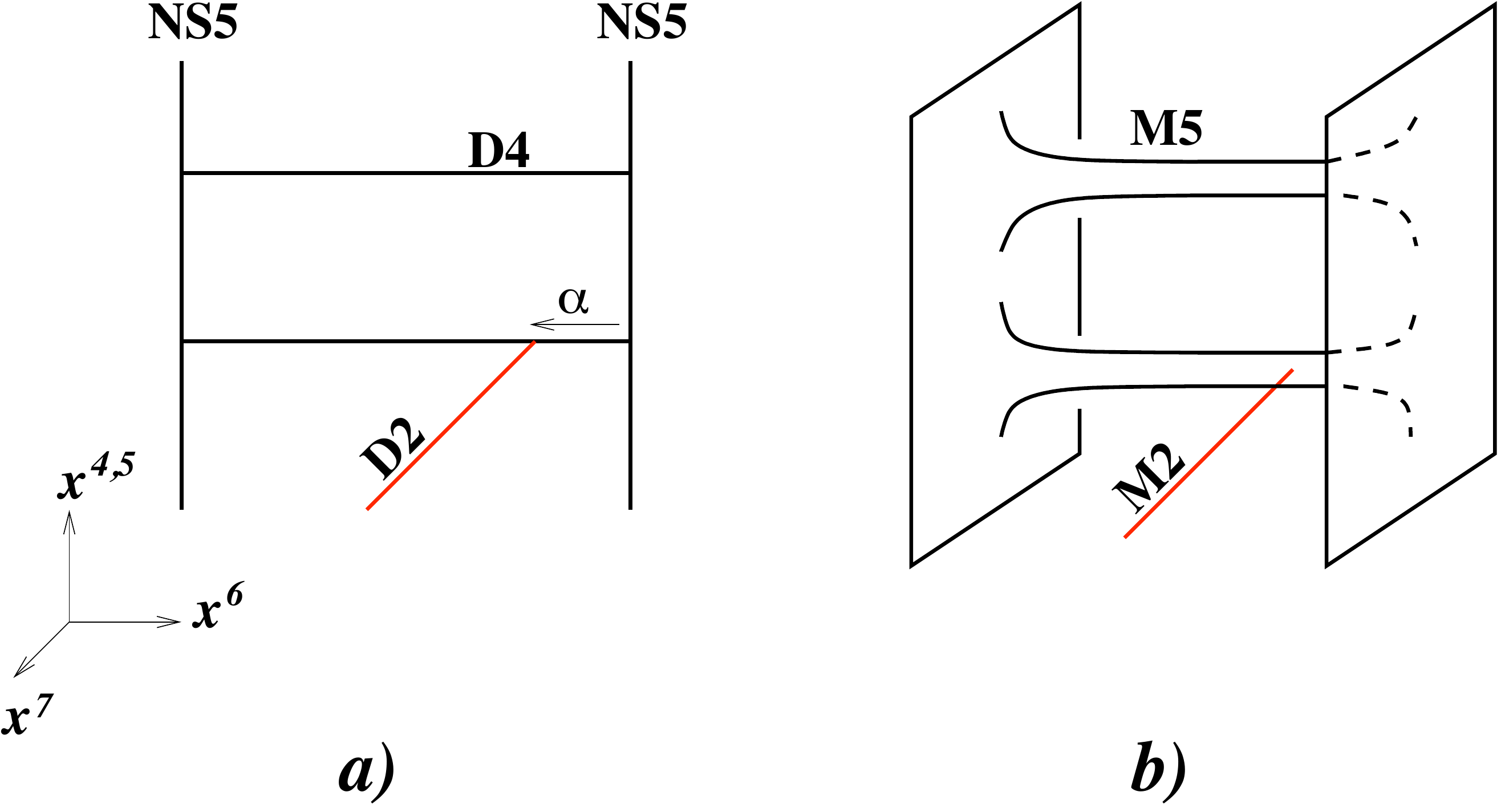}
\caption{The brane construction of $\CN=2$ super Yang-Mills theory
with a half-BPS surface operator in type IIA string theory $(a)$ and its M-theory lift $(b)$.
\label{branefig}} \end{figure}

In this construction, the M2-brane is localized along $\Sigma$
(the choice of the point $t \in \Sigma$ corresponds to the IR parameters
of the surface operator) and has a semi-infinite extent along the direction $x^7$,
as described in \eqref{iiabranes}.


\subsection{Two-dimensional Gauge Theory and Vortex Counting}
\label{sec:vortex}

Surface operators in gauge theory can be also defined by considering
two-dimensional field theory on a surface $D$ coupled to gauge theory
in four dimensions.

A basic example of a half-BPS surface operator is a surface operator
of the next-to-maximal Levi type. In a theory with gauge group $G=SU(N)$
this means the surface operator with Levi subgroup $\LL = SU(N-1) \times U(1)$,
and similarly for $G=U(N)$.
For this surface operator, the extra 2d degrees of freedom on the surface $D$
can be described by a quiver-like theory with gauge group factors
$U(N)$ and $U(1)$ representing 4d and 2d gauge bosons, respectively,
coupled through 2d chiral matter multiplets in the bifundamental
representation of $U(1) \times U(N)$,
\medskip
\begin{equation}
\label{4d2dqiver}
{\,\raisebox{-.08cm}{\includegraphics[width=3cm]{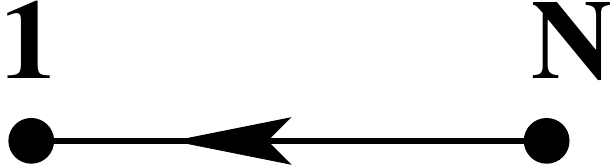}}\,}
\end{equation}
\smallskip

The path integral of the combined system of the 4d gauge theory
coupled to the extra 2d degrees of freedom on the surface operator
localizes to solutions of the BPS equations.

For a moment, let us focus on the two-dimensional part of this combined system,
where the four-dimensional gauge symmetry group $G$ plays the role of the global symmetry.
In our example of gauge theory with $G = U(N)$ and the simplest surface operators
(with ``next-to-maximal'' $\LL$), this two-dimensional sector
is described by the abelian gauge theory with $N$ massless chiral multiplets $\psi_i$.
The chiral fields $\psi_i$ have charge $+1$ under $U(1)$ gauge group
of the two-dimensional theory and transform in the (anti-)fundamental
representation of the $U(N)$ symmetry group. (From the vantage point of
the two-dimensional theory, this is equivalent to $N$ copies of charges
massless chiral multiplets.)
The corresponding BPS equations are the standard vortex equations:
\begin{eqnarray}
\label{vortexeq}
* F_{A} & = & i \sum_{i=1}^N \psi_i \bar \psi_i - i t \\
\bar \partial_A \psi_i & = & 0 \nonumber
\end{eqnarray}
where we introduced the FI term $t$.

We are interested in the moduli space, $\CM_{{\rm vortex}} (\m,N)$,
of solutions to the vortex equations \eqref{vortexeq}
with the magnetic flux \eqref{mnumber} through the surface $D$ equal to $\m$.
In the present example of surface operators with next-to-maximal $\LL$,
the gauge group in a two-dimensional theory on $D$ is the abelian group $U(1)$, so that $\m \in \Z$.
(Notice, the same conclusion follows more directly from the fact that $\Lambda_{\LL} = \Z$.)
In this case, the vortex moduli space $\CM_{{\rm vortex}} (\m,N)$
is a K\"ahler manifold of (real) dimension
\be
\dim~ \CM_{{\rm vortex}} (\m,N) = 2 \m N \,,
\label{mmndimension}
\ee
with a $U(1)_E \times U(N)$ isometry that we wish to use for equivariant integration
over $\CM_{{\rm vortex}} (\m,N)$.
Explicitly, the K\"ahler form on $\CM_{{\rm vortex}}$ can be written as
\be
\omega = \frac{i}{2\pi} \int_D d^2z \left( \delta A_{\bar w} \wedge \delta A_w + \delta \psi_w \wedge \delta \bar \psi_{\bar w} \right) \,.
\ee

Let us start our analysis of the vortex equations \eqref{vortexeq} with the familiar
case of the abelian Higgs model, {\it i.e.} a theory\footnote{Note, that
$N=1$ means considering gauge group $G=U(1)$ in four dimensions.} with $N=1$.
As is very well known (see {\it e.g.} \cite{JaffeT}),
the moduli space of $\m$ vortices in the abelian Higgs model on $D = \C$
is\footnote{More generally, $\CM_{{\rm vortex}} (\m,1) = {\rm Sym}^{\m} (D)$.}
\be
\label{vkone}
\CM_{{\rm vortex}} (\m,1) = {\rm Sym}^{\m} (\C) \equiv \C^{\m} / S_{\m} \,,
\ee
parametrized essentially by the positions of $\m$ vortices,
where the quotient by the symmetric group $S_{\m}$
reflects the fact that vortices are indistinguishable.
The rotation group $U(1)_E$ acts on this space in a natural way,
by equal phase rotations on all factors in the symmetric product.
In order to compute its character, $Ch_q (\CM_{{\rm vortex}})$,
it is convenient to note that ${\rm Sym}^{\m} (\C) \cong \C^{\m}$.
An easy way to see this is to realize the space ${\rm Sym}^{\m} (\C)$
as the space of monic polynomials of degree $\m$,
$$
f(x) = \prod_{j=1}^{\m} (x - x_j) = x^{\m} + a_1 x^{\m-1} + \ldots + a_{\m} \,.
$$
On the one hand, the space of such polynomials is a space of all $x_i$'s
modulo the action of the symmetric group $S_{\m}$, so that
$$
U(1)_E : \quad x_j \to e^{i \theta} x_j \,.
$$
On the other hand, the same space is parametrized by the values of
the coefficients $\{ a_1, \ldots , a_{\m} \} \in \C^{\m}$ on which $U(1)_E$ acts as
$$
U(1)_E : \quad a_n \to e^{i n \theta} a_n \,.
$$
Therefore, in a theory with gauge group $G = U(1)$, the moduli space
of abelian vortices localized {\it on} the surface operator is
a vector space, $\CM_{{\rm vortex}} (\m,1) \cong \C^{\m}$,
on which $U(1)_E$ acts with eigenvalues $w_n = n$, where $n = 1, \ldots \m$.

Now we are ready to compute the equivariant character of the $U(1)_E$ action
on the moduli space $\CM_{{\rm vortex}}$.
In general, if $\CV = \C^{\m}$ is a vector space on which $U(1)_E$ acts with
eigenvalues $(w_1, \ldots, w_{\m})$, its character has the form
(see \cite{ABCD} for a very nice exposition in the context of the equivariant instanton counting):
$$
Ch_q (\CV) \equiv \Tr_{\CV} (q)
= \frac{1}{(1-q^{w_1}) \ldots (1-q^{w_{\m}})} \,.
$$
Applying this to the problem at hand we find
$$
Ch_q (\CM_{{\rm vortex}} (\m,N=1))
= \frac{1}{(1-q)(1-q^2) \ldots (1-q^{\m})} \,.
$$

Introducing the generating function,
$$
Z^{{\rm vortex}} (z;q) := \sum_{\m} z^m Ch_q (\CM_{{\rm vortex}} (\m)) \,,
$$
we obtain
\be
\label{vortexqdilog}
Z^{{\rm vortex}} (z;q)
= \sum_{\m=0}^{\infty} \frac{z^{\m}}{(1-q) \ldots (1-q^{\m})}
= \prod_{i=1}^{\infty} (1 - q^{-i} z) \,.
\ee

To be more precise, what we computed is the K-theory version of the partition function.
(In the context of instanton counting, it is also known as the 5d version.)
The usual, homological version can be obtained by writing
\be
\label{qlimit}
q = e^{-\beta \hbar}
\ee
and taking the limit $\beta \to 0$ (along with a redefinition $z \to \beta z$).
In this limit, we find
\be
\label{zvortexuone}
Z^{{\rm vortex}} (z;\hbar) = \sum_{\m} z^{\m} Z^{{\rm vortex}}_{\m}
\ee
where
\be
\label{zuonem}
Z^{{\rm vortex}}_{\m} = \frac{1}{\hbar^{\m} \m !} \,.
\ee

Notice, the partition function \eqref{zvortexuone} can be also written as
a multiple contour integral
\be
\label{zvortexoint}
Z^{{\rm vortex}} (z;\hbar)
= \sum_{\m=0}^{\infty} \frac{z^{\m}}{\m! \hbar^{\m}}
\oint \prod_{j=1}^{\m} {d \varphi_j \over 2\pi i}
~\prod_{i \ne j}^{\m} {\varphi_i - \varphi_j \over \varphi_i - \varphi_j - \hbar}
\prod_{i=1}^{\m} {1 \over \varphi_i}
\ee
that has a simple interpretation in the two-dimensional gauge theory
on the surface operator. Indeed, the contour integral \eqref{zvortexoint}
is a result of localizing the integral
\begin{equation}
\label{zvortex}
Z^{{\rm vortex}} (z;\hbar)
= \sum_{\m} z^{\m}
\oint_{\CM_{{\rm vortex}} (\m)} 1
\end{equation}
further with respect to the action of~$U(\m)$
and then integrating over $U(\m)$ eigenvalues~$\varphi_i$.
The resulting contour integral \eqref{zvortexoint} receives contributions from the poles at
\be
\label{polesuone}
\varphi_{i_j} = (j-1) \hbar \,, \qquad\qquad j=1,\ldots,\m
\ee
that can be identified with the fixed points of the $U(\m)$ symmetry acing on $\CM_{{\rm vortex}} (\m)$.

The integrand of \eqref{zvortexoint} is a product of two factors, which come from
the basic ingredients of the abelian Higgs model, namely from a gauge multiplet
and a charged chiral multiplet. In general, the vortex partition function of
a two-dimensional $\CN=(2,2)$ supersymmetric gauge theory can be expressed as
as a contour integral like \eqref{zvortexoint} where the gauge multiplet contributes
to the integrand a factor
\be
\label{zintgauge}
\CZ^{2d,{\rm gauge}} (\varphi) = {1 \over \m!} {1 \over \hbar^{\m}}
\prod_{i \ne j}^{\m} {\varphi_i - \varphi_j \over \varphi_i - \varphi_j - \hbar}
\ee
and a massive chiral multiplet in the fundamental representation
contributes a factor
\be
\CZ^{2d,{\rm fund}} (\varphi) = \prod_{i=1}^{\m} {1 \over \varphi_i + m}
\ee
Here, $m$ is the mass parameter.
It is easy to see that in the abelian Higgs model, which is a theory
of a single vector multiplet and one massless chiral multiplet,
the integrand of \eqref{zvortexoint} is indeed a product of
$\CZ^{2d,{\rm gauge}}$ and $\CZ^{2d,{\rm fund}}$, with $m=0$.
Likewise, the contribution of the anti-fundamental chiral multiplet is
\be
\CZ^{2d,{\rm anti-fund}} (\varphi) = \prod_{i=1}^{\m} (\varphi_i + m - \hbar) \,.
\ee
After summing over the residues of the poles \eqref{polesuone},
such fundamental matter multiplets contribute to the vortex partition
function $Z^{{\rm vortex}}_{\m}$ the following products
\be
\label{vortfund}
{\rm fundamental} : \qquad \prod_{j=1}^{\m} \frac{1}{m + j \hbar}
\ee
and
\be
\label{vortantifund}
{\rm anti-fundamental} : \qquad \prod_{j=0}^{\m-1} (m + j \hbar)
\ee
respectively.
For instance, including an extra anti-fundamental chiral multiplet of mass $m$
in the abelian Higgs model leads to the following contour integral
\be
\label{zzzzanti}
Z^{{\rm vortex}} (z;\hbar)
= \sum_{\m=0}^{\infty} z^{\m} \oint \prod_{j=1}^{\m} {d \varphi_j \over 2\pi i}
~\CZ^{2d,{\rm gauge}} \cdot \CZ^{2d,{\rm fund}} \cdot \CZ^{2d,{\rm anti-fund}}
\ee
which integrates to the partition function
\be
\label{U1massvortex}
Z^{{\rm vortex}} (z;\hbar)
= \sum_{\m=0}^{\infty} \frac{z^{\m}}{\m! \hbar^{\m}} \prod_{j=0}^{\m-1} (m + j \hbar)
= (1 - z)^{-\frac{m}{\hbar}} \,.
\ee

Returning to our basic example of surface operators
in four-dimensional $U(N)$ gauge theory, we wish to perform
vortex counting in the ``two-dimensional part'' of the combined
4d-2d theory described by the quiver diagram \eqref{4d2dqiver}.
The relevant two-dimensional theory in this case
is a simple generalization of the abelian Higgs model,
namely a two-dimensional $U(1)$ gauge theory with $N$ charged chiral multiplets.
The contribution to the vortex partition function $Z^{{\rm vortex}}_{\m}$
is a product of factors of the form~\eqref{vortfund}:
\be
\label{zmforcpn}
Z^{{\rm vortex}}_{\m} = \prod_{i=1}^N \prod_{j=1}^{\m} \frac{1}{a_i + j \hbar}
\ee
where we used the fact that the Coulomb branch parameters $a_i$
of the four-dimensional gauge theory play the role of mass
parameters for the bi-fundamental fields in the 2d theory
on the surface operator.
Note, that specializing to $N=1$ and $a_i=0$ gives
the vortex partition function
in the abelian Higgs model.\\



\subsection{Vortex Counting from Instanton Counting}

So far we discussed vortex counting in a two-dimensional gauge theory
that describes the physics of a surface operator in four-dimensional gauge theory.
In particular, we defined the vortex partition function $Z^{{\rm vortex}} (\hbar)$
by focusing our attention on the two-dimensional part of the combined 2d-4d system,
where the four-dimensional gauge symmetry group $G$ plays the role of the global symmetry.

At least in some cases, the vortex partition function of this two-dimensional
gauge theory on $D$ can be computed as a certain limit of another, auxiliary
$\CN=2$ four-dimensional gauge theory on $D \times \R^2$ with the same gauge
group as in the two-dimensional $\CN=(2,2)$ theory that we wish to study.
(In particular, we emphasize that the gauge group of this {\it parent}
four-dimensional gauge theory has nothing to do with the gauge group $G$
of the original $\CN=2$ theory whose surface operators we study.)
For example, for the abelian Higgs model the auxiliary four-dimensional
gauge theory is a supersymmetric Maxwell theory (with gauge group $U(1)$).

The relation between $Z^{{\rm vortex}} (\hbar)$ in a two-dimensional gauge theory
and $Z^{{\rm inst}} (\epsilon_1, \epsilon_2)$ in a parent four-dimensional theory
has the form
\be
\label{4d2dreduction}
Z^{{\rm vortex}} (\hbar) =
\exp \Big[ \lim_{\epsilon_2 \to 0} \big( \epsilon_2 \cdot \log Z^{{\rm inst}} (\epsilon_1 = \hbar, \epsilon_2) \big) \Big]
\ee
and admits a very simple interpretation.
Indeed, equivariant integration can be thought of as a way to regularize
a two-dimensional (resp. four-dimensional) gauge theory on $\R^2$ (resp. $\R^4$),
such that
$$
{\rm Vol} (\R^2) = \int_{\R^2} 1 = \frac{1}{\hbar}
\qquad {\rm and} \qquad
{\rm Vol} (\R^4) = \int_{\R^4} 1 = \frac{1}{\epsilon_1 \epsilon_2} \,.
$$
As a result, the path integral in a two-dimensional (resp. four-dimensional)
gauge theory in the $\Omega$-background looks like
\be
\label{zw}
Z^{{\rm vortex}} (\hbar) = \exp \Bigl(\, \frac{1}{\hbar} \CW (\hbar) \, \Bigr)
\ee
and
\be
\label{zf}
Z^{{\rm inst}} (\epsilon_1, \epsilon_2)
= \exp \Bigl(\, \frac{1}{\epsilon_1 \epsilon_2} \CF (\epsilon_1, \epsilon_2) \, \Bigr)
\ee
where the twisted superpotential $\CW (\hbar)$ and the prepotential $\CF (\epsilon_1, \epsilon_2)$
are regular functions of $\hbar$, $\epsilon_1$, and $\epsilon_2$.
By comparing \eqref{zw} and \eqref{zf} it is easy to see that the limit $\epsilon_2 \to 0$
corresponds to the dimensional reduction along $\R^2_{\epsilon_2}$,
and the prepotential $\CF$ of the four-dimensional theory essentially
becomes the superpotential $\CW$ of the corresponding two-dimensional theory,
up to a universal factor $\frac{1}{\epsilon_2} = {\rm Vol} (\R^2_{\epsilon_2})$
which is removed in the limit \eqref{4d2dreduction}.
A closely related version of \eqref{4d2dreduction}
was studied earlier by Braverman and Etingof \cite{BravermanE}
and, more recently, by Nekrasov and Shatashvili \cite{NShatashvili}
in connection to quantum integrable systems (see also \cite{NW,AT}).

Let us illustrate the relation \eqref{4d2dreduction} in a basic example
of the abelian Higgs model in two dimensions. In this case, the parent four-dimensional
gauge theory is a pure $\CN=2$ supersymmetric gauge theory with gauge group $U(1)$.
The instanton partition function of this theory has been extensively
studied (see {\it e.g.} \cite{LMN,NO}) and can be expressed as a sum over
a single 2d partition / Young tableaux $\lambda$:
\be
Z^{{\rm inst}} (\epsilon_1, \epsilon_2)
= \exp \Bigl(\, \frac{\Lambda^2}{\epsilon_1 \epsilon_2} \, \Bigr)
= \sum_{k=0}^{\infty} \Lambda^{2k} \sum_{|\lambda| = k} \frac{\mu^2 (\lambda)}{(\epsilon_1 \epsilon_2)^k}
\label{zfourduxz}
\ee
where $\mu^2 (\lambda)$ denotes the Plancherel measure of $\lambda$.
Substituting this into the relation \eqref{4d2dreduction} we obtain
the vortex partition function of the abelian Higgs model:
\be
Z^{{\rm vortex}} (\hbar) = \exp \Bigl(\, \frac{\Lambda^2}{\hbar} \, \Bigr)
= \sum_{\m=0}^{\infty} \Lambda^{2\m} \frac{1}{\m! \hbar^{\m}}
\label{ztwoduxz}
\ee
which is a sum over a ``1d partition'' $\m$,
{\it i.e.} a single sum over $\m = 0,1,2, \ldots$.
This result is in complete agreement with \eqref{zvortexuone} -- \eqref{zuonem}
found earlier via direct analysis of the vortex moduli spaces.

Note that when we go from the instanton partition function $Z^{{\rm inst}}$
to the vortex partition function $Z^{{\rm vortex}}$ via the relation \eqref{4d2dreduction}
the instanton number $k$ turns into the monopole number $\m$ (and, correspondingly,
the parameter $\Lambda^N$ in the instanton partition function transforms into
the exponentiated FI parameter $z \sim \exp (it)$ of the two-dimensional theory).

In general,
assuming that the limits $\Lambda \to 0$ and $\epsilon_2 \to 0$ commute,
we can implement \eqref{4d2dreduction} directly in \eqref{zinst}:
\begin{eqnarray}
Z^{{\rm vortex}}
& = & \lim_{\epsilon_2 \to 0} \Bigl(\, Z^{{\rm inst}} \, \Bigr)^{\epsilon_2} \nonumber \\
& = & \lim_{\epsilon_2 \to 0} \Bigl(\, 1 + \epsilon_2 \Lambda^{2N} Z^{{\rm inst}}_1 +
\Bigl( \frac{\epsilon_2 (\epsilon_2-1)}{2} (Z^{{\rm inst}}_1)^2 + \epsilon_2 Z^{{\rm inst}}_2 \Bigr) \Lambda^{4N}
+ \ldots \, \Bigr) \nonumber \\
& = & 1 + \lim_{\epsilon_2 \to 0} \epsilon_2 Z^{{\rm inst}}_1 \Lambda^{2N}
+ \lim_{\epsilon_2 \to 0} \Bigl( \frac{\epsilon_2 (\epsilon_2-1)}{2} (Z^{{\rm inst}}_1)^2
+ \epsilon_2 Z^{{\rm inst}}_2 \Bigr) \Lambda^{4N} + \ldots \nonumber
\end{eqnarray}
It is easy to see that applying this {\it e.g.} to \eqref{zfourduxz}
gives the same vortex partition function \eqref{ztwoduxz} as in the previous analysis.


\subsection{Non-linear Sigma Model and Equivariant Quantum Cohomology}
\label{sec:NLSM}

In the above discussion, we used two different ways of describing
half-BPS surface operators in $\CN=2$ four-dimensional gauge theories:
one as a singularity for the gauge field along a surface $D$,
and another in terms of additional degrees of freedom on $D$ that
couple to the gauge field of the four-dimensional bulk theory.
Specifically, in the previous subsection we discussed a special
class of surface operators for which the extra degrees of freedom on $D$
in turn can be described by a gauge theory, now in two dimensions.

For example, in a four-dimensional gauge theory with a gauge group $G=SU(N)$
a basic half-BPS surface operator of Levi type $\LL = SU(N-1) \times U(1)$
can be described by a linear sigma-model on $D$ with gauge group $U(1)$
and $N$ chiral multiplets that form a fundamental $N$-dimensional
representation of the group $G$.
The field content of this combined 4d-2d system
can be conveniently summarized by a quiver diagram~\eqref{4d2dqiver}.
Note, in this description the parameter $\a$ of the surface operator is
the FI term for the $U(1)$ gauge field.
What happens if we try to integrate out the $U(1)$ gauge field?

As one very well knows, at low energies the resulting theory is
equivalent to a non-linear sigma-model on $D$
with the target space $\C^N / \! / U(1) \cong \cp^{N-1}$.
Furthermore, the FI parameter $\a$ of the linear sigma-model becomes
the K\"ahler parameter of the $\cp^{N-1}$.
More generally, a surface operator of Levi type $\LL$ can be similarly
represented by a non-linear sigma-model on $D$ with target space $G/ \LL$.
Note, the flag manifold $G/ \LL$ admits a K\"ahler metric
which allows to define a sigma-model with $\CN=(2,2)$ supersymmetry.
It also has a global symmetry group $G$ that can be gauged in coupling
to the gauge field of the four-dimensional bulk theory.
As a result, the corresponding surface operator supported on $D$
preserves half of the $\CN=2$ supersymmetry in four dimensions,
{\it i.e.} it is half-BPS.

To summarize, we obtain yet another description of surface operators
via non-linear sigma-model on $D$, {\it i.e.} a theory of maps from
the surface $D$ to a K\"ahler target manifold, such as $G/ \LL$.
{}From the vantage point of this theory, the partition function $Z^{{\rm vortex}}$
counts BPS field configurations of finite energy.
In fact, since in such configurations the fields at infinity (along $D$)
approach a constant value, $Z^{{\rm vortex}}$ effectively counts maps from
$\bar D = \cp^1$, a one-point compactification of $D$, into the target space $G/ \LL$.
Such maps are classified by the second homology class (or degree) $\m$
which, according to \eqref{mlattice}, can be identified with the monopole number
in our previous discussion.
Hence, the expansion \eqref{zvortexuone} can be viewed
as a generating function that counts maps
\be
\cp^1 \;\overset{\m}{\longrightarrow}\; G / \LL \,.
\ee

In Gromov-Witten theory, such generating function is usually called
the equivariant $J$-function (see {\it e.g.} \cite{GiventalKim,GiventalLee}).
Thus, in our basic example of surface operators in $SU(N)$ gauge theory
with $\LL = SU(N-1) \times U(1)$ the flag manifold is $G / \LL \cong \cp^{N-1}$,
and the K-theoretic version of the equivariant $J$-function has the form \cite{GiventalLee}:
\be
\label{jforcpn}
J (z; \hbar) = ~\sum_{\m}~ \frac{z^{\m}}{\prod_{i=1}^N \prod_{j=1}^{\m} (1-Q_i q^j)} \,.
\ee
The non-equivariant limit of this formula is obtained by setting $Q_1 = \ldots = Q_N = Q$,
whereas the cohomological limit is obtained by writing
\be
q = e^{- \beta \hbar} \,, \qquad Q_i = e^{- \beta a_i} \,,
\ee
and taking the limit $\beta \to 0$
(along with a redefinition $z \to \beta z$), as in eq. \eqref{qlimit}.
It is easy to see that, in the latter case, the equivariant $J$-function \eqref{jforcpn}
reduces to the vortex partition function \eqref{zmforcpn} computed earlier.


\section{Geometric engineering of surface operators}
\label{sec:bps}

We are now ready to begin to make contact with the geometric side of our story.
For four-dimensional gauge theory \emph{without} surface operators,
it is well known that the relation between graviphoton
and $\Omega$ backgrounds allows the K-theoretic (or five-dimensional)
instanton partition function to be re-expressed as a partition function
of BPS states in five-dimensional gauge theory \cite{HIV}.
In turn, these BPS states can be realized as BPS states of D-branes in a Calabi-Yau
threefold --- the noncompact Calabi-Yau threefold that is used
to \emph{geometrically engineer} the original $\CN=2$ supersymmetric gauge theory.
Our goal in this section is to extend this correspondence to surface operators.

\subsection{Surface operators from Lagrangian 3-manifolds}
\label{sec:GE}

Let us consider a four-dimensional $\CN=2$ gauge theory that can be geometrically
engineered via type IIA string ``compactification'' on a Calabi-Yau space $X$.
In other words, we take the ten-dimensional space-time to be $W \times X$,
where $W$ is a 4-manifold (where $\CN=2$ gauge theory lives) and $X$ is a Calabi-Yau space.
We recall that $X$ is non-compact
and toric, and that its toric polygon coincides with the Newton polygon of the Seiberg-Witten curve $\Sigma$.
In our applications we will simply take $W = \R^4$.

Aiming to reproduce half-BPS surface operators supported on $D = \R^2$,
we need an extra object that breaks part of the Lorentz symmetry (along $W = \R^4$)
and half of the supersymmetry.
It is easy to see that D4-branes supported on supersymmetric 3-cycles in $X$
provide just the right candidates \cite{OV}.
Indeed, if the world-volume of a D4-brane is $\R^2 \times \Lag$, where
\be
\begin{matrix}
{\mbox{\rm space-time:}} & \qquad & \R^4 & \times & X \\
& \qquad & \cup &  & \cup \\
{\mbox{\rm D4-brane:}} & \qquad & \R^2 & \times & \Lag
\end{matrix}
\label{surfeng}
\ee
and $\Lag$ is a special Lagrangian submanifold of $X$,
then such a D4-brane preserves exactly the right set of symmetries
and supersymmetries as the half-BPS surface operators discussed in section \ref{sec:gauge}.

A nice feature of this construction is that it is entirely geometric:
all the parameters of a surface operators (discrete and continuous)
are encoded in the geometry of $\Lag \subset X$.
In particular, it among the different choices of $\Lag$ we should be
able to find those which correspond to half-BPS surface operators of Levi type $\LL$
with the continuous parameters $\alpha$ and $\eta$.

We claim that a basic surface operator (with next-to-maximal $\LL$)
corresponds to a simple
special Lagrangian submanifold $\Lag \cong {\bf S}^1 \times \R^2$
invariant under the toric symmetry of $X$.
Such toric Lagrangian submanifolds
have been extensively used in the physics literature (\emph{cf.}\ \cite{OV, AV-discs, AKMV}),
so we can borrow many results to apply to our problem.
One way to see this is to start with a D-brane construction \eqref{iiabranes}
of such surface operator and apply a chain of dualities (see {\it e.g.} \cite{Karch}):

\begin{itemize}

\item[$a)$]
First, as in section \ref{sec:gauge}, we lift this configuration to M-theory (see Figure~\ref{branefig}$b$)
and then reduce it back to type IIA theory along the dimension $x^9$.
(As a result, we perform what is sometimes called a ``11-9 flip.'')
This yields a configuration of a NS5-brane on $\R^4 \times \Sigma$
and a D2-brane on $\R^2 \times \R_+$.

\item[$b)$]
A combination of T-duality and mirror symmetry maps type IIA theory back to type IIA theory,
and transforms the NS5-brane on $\R^4 \times \Sigma$ into a purely geometric background
of $\R^4 \times X$, where $X$ is a toric Calabi-Yau manifold (associated with $\Sigma$).
Since the net effect of T-duality and mirror symmetry is equivalent to dualizing
only two out of three directions of the SYZ fiber, this transformation
maps a D2-brane on $\R^2 \times \R_+$ into a D4-brane on $\R^2 \times \Lag$,
where $\Lag \subset X$ is a Lagrangian submanifold.
\end{itemize}

\begin{figure}[ht] \centering
\includegraphics[width=2in]{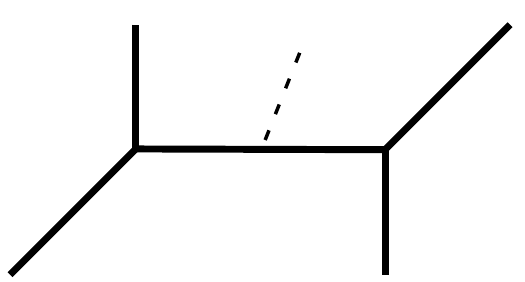}
\caption{\label{fig:U1basic} $U(1)$ toric geometry with a single Lagrangian brane.}
\end{figure}

Notice that, just like the original surface operator in $\CN=2$ gauge theory,
the final D4-brane is localized along the two-dimensional subspace, $\R^2 \subset \R^4$,
of the four-dimensional space-time, and preserves half of the supersymmetry.
Moreover, since it was obtained by dualizing two directions of the SYZ fiber
the Lagrangian submanifold $\Lag \subset X$ has two dimensions
in the toric fiber of $X$ and one dimension along the base, as in Figure \ref{fig:U1basic}.
As a result, we map the brane construction of $\CN=2$ gauge theory
with a half-BPS surface operator into geometric engineering of that gauge
theory with a Lagrangian D4-brane on~$\R^2 \times \Lag$.

This geometric engineering construction with surface operators allows us to relate instanton and vortex counting in gauge theory with BPS generating functions --- or topological string partition functions --- on a Calabi-Yau. Let us recall for a moment how this works in the absence of surface operators \cite{HIV}. Upon lifting the type IIA geometric engineering compactification on $\R\times X$ to M-theory on $\R\times \S^1_\beta\times X$, instantons in four dimension become associated with 1/2-BPS states in $\R\times \S^1_\beta$. In turn, these BPS states are engineered by M2-branes wrapping various 2-cycles in $X$. In five dimensions, BPS states transform under the little group $SO(4)\simeq SU(2)_L\times SU(2)_R$, and the maximal torus of this group, $\mathbf{T}^2 = SO(2)_L\times SO(2)_R$, can be identified with the rotation group $\mathbf{T}^2_E$ that defines the $\Omega$-deformation. In particular, $SO(2)_1\subset \mathbf{T}^2_E$ acts as the diagonal combination of $SO(2)_L\times SO(2)_R$ with charges $(+1,+1)$, and $SO(2)_2 \subset \mathbf{T}^2_E$ acts as the complementary combination with charges $(-1,1)$. If we then set $q_1=e^{\beta\epsilon_1}$ and $q_2=e^{-\beta\epsilon_2}$, and define $N^{j_L,j_R}_\beta$ to be the degeneracy of states with charge $\beta\in H_2(X;\Z)$ and $SU(2)_L\times SU(2)_R$ weights $(j_L,j_R)$, the generating function of BPS states
\be Z_{\rm BPS}^{\rm closed} := \prod_{\beta>0}\prod_{j_L,j_R\in \Z/2}\prod_{n_1,n_2=0}^\infty \left(1-q_1^{j_L+j_R+n_1+\frac12}q_2^{j_L-j_R+n_2+\frac12}Q^\beta\right)^{(-1)^{2j_L+2j_R}N^{j_L,j_R}_{\beta}}\, \label{closedBPS}
\ee
should reproduce K-theoretic instanton counting in the original 4-dimensional gauge theory. In other words,
\be Z_{\rm BPS}^{\rm closed}(Q;q_1,q_2) = Z_{K}^{\rm inst}(\Lambda,a_i;q_1,q_2)\,. \ee

Note that here $Q$ is a vector of the exponentiated, complexified K\"ahler parameters of $X$, \emph{i.e.}\ the masses of possible M2-brane states. In geometric engineering for a single gauge group, the Calabi-Yau space $X$ has the structure of a local surface that itself is fibered over a distinguished base $\cp^1_b$. Then there is a correspondence of parameters
\begin{equation}
\renewcommand{\arraystretch}{1.3}
\begin{tabular}{|@{\quad}c@{\quad}|@{\quad}c@{\quad}| }
\hline  {\bf gauge theory} & {\bf geometry of} $X$
\\
\hline
\hline $\Lambda$ & $Q_\Lambda \sim \exp(-{\rm Vol}(\mathbb{C}\mathbf{P}^1_b))=(\beta\Lambda)^{2N_c-N_f}$ \\
\hline \multirow{2}{*}{instanton number $k$}
& exponent of $Q_\Lambda$
\\
 &  (IIA worldsheet instanton number on $\cp^1_b$) \\
\hline Coulomb parameters $a_i$\,, & other K\"ahler moduli
\\
   bare masses $m_i$ & $Q_{a_i}=e^{-\beta a_i}$, $Q_{m_i}=e^{-\beta m_i}$
\\
\hline
\end{tabular}
\label{closed_params}
\end{equation}

We want to extend the relation between $Z^{\rm inst}$ and $Z^{\rm BPS}$ to configurations involving surface operators. The basic idea is the same. Upon a lift to M-theory on $\R^4\times \S^1_\beta\times X$, the D4-brane wrapping $\R^2\times\Lag$ becomes an M5-brane wrapping $\R^2\times \S^1_\beta\times \Lag$. The presence of this brane breaks $SU(2)_L\times SU(2)_R$ to the maximal torus $\mathbf{T}^2 =SO(2)_L\times SO(2)_R$. Vortices on the surface operator are lifted to BPS states in three dimensions (on the $\R^2\times \S^1_\beta$ part of the M5-brane), and these BPS states are realized by M2-branes with boundary on the M5-brane in $X$. The three-dimensional BPS states transform as representations of $SO(2)_L\times SO(2)_R$. Choosing%
\footnote{Recall that the surface operator in the $\Omega$-background must lie in either the $\R^2_{\epsilon_1}$ plane or the $\R^2_{\epsilon_2}$ plane.} %
the original surface operator to lie along $\R^2_{\epsilon_1}$, we find that the $SO(2)_1$ charge of a BPS state determines its three-dimensional spin $s$, while the $SO(2)_2$ charge is its \emph{R-charge} $r$.

The full BPS generating function factorizes as
\be Z^{\rm BPS} = Z^{\rm closed}_{\rm BPS} \times Z^{\rm open}_{\rm BPS}\,,\ee
where $Z^{\rm closed}_{\rm BPS}$ is as in \eqref{closedBPS}, and $Z^{\rm open}_{\rm BPS}$ counts three-dimensional states on the M5-brane. For a simple surface operator corresponding to a single D4 or M5-brane (\emph{i.e.}\ with a $U(1)$ worldvolume gauge theory), $Z^{\rm open}_{\rm BPS}$ takes the form
\be Z^{\rm open}_{\rm BPS} = \prod_{\mu>0}\prod_{s,r\in\IZ/2}\prod_{n=0}^\infty \left(1-q_1^{s+n+\frac12}q_2^{r+\frac12} z^\mu\right)^{-(-1)^{2s}N_{\mu}^{s,r}}\,.
\label{openBPS}
\ee
Here, $N^{s,r}_{\mu}$ is the degeneracy of states with M2-charge $\mu \in H_2(X,\Lag;\Z)$%
, and with spin $s$ and R-charge $r$. The quantity $z^\mu$ can be thought of as the exponentiated mass of the BPS state, determined classically by the volume of the appropriate M2-brane.
Note that there is only one extra product over $n$ in \eqref{openBPS}, since the BPS states are fixed to the M5-brane (\emph{cf.}\ \cite{HIV, AY}).

Similar to the closed correspondence \eqref{closed_params}, there is now a relation in the open sector:
\begin{equation}
\renewcommand{\arraystretch}{1.3}
\begin{tabular}{|@{\quad}c@{\quad}|@{\quad}c@{\quad}| }
\hline  {\bf surface operators} & {\bf geometry of} $\Lag \subset X$
\\
\hline
\hline $\Lambda_{\LL}$ & $H_1 (\Lag; \Z)/$torsion
\\

\hline cts. (FI) parameters \quad\quad & open string moduli $z \sim \beta^{\#}e^{it}$ \\
~~~~~$t = \eta_{{\rm eff}} + i \alpha_{{\rm eff}}$ & \quad(complexified holonomy eigenvalues)
\\

\hline \multirow{2}{*}{monopole number $\m_i$} & exponent of $z_i$ \\ & \quad(IIA disk instanton number)
\\
\hline
\end{tabular}
\label{open_params}
\end{equation}
The full BPS generating function $Z^{\rm BPS} = Z^{\rm closed}_{\rm BPS} \times Z^{\rm open}_{\rm BPS}$ is then expected to reproduce the K-theoretic version of the equivariant instanton partition function for a full four-dimensional theory with surface operators,
\be Z^{\rm BPS}(Q,z;q_1,q_2) = Z^{\rm inst}_{\rm K-theory}(\Lambda,a_i,t;q_1,q_2)\,. \label{BPSinst} \ee
An important special case of this relation is the limit $\Lambda\to 0$ (\emph{i.e.}\ $Q_\Lambda\to 0$) and $(q_1,q_2)\to (q,1)$. On the gauge theory side, this decouples the four-dimensional theory from the surface operator, and counts vortices on the surface operator with respect to two-dimensional rotations (but not R-charge), as discussed above \eqref{zvortex-limit}. We therefore expect that
\be Z^{\rm open}_{\rm BPS}(Q_\Lambda=0,Q_{a_i},z;q,1) = Z^{\rm vortex}_{\rm K-theory}(z,a_i;q)\,. \label{BPSvortex} \ee

In addition to the parameters listed in \eqref{closed_params}-\eqref{open_params}, there are two more discrete quantities that can enter the two sides of \eqref{BPSinst} and \eqref{BPSvortex}. When considering K-theoretic (or five-dimensional) instanton counting, a five-dimensional Chern-Simons term may be introduced, with a discrete coupling. Different choices of the coupling correspond to different fibrations of the local surface in $X$ over $\cp^1_b$. Similarly, a three-dimensional Chern-Simons term can be introduced when lifting a surface operator to five dimensions. Its discrete coupling corresponds to the \emph{framing} of the Lagrangian brane $\Lag$: a choice of IR regularization that compactifies $\Lag$ and is needed to properly define the BPS generating function \cite{AKV-framing, MV-framed}. Both of these discrete parameters disappear in the homological limit $\beta\to 0$.



\subsection{Vortices and BPS states}
\label{sec:GEexamples}

We next present some precision tests of the proposal $Z^{\rm BPS}=Z^{\rm inst}$, especially in the limit $Z^{\rm open} = Z^{\rm vortex}$. We begin by outlining some further details and conventions in the Lagrangian-brane construction of surface operators, and then continue to a selection of examples.

For the interested reader, a review of toric geometries and conventions pertaining to gauge theories \emph{without} surface operators can be found in Appendix \ref{app:GE}.

\begin{figure}[htb]
\centering
\includegraphics[width=5in]{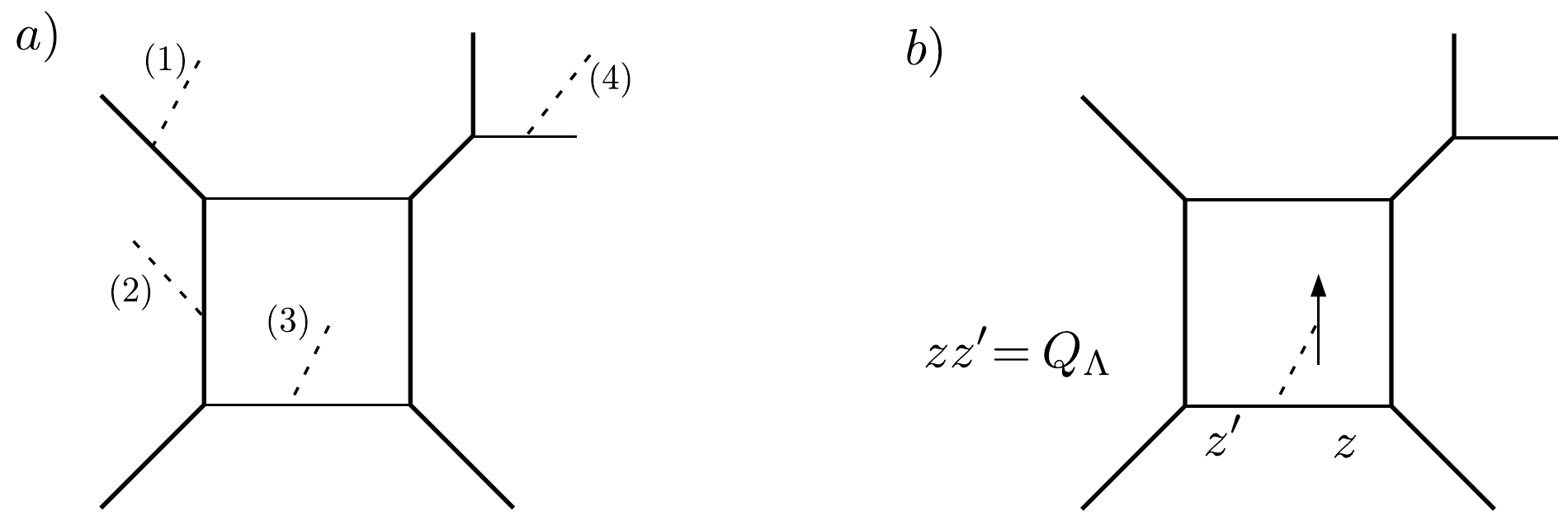}
\caption{a) Possible phases (placements) of a Lagrangian brane. b) The phase for instanton counting.}
\label{fig:branechoices2}
\end{figure}

If we are to add a Lagrangian brane to a toric geometry, like the one for SU(2) theory shown in Figure \ref{fig:branechoices2}(a), we can attach it in several different places. By considering the type IIB mirror to such a toric geometry, as in \cite{AV-discs, AKV-framing}, it becomes clear that the different placements should be smoothly connected. However, from the BPS-counting perspective, different choices give inequivalent partition functions --- essentially because inequivalent (and in a sense non-commuting) expansion parameters are involved.

For a surface operator in a $U(N)$ or $SU(N)$ gauge theory, the natural placement of a Lagrangian brane is on a horizontal, internal ``gauge leg'' (toric degeneration locus), as indicated in Figure \ref{fig:branechoices2}(b). The resulting BPS/instanton expansion will depend on two open parameters $z$ and $z'$ that satisfy $zz'=Q_\Lambda$. Either one can be identified with the exponentiated FI parameter $t$ of a two-dimensional gauge theory living on the surface operator, depending on how 2d-4d couplings are defined. Choosing $z=e^{-t}$, the instanton expansion can be rewritten as a series in $z$ and $z'=Q_\Lambda z^{-1}$. This is exactly the form expected for an instanton partition function, with powers of both $z$ and $z^{-1}$, as discussed for example in \cite{AGGTV, Gaiotto-surf}.

A brane placed on a gauge leg as in Figure \ref{fig:branechoices2}(b) also leads to a proper homological (or 4d) limit of K-theoretic instanton counting (or 5d BPS counting). For a $SU(N)$ gauge theory with $N_f$ fundamental flavors and $\bar{N}_f$ antifundamental flavors, the homological limit is accompanied%
\footnote{In the superconformal case $2N=N_f+\bar{N}_f$, $Q_\Lambda\sim $``$\Lambda$'' is simply identified with the UV coupling of the theory, $e^{2\pi i\tau}$; there is no additional rescaling.} %
by a rescaling $Q_\Lambda \to (\beta\Lambda)^{2N-N_f-\bar{N}_f}$. Including a surface operator with parameters $z$ and $z'$, we must also independently scale $z\to \beta^{N-N_f}z$ and $z'\to \beta^{N-\bar{N}_f}z'$ to obtain a nontrivial limit. This is completely consistent with the relation $zz'=Q_\Lambda$. For comparison, had we placed a Lagrangian brane on a ``flavor'' leg (position (4) in Figure \ref{fig:branechoices2}(a)), there would again be two parameters $z,z'$, now with $z'z^{-1}=Q_\Lambda$. It would be impossible to retain both of them in a homological limit --- in effect, the 2d theory on the surface operator would completely decouple from the 4d gauge theory.

\begin{figure}[htb]
\centering
\includegraphics[width=1.8in]{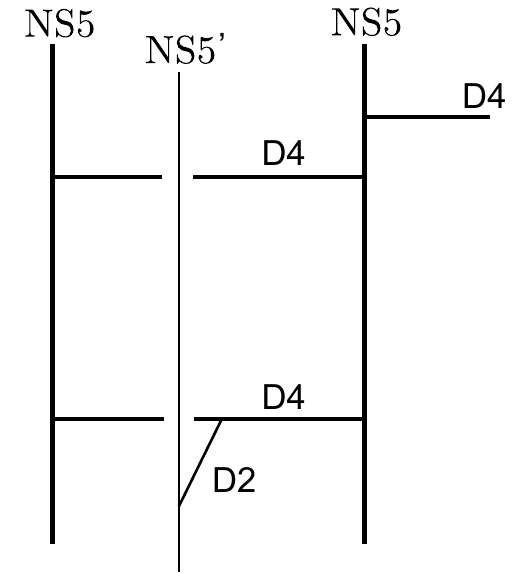}
\caption{Regulating the surface-operator theory in the brane construction. (This setup is dual to the geometry in Figure \protect\ref{fig:branechoices2}(b).)}
\label{fig:D2reg}
\end{figure}

Along with brane placement, we must also make a choice of brane framing for open toric geometries, related to the choice of three-dimensional Chern-Simons coupling. This is done by including an extra toric degeneration locus, in a plane parallel to but disjoint from that of the initial toric diagram \cite{AKV-framing, AKMV}. The extra degeneration locus compactifies the Lagrangian brane to an $\S^3$. We will always chose a vertical framing, as in Figure \ref{fig:branechoices2}(b), which should correspond to zero Chern-Simons coupling. It is useful to note that a similar regularization is typically introduced in the brane-engineering of surface operators, as in Figure \ref{fig:D2reg}: an extra NS5' brane, parallel but disjoint from the gauge theory branes, is added to give the 2d surface-operator theory a finite gauge coupling (\cf\ \cite{Hanany-Hori, AGGTV}).

\begin{figure}[htb]
\centering
\includegraphics[width=5in]{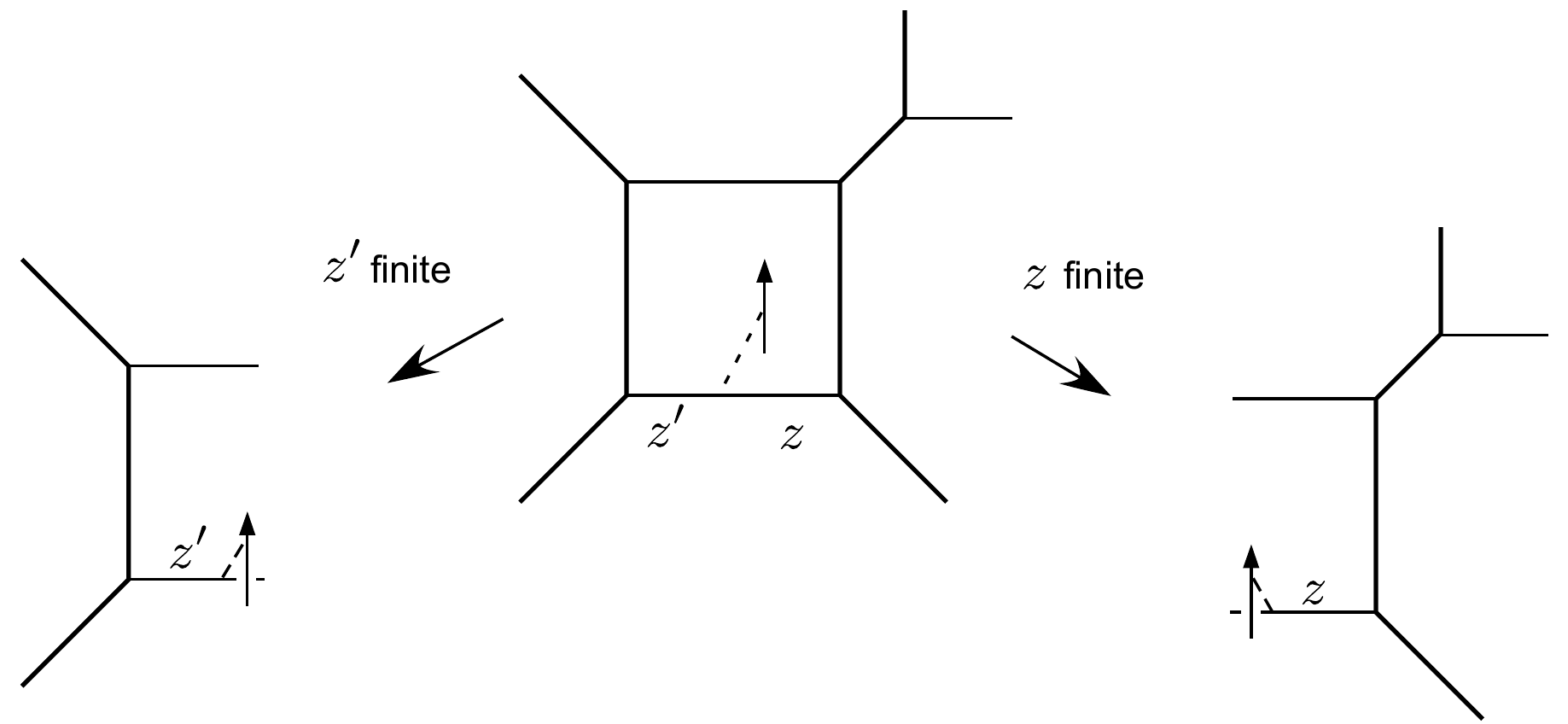}
\caption{Two possible ways to take the 2d-4d decoupling limit $Q_\Lambda=zz'\to0$.}
\label{fig:decouple}
\end{figure}

In many cases, we will only be interested in the two-dimensional $\CN=(2,2)$ gauge theory living on a surface operator itself, and the associated vortex partition function $Z^{\rm vortex}$. For a Lagrangian brane placed on a gauge leg as in Figure \ref{fig:branechoices2}(b), we can take the decoupling limit $Q_\Lambda\to 0$ in two different ways: keeping either $z$ or $z'$ fixed and finite. The appropriate choice depends on what we have identified as the FI parameters of the two-dimensional theory. If $z=e^{-t}$, then keeping $z$ finite --- calculating $Z^{\rm open}_{\rm BPS}$ for the toric geometry on the right of Figure \ref{fig:decouple} --- will reproduce K-theoretic vortex counting.

A central computational tool in almost all of the following examples is the topological vertex, which warrants a few final remarks.

In the unrefined limit $q_1=q_2=q$, the BPS partition function in any toric geometry with any number of branes may be calculated using the original topological vertex of \cite{AKMV}.
When $q_1\neq q_2$, the refined topological vertex \cite{IKV} was developed to provide a corresponding construction of refined BPS amplitudes.\footnote{An alternative formulation of the refined vertex appears in \cite{Awata-refvx}, but its applicability is apparently no broader than the vertex of \cite{IKV}.} However, in its present formulation, the refined topological vertex is merely a combinatorial tool rather than an object derived from fundamental physics. It is known to give correct BPS state counting precisely for the closed toric geometries that geometrically engineer gauge theories. We will \emph{cautiously} attempt to extend its use to open amplitudes involving simple Lagrangian branes in geometric-engineering geometries. In many cases (in particular, for vortex counting on single branes), the results agree perfectly with expected instanton-counting expressions.

\subsubsection{$U(1)$ theory}
\label{sec:U1surf}

\begin{figure}
\centering
\includegraphics[width=4in]{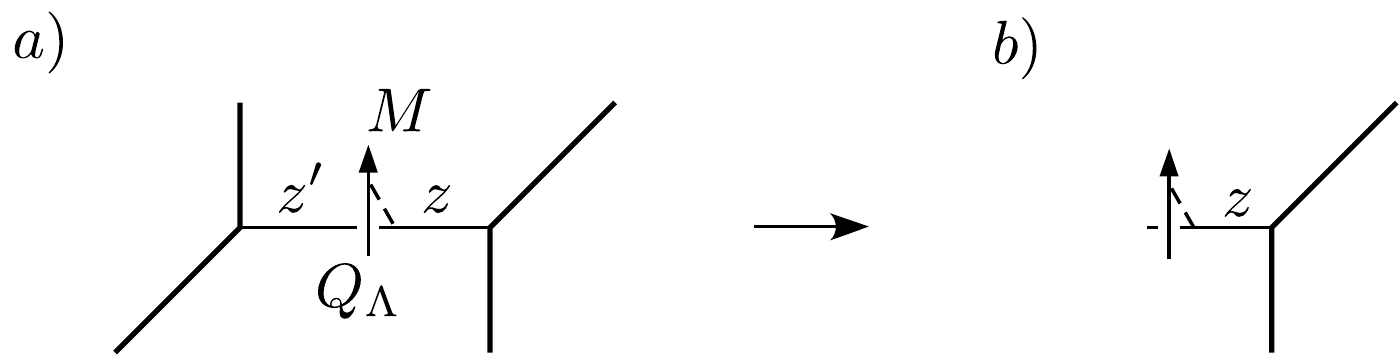}
\caption{A simple surface operator in $U(1)$ theory, and the decoupling limit $Q_\Lambda\to 0$.}
\label{fig:U1brane}
\end{figure}

We begin with the simplest possibility: an elementary surface operator in pure $\CN=2$ Maxwell theory, \ie\ with abelian gauge group $G=U(1)$. The gauge theory is engineered by type IIA string compactification on the resolved conifold
\be
\label{conifold}
X ~= \quad \CO (-1) \oplus \CO (-1) \to \cp^1 \,,
\ee
whose toric diagram appears in Figure \ref{fig:U1brane}(a). The complexified K\"ahler parameter of the $\cp^1$ in this geometry is $Q_\Lambda = \beta^2\Lambda^2$.

For $G=U(1)$, the only choice of Levi subgroup is $\LL = U(1)$. Since $\LL$ has a single abelian component, the lattice $\Lambda_\LL \simeq \Z$ is one-dimensional. The elementary surface operator corresponding to this Levi subgroup is engineered by placing a single Lagrangian brane $\Lag$ in the $U(1)$ geometry, as in Figure \ref{fig:U1brane}(a).

Since the resolved conifold geometry is very simple, we can obtain the BPS partition function $Z^{\rm BPS}$ directly from the fundmamental expressions \eqref{closedBPS}-\eqref{openBPS}. The $\cp^1$ is the only 2-cycle and it is rigid, with trivial first homology, so
\be N_\beta^{j_L,j_R} = \left\{ \begin{array}{c@{\quad}l}
  1 & \beta = [\cp^1],\,j_L=j_R=0\,, \\
  0 & \text{otherwise}\,. \end{array}\right.
\ee
Therefore (as is very well known) \eqref{closedBPS} leads to,
\be Z^{\rm closed}_{\rm BPS} = \prod_{i,j=1}^\infty\left(1-q_1^{i-\frac12}q_2^{j-\frac12}Q_\Lambda\right)\,. \ee
Similarly, the Lagrangian brane $\Lag$ intersects the $\cp^1$ in a circle, cutting it into two discs $D,D'$ of (exponentiated) areas $z$ and $z'$. These discs are rigid and have trivial first homology, so we expect
\be N_\mu^{s,r} = \left\{\begin{array}{c@{\quad}l}
 1 & \mu=[D],\,r=s=0\,,\\
 1 & \mu=[D'],\,r=s=0\,,\\
 0 & \text{otherwise}\,. \end{array}\right.
\ee
Then \eqref{openBPS} gives
\begin{align} Z^{\rm open}_{\rm BPS} &= \prod_{i=1}^\infty\left(1-q_1^{i-\frac12}q_2^{\,\frac12} z\right)^{-1}\left(1-q_1^{i-\frac12}q_2^{\,\frac12} z'\right)^{-1} \label{U1openprod} \\
&= \prod_{i=1}^\infty\left(1-q_1^{i-\frac12}q_2^{\,\frac12} z\right)^{-1}\left(1-q_1^{i-\frac12}q_2^{\,\frac12} Q_\Lambda z^{-1}\right)^{-1}\,. \notag
\end{align}
In the homological limit
\be z\to \beta z\,,\quad z'\to\beta z'\,,\quad Q_\Lambda = \beta^2\Lambda^2\,,\quad q_1=e^{-\beta \epsilon_1}\,,\quad q_2=e^{\beta\epsilon_2}\,; \notag \ee
\be \qquad \beta\to 0\,, \notag \ee
the full BPS partition function should correspond to instanton counting in the presence of a surface operator \eqref{zinstkm}:
\begin{align} Z^{\rm BPS, 4d} &= \lim_{\beta\to0} Z^{\rm closed}_{\rm BPS}(Q_\Lambda)\,Z^{\rm open}_{\rm BPS}(Q_\Lambda,\beta z,\beta z') \notag \\
&= Z^{\rm inst}_{U(1),\,{\rm closed,\,4d}}\times \sum_{m,n=0}^\infty \frac{z^m(z')^n}{m!n!\epsilon_1^{m+n}} = Z^{\rm inst}_{U(1),\,{\rm closed,\,4d}} \times \exp\left(\frac{z+\Lambda^2/z}{\epsilon_1}\right) \label{U1openlim} \\
&= Z^{\rm inst}_{\rm 4d}(\Lambda,z)\,. \notag
\end{align}
Here $Z^{\rm inst}_{U(1),\,{\rm closed,\,4d}}(\Lambda)$ denotes the usual equivariant partition function of $\CN=2$ Maxwell theory in the absence of surface operators. Note that both terms in the product \eqref{U1openprod} persist nontrivially in the homological limit \eqref{U1openlim} due to the special scaling $z,z'\to \beta z,\beta z'$.

To compare to vortex counting on the surface operator itself,
we decouple the four-dimensional theory by sending $Q_\Lambda\to 0$
while keeping $z$ fixed, as in Figure \ref{fig:U1brane}(b).
The remaining two-dimensional theory on the surface operator can be described
as a $\CN=(2,2)$ $U(1)$ gauge theory with a single massless fundamental
chiral multiplet --- \ie\ the abelian Higgs model of section \ref{sec:vortex}:
\begin{eqnarray}
{\bf 2D\ Theory:} && \CN=(2,2) ~~U(1) \mbox{\ theory\ with\ a\ charged\ chiral\ multiplet\ }
\nonumber
\end{eqnarray}
The decoupled open BPS partition function is given by the first product in \eqref{U1openprod}.
Upon setting $q_2\to 1$ and $q_1\to q=e^{-\beta\hbar}$ (to count spin but not R-charge),
and shifting $z\to zq^{-\frac12}$, we obtain agreement with the K-theoretic vortex partition function \eqref{vortexqdilog}:
\be Z^{\rm open}_{\rm BPS}(z;q) = \prod_{i=0}^\infty \frac1{1-q^iz} = \sum_{\m=0}^\infty \frac{z^\m}{(1-q)\cdots(1-q^\m)} = Z^{\rm vortex}_{\rm K-theory}(z;q)\,. \label{vortexqdilog2} \ee
Furthermore, taking the homological limit $\beta \to 0$ (with $z\to \beta z$), we find
\be Z^{\rm open}_{\rm BPS}(\beta z;q)\overset{\beta\to0}{\longrightarrow} \sum_{\m=0}^\infty \frac{z^\m}{\m!\hbar^\m} = e^{z/\hbar}\,,\ee
reproducing the vortex-counting contribution from a 2d fundamental chiral \eqref{vortfund} with $m=0$.

\begin{figure}
\centering
\includegraphics[width=1in]{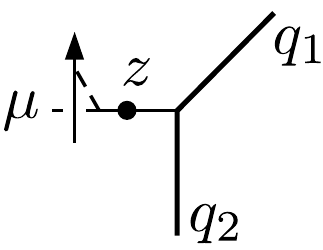}
\caption{Refined topological vertex labelings for a $U(1)$ surface operator in the decoupled limit $\Lambda\to 0$. The dot indicates a ``preferred direction.''}
\label{fig:U1topvx}
\end{figure}

The decoupled partition function $Z^{\rm open}_{\rm BPS}$ can also be obtained with a refined topological vertex computation.%
\footnote{In contrast, it is not completely clear at the moment how an internal brane in the full resolved conifold geometry should be analyzed with the refined vertex. The unrefined computation in this case is discussed in the original paper \cite{AKMV}.} The relevant diagram is shown in Figure \ref{fig:U1topvx}.
We place the preferred direction of the refined vertex (indicated with a dot
) on what would have been the four-dimensional $U(1)$ gauge leg. For the indicated brane framing, we should consider a single-row partition on the brane $\mu = (1,1,...,1)=(1^\m)$, which is incorporated in the computation via the factor
\be s_{\mu^t}(z) = \left\{\begin{array}{c@{\quad}l} z^\m & \mu = (1^\m) \\
0 & \text{otherwise}\,. \end{array}\right.
\ee
Then
\begin{align} \raisebox{-.2in}{\includegraphics[width=.7in]{U1topvx}}\;\; &= \sum_\mu s_{\mu^t}(z)\,C_{\circ\circ\mu}(q_2,q_1) \notag \\
 &= \sum_{\m=0}^\infty \frac{q_2^{\,\m/2}z^\m}{(1-q_1)\cdots(1-q_1^\m)} = \prod_{i=0}^\infty\frac{1}{1-q_1^{\,i}q_2^{\,\frac12}z}\,.
\label{U1vxopenprod}
\end{align}
Upon setting $q_2\to 1$ and $q_1\to q$, this agrees directly with the K-theoretic vortex partition function \eqref{vortexqdilog2}. (A shift $z\to q_1^{\,\frac12}z$ is needed for agreement with \eqref{U1openprod}.)


\subsubsection{$U(1)$ theory with matter}
\label{sec:U1mbrane}

To the four-dimensional $U(1)$ theory of the previous example, we can consider adding fundamental or antifundamental matter. This leads to the next-simplest example of a surface operator, now comprising a two-dimensional $U(1)$ $\CN=(2,2)$ theory with antifundamental chiral matter.

\begin{figure}[hbt]
\centering
\includegraphics[width=2.1in]{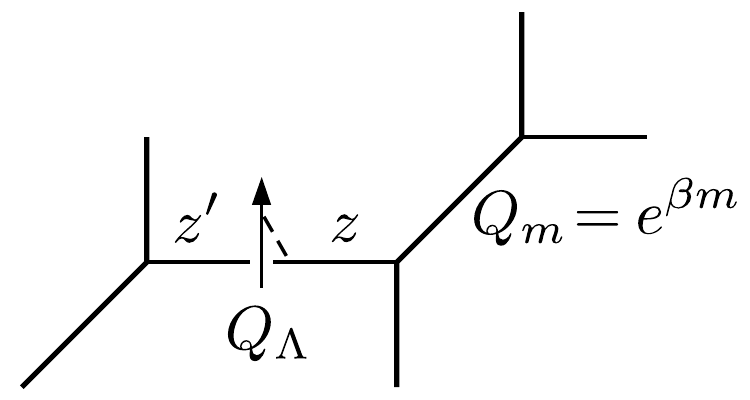}
\caption{$U(1)$ theory with a massive hypermultiplet and a surface operator.}
\label{fig:U1matter}
\end{figure}

\begin{figure}[ht]
\centering
\includegraphics[width=4.3in]{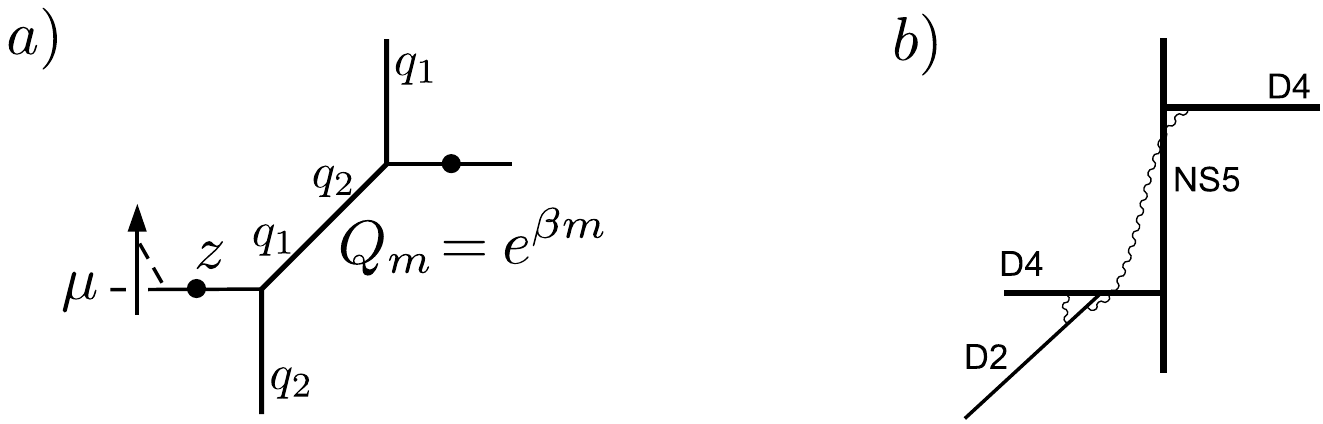}
\caption{The geometry of a $U(1)$ theory with a massive hypermultiplet in the $Q_\Lambda\to 0$ limit, and the corresponding brane engineering picture of the two-dimensional theory on the surface operator.}
\label{fig:U1matter2d}
\end{figure}

Suppose that we add a four-dimensional fundamental hypermultiplet as in Figure \ref{fig:U1matter}. Again, the unique choice of Levi subgroup is $\LL=U(1)$ (hence $\Lambda_\LL\simeq \Z$), and we incorporate the corresponding elementary surface operator via a single framed Lagrangian brane $\Lag$. In the decoupling limit $Q_\Lambda\to 0$ ($z$ finite), to which we will pass automatically for the remainder of this section, we obtain the geometry in Figure \ref{fig:U1matter2d}(a). (Note that the same decoupled geometry could have been obtained from a $\CN=2^*$ $U(1)$ theory with adjoint matter as well.)

The decoupled theory on the surface operator can also be realized via  a brane construction
as in Figure \ref{fig:U1matter2d}(b). It was shown in \cite{Hanany-Hori}
that this is a two-dimensional $U(1)$ theory with a massless fundamental chiral multiplet,
and an antifundamental chiral multiplet of (classical) twisted mass\footnote{In what follows,
we shall denote all two-dimensional mass parameters with a tilde.} $\tilde{m}=-m$:
\begin{eqnarray}
{\bf 2D\ Theory:} && \CN=(2,2) ~~U(1) \mbox{\ with\ chiral\ multiplets\ of\ charge\ } +1 \mbox{\ and\ } -1
\nonumber
\end{eqnarray}
The two-dimensional matter comes from open strings stretched between the D2-brane and D4-branes.
Equivariant vortex counting of Section \ref{sec:vortex} then predicts
\be Z^{\rm vortex}(z,\tilde{m};\hbar) = \sum_{\m} \frac{\tilde{m}(\tilde{m}+\hbar)\cdots(\tilde{m}+(\m-1)\hbar)}{\m!\hbar^\m}z^\m\,. \label{ZvxU1m} \ee

The refined topological vertex calculates
\begin{align} Z^{\rm BPS}_{\rm open}(z,Q_m;q_1,q_2) &=
 \sum_\mu s_{\mu^t}(z)\,C_{\lambda\circ\mu}(q_1,q_2)(-Q_m)^{|\lambda|}C_{\lambda^t\circ\circ}(q_2,q_1) \\
  &= \prod_{r=1}^\infty \frac{1-q_1^{r-1/2}Q_mz}{1-q_1^{r-1}q_2^{1/2}z} \notag \\
 &= \sum_{\m=0}^\infty \frac{\big(1-q_1^{1/2}q_2^{-1/2}Q_m\big)\cdots\big(1-q_1^{\m-1/2}q_2^{-1/2}Q_m\big)}{\big(1-q_1\big)\cdots\big(1-q_1^\m\big)}\big(q_2^{1/2}z\big)^\m\,.
\end{align}
After shifting the mass of the antifundamental chiral $\tilde{m}\to\tilde{m}-(\epsilon_1+\epsilon_2)/2$ (or $Q_m\to q_1^{-\frac12}q_2^{\,\frac12}Q_m$), sending $q_2\to 1$, $q_1\to q$, and taking the homological limit $\beta\to 0$, we find agreement
\be Z^{\rm BPS}_{\rm open}(z,\tilde{m};\hbar) = Z^{\rm vortex}(z,\tilde{m};\hbar)\,.\ee


\subsubsection{$U(2)$ theory}

Let us now consider four-dimensional pure $\CN=2$ super-Yang-Mills theory with gauge group $G=SU(2)$.
In this case, the next-to-maximal choice of Levi subgroup is $\LL = U(1)$ and,
according to \eqref{mlattice}.
(Similarly, we could have considered a close cousin of this theory with $G = U(2)$.)

\begin{figure}[ht]
\centering
\includegraphics[width=6in]{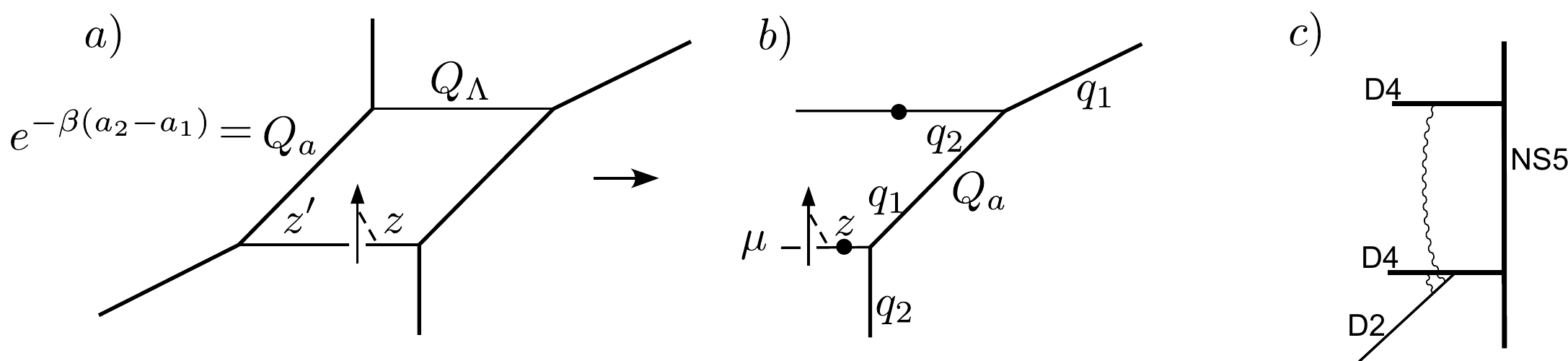}
\caption{a) The toric geometry for $SU(2)$ theory with a surface operator. b) The $Q_\Lambda\to0$ decoupling limit. c) The brane construction of the 2d surface operator theory.}
\label{fig:SU2}
\end{figure}

The toric geometry that engineers the four-dimensional theory is
the local Hirzebruch surface $\mathbb{F}_0=\cp^1_b\times\cp^1_f$, as shown in Figure \ref{fig:SU2}(a).%
\footnote{The slightly strange orientation of this toric diagram --- related by a simple $SL(2,\Z)$ transformation to the more standard ``upright'' picture of $\mathbb{F}_0$ --- is chosen to give the most natural framing to the brane.} %
 The ``base'' $\cp^1_b$ has K\"ahler parameter $Q_\Lambda = \beta^4\Lambda^4$, while the ``fiber'' $\cp^1_f$ has K\"ahler parameter $Q_a = e^{-\beta(a_2-a_1)}$, where $a_{1,2}$ are the adjoint scalar eigenvalues on the Coulomb branch as explained in Appendix \ref{app:GE}. (For $SU(2)$ theory we simply set $a_1=-a_2=a$.)

To realize an elementary surface operator, we place a single Lagrangian brane in our canonical location,
on the bottom gauge leg of the toric diagram. Decoupling the four-dimensional theory leads to
the geometry in Figure \ref{fig:SU2}(b), and the corresponding brane construction in Figure \ref{fig:SU2}(c).
The two-dimensional theory on the surface operator has the usual massless fundamental chiral multiplet,
plus a second fundamental chiral of (classical) twisted mass $\tilde{m} = a_2-a_1$:
\begin{eqnarray}
{\bf 2D\ Theory:} && \CN=(2,2) ~~U(1) \mbox{\ with\ two\ chiral\ multiplets\ of\ charge\ } +1
\nonumber
\end{eqnarray}
{}From vortex counting of section \ref{sec:vortex}, we then expect
\be Z^{\rm vortex}(z,\tilde{m};\hbar) = \sum_{\m=0}^\infty \frac{1}{\m!\hbar^\m(\tilde{m}+\hbar)\cdots(\tilde{m}+\m\hbar)}z^\m\,.\ee
In this case, we can also obtain a K-theoretic expression by using the non-linear sigma model description of the surface operator. The equivariant $J$-function \eqref{jforcpn} predicts
\be Z^{\rm vortex}_{\rm K-theory}(z,Q_a;q) = \sum_{\m=0}^\infty \frac{1}{(1-q)\cdots(1-q^\m)(1-Q_aq)\cdots(1-Q_aq^\m)}z^\m\,, \label{vortexSU2} \ee
with $Q_a = e^{-\beta\tilde{m}}$.

Correspondingly, the refined topological vertex calculates a normalized partition function
\begin{align}\label{eqn:opensu2onebranebottom}
Z^{\rm open}_{\rm BPS}(z,Q_a;q_1,q_2) &=  \frac{\sum_{\mu,\lambda} s_{\mu^t}(z) C_{\circ\lambda\circ}(q_1,q_2)(-Q_a)^{|\lambda|}f_{\lambda}(q_1,q_2)C_{\lambda^t\circ\mu}(q_1,q_2)}
{\sum_\lambda C_{\circ\lambda\circ}(q_1,q_2)(-Q_a)^{|\lambda|}f_{\lambda}(q_1,q_2)C_{\lambda^t\circ\circ}(q_1,q_2) } \notag \\
 &= \sum_{\m=1}^\infty \frac{1}{\prod_{j=1}^\m (1-q_1^j)(1-Q_aq_1^{j-1})}
  \big(q_2^{1/2}z\big)^\m\,.
\end{align}
After setting $q_2\to 1$, $q_1\to q$ and shifting $Q_a\to Q_aq$, we find complete agreement
\be Z^{\rm open}_{\rm BPS}(z,Q_a;q,1) = Z^{\rm vortex}_{\rm K-theory}(z,Q_a;q)\,.
\ee

\begin{figure}[ht]
\centering
\includegraphics[width=4in]{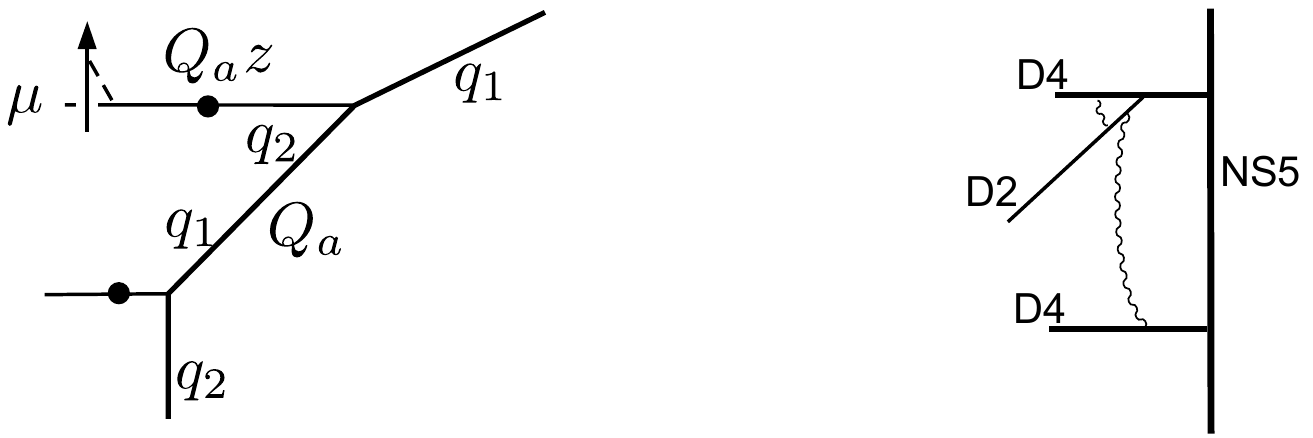}
\caption{$SU(2)$ geometry with alternative placement of the brane. In order to correlate with the initial choice in Figure \protect\ref{fig:SU2}(b), the disk instanton parameter is now $Q_az$, and a framing factor $f_\mu(t,q)^{-1}$ must be included in the calculation.}
\label{fig:SU2top}
\end{figure}

In addition to the setup of Figure \ref{fig:SU2}(a-b), we could also have placed the Lagrangian brane on the top gauge leg, as in Figure \ref{fig:SU2top}. This then corresponds to a two-dimensional theory with fundamental chirals of masses zero and $\tilde{m}=-(a_2-a_1)$. (Note that the nonzero twisted mass has changed sign.) The normalized BPS partition function is
\begin{align} {Z^{\rm open}_{\rm BPS}}'(z,Q_a;q_1,q_2) &=
\frac{\sum_{\mu,\lambda} s_{\mu^t}(Q_az) f_\mu(q_1,q_2)^{-1} C_{\circ\lambda\mu}(q_1,q_2)(-Q_a)^{|\lambda|}f_{\lambda}(q_1,q_2)C_{\lambda^t\circ\circ}(q_1,q_2)}
{\sum_\lambda C_{\circ\lambda\circ}(q_1,q_2)(-Q_a)^{|\lambda|}f_{\lambda}(q_1,q_2)C_{\lambda^t\circ\circ}(q_1,q_2) } \notag \\
&= \sum_{\m=0}^\infty \frac{1}{\prod_{j=1}^\m(1-q_1^j)(1-Q_a^{-1}q_2^{-1}q_1^j)}\big(q_1q_2^{-1/2}z\big)^\m\,.
\end{align}
We now find that after setting $q_2\to1$ and $q_1\to q$ we should also rescale $z\to q^{-1}z$ in order to match the K-theoretic vortex partition function \eqref{vortexSU2}.

It is interesting to note that the choice of brane placement in a toric geometry is completely mirrored by a choice of pole prescription in a contour-integral expression for the equivariant vortex partition function. In the present case, the discussion in Section \ref{sec:vortex} produces the contour integral
\begin{align}\label{eqn:U2vortexcontour}
& Z^{\mathrm{vortex}}_{\m} = \frac{1}{\m! \hbar^{\m}} \oint \frac{d \varphi_i}{2 \pi i} ~ \prod_{i \neq j}^{\m} \frac{\varphi_i - \varphi_j}{\varphi_i - \varphi_j - \hbar} \prod_{i=1}^{\m} \frac{1}{\varphi_i(\varphi_i - \tilde{m})}\,.
\end{align}
In the second product, one of the chiral multiplets is massless while the other has mass $\tilde{m}$. The BPS partition function for a brane on the bottom gauge leg is reproduced by including only residues from the terms $\varphi_i$ in the denominator, while the partition function for a brane on the top leg (with $\tilde{m}\mapsto-\tilde{m}$) is reproduced by including only residues from the terms $\varphi_i-\tilde{m}$.


\subsubsection{The general case}
\label{sec:generalsurf}

It is straightforward to generalize the previous examples to construct an elementary surface
operator in a four-dimensional $U(N)$ or $SU(N)$ theory with arbitrary matter content.
In either case, such surface operator can be described by a $\CN=(2,2)$ $U(1)$ gauge theory
in two dimensions and corresponds to the maximal nontrivial Levi subgroup which gives $\Lambda_\LL\simeq \Z$.
After passing to the decoupling limit $Q_\Lambda\to\infty$,
we are left with only this 2d $U(1)$ gauge theory,
which has $N$ chiral fundamental multiplets (one of which is always massless)
and $N_f$ chiral antifundamentals (where $N_f$ is the number of original 4d fundamental hypers).

\begin{figure}[htb]
\centering
\includegraphics[width=5in]{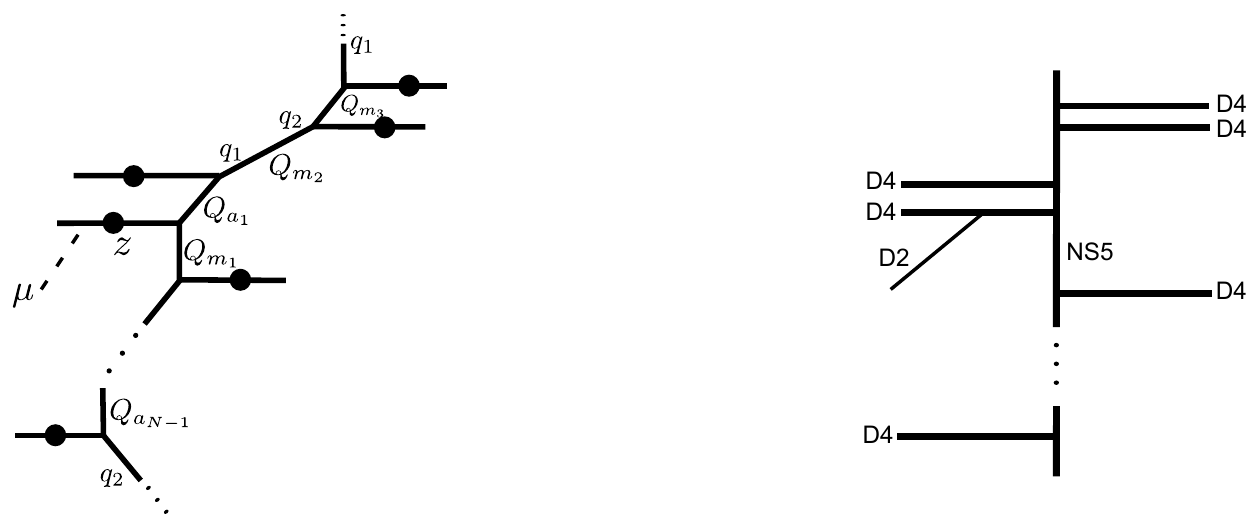}
\caption{Surface operators in $U(N)$ theory.}
\label{fig:SUN2d}
\end{figure}

A typical setup of this type is illustrated in Figure \ref{fig:SUN2d}.
The classical twisted masses of two-dimensional matter can be easily
read off from four-dimensional Coulomb parameters and bare masses,
keeping in mind that 2d matter comes from string stretched between
a D2-brane (the surface operator) and various D4's.
After appropriate quantum shifts of 2d masses,
which will depend on the precise form of the 4d engineering geometry,
we obtain a BPS/vortex partition function
\be Z^{\rm open}_{\rm BPS} = Z^{\rm vortex}_{\rm K-theory} = \sum_{\m=1}^\infty \frac{\prod_{i=1}^{N_f} \prod_{j=1}^\m (1-Q_{\tilde{\bar{m}}_i}q^{j-1})}{\prod_{j=1}^\m(1-q^j)\,\prod_{i=1}^{N-1}\prod_{j=1}^\m(1-Q_{\tilde{m}_i}q^j)}z^\m \,,
\label{gen2d}
\ee
where $Q_{\tilde{m}_i}=e^{-\beta \tilde{m}_i}$, $Q_{\tilde{\bar{m}}_i}=e^{-\beta \tilde{\bar{m}}_i}$ for fundamental (resp. antifundamental) masses $\tilde{m}_i$ (resp, $\tilde{\bar{m}}_i$). (As usual, we have taken $q_2\to1$, $q_1\to q$ here.)

The precise placement of the Lagrangian brane in the toric geometry is unimportant,
as long as one remembers to count 2d matter arising from strings ``below'' the brane with negative mass.
The contour integral \eqref{zzzzanti} that reproduces $Z^{\rm vortex}$ in the homological limit $z\to \beta^{N-N_f}z$,\, $\beta\to 0$ is
\be
Z^{\rm vortex}_{\m} = \frac{1}{\m! \hbar^{\m}} \oint \frac{d \varphi_i}{2 \pi i} ~ \prod_{i \neq j}^{\m} \frac{\varphi_i - \varphi_j}{\varphi_i - \varphi_j - \hbar} \prod_{i=1}^{\m} \frac{(\varphi_i + \tilde{\bar{m}}_1)\cdots (\varphi_i + \tilde{\bar{m}}_{N_f})}{\varphi_i (\varphi_i+\tilde{m}_{1})\cdots (\varphi_i + \tilde{m}_{N-1})}. \label{gencontour}
\ee
We take into account the poles created by only one of the $N$ terms in the denominator of the second product, and the choice of term is directly related to the choice of brane placement. \\

\begin{figure}[htb]
\centering
\includegraphics[width=5.5in]{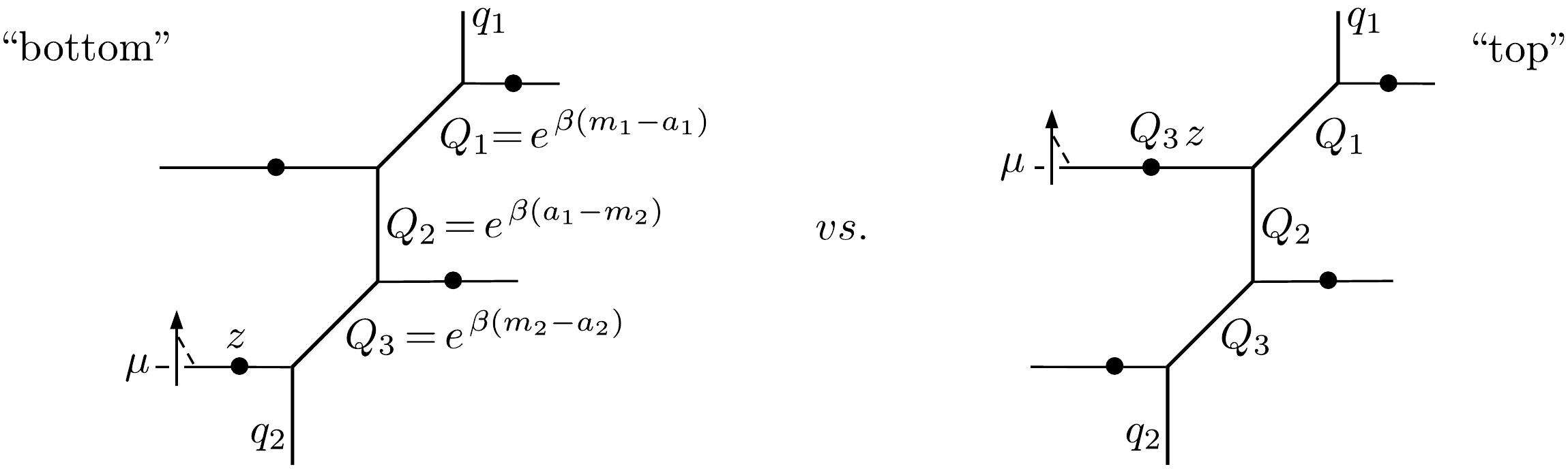}
\caption{Brane placements in (decoupled) superconformal SU(2) theory.
The 4d bare masses are $m_{1,2}$, and the Coulomb parameter is $a=a_1=-a_2$. Note $Q_a = Q_2Q_3=e^{2\beta a}$.
See Appendix \protect\ref{app:GE} for further discussion of the 4d parameters.}
\label{fig:SU2hypers}
\end{figure}

As a final variation, which we will be important later in the paper, let's consider a surface operator in $SU(2)$ theory with two fundamental hypermultiplets. (Adding an additional two antifundamental hypers, this would be superconformal $\CN=2$ theory.) The two possible brane placements in the $Q_\Lambda\to0$ limit are shown in Figure \ref{fig:SU2hypers}.

For a brane on the bottom leg, the (classical) 2d fundamental mass is $\tilde{m}=a_2-a_1=-2a$
and the antifundamental masses are $\tilde{\bar{m}}_{1,2}=a_2-m_1$, $a_2-m_2$.
For a brane on the top leg, these masses are $\tilde{m}=+2a$ and $\tilde{\bar{m}}_{1,2}=\,a_1-m_1,\,a_1-m_2$.
The normalized BPS partition functions turn out to be
\begin{align} Z^{\rm bottom}_{\rm BPS} &= \sum_{\m=0}^\infty \frac{\prod_{j=1}^\m(1-Q_{\tilde{\bar{m}}_1}q_2^{-\frac12}q_1^{j-\frac12})(1-Q_{\tilde{\bar{m}}_2}q_2^{-\frac12}q_1^{j-\frac12})}{\prod_{j=1}^\m(1-q_1^{\,j})(1-Q_{\tilde{m}}q_2^{-1}q_1^{\,j})}q_2^{\,\frac{\m}{2}}z^\m\,, \label{Zbottom} \\
 Z^{\rm top}_{\rm BPS} &= \sum_{\m=0}^\infty \frac{\prod_{j=1}^\m(1-Q_{\tilde{\bar{m}}_1}q_2^{-\frac12}q_1^{j-\frac12})(1-Q_{\tilde{\bar{m}}_2}q_2^{-\frac12}q_1^{j-\frac12})}{\prod_{j=1}^\m(1-q_1^{\,j})(1-Q_{\tilde{m}}q_1^{\,j-1})}q_1^{-\frac{\m}{2}}q_2^{\,\m}z^\m\,, \label{Ztop}
\end{align}
with $Q_{m}\equiv e^{-\beta m}$ for each respective 2d mass parameter. They can both clearly be put into the form \eqref{gen2d} at $(q_1,q_2)=(q,1)$.


\subsubsection{Multiple branes}
\label{sec:multiple}

So far, we have seen that the relation $Z^{\rm BPS}=Z^{\rm vortex}$ holds up for
the $U(1)$ two-dimensional gauge theory on an elementary surface operator embedded in an arbitrary four-dimensional theory.
In this last section, we will consider the simplest extensions of this relation
to more general surface operators: ensembles of elementary $U(1)$ surface operators
in 4d $U(N)$ theory.
(For some comments on interacting $U(p)$ surface operators in $U(N)$ theory, see Section \ref{sec:GT}.)
Refined topological vertex calculations become somewhat tenuous when such non-elementary surface operators are introduced.
Nevertheless, it is still possible to discern some main features of expected vortex partition functions.

\begin{figure}[hbt]
\centering
\includegraphics[width=5.5in]{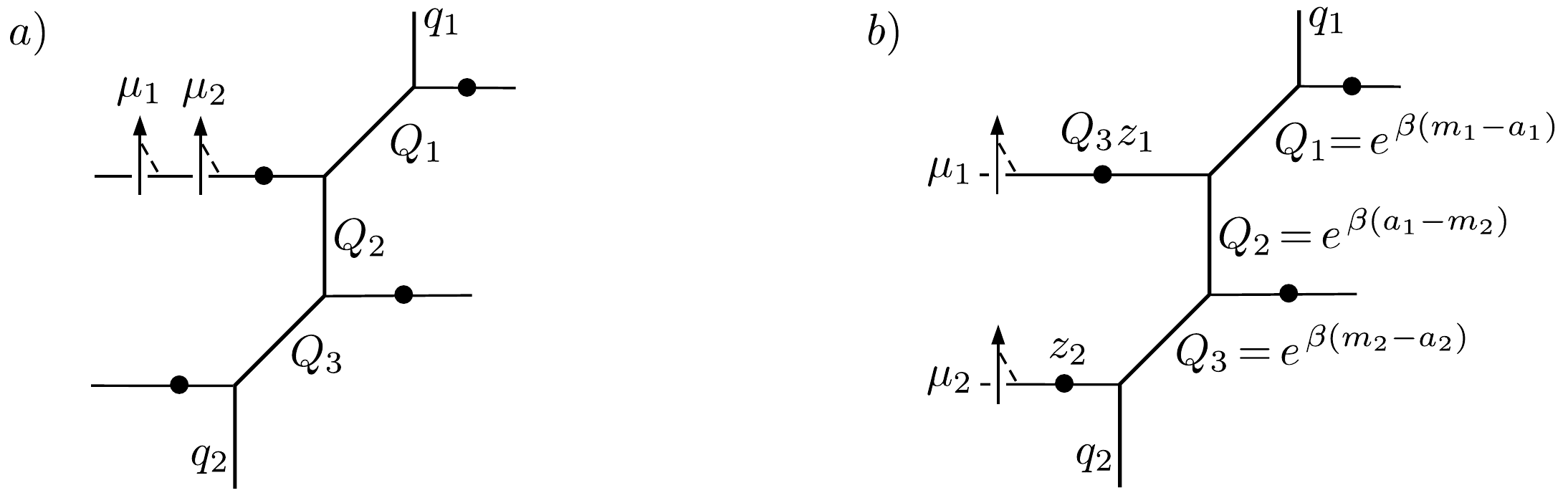}
\caption{Possible placements of multiple elementary branes in a decoupled $SU(2)$ $N_f=2$ geometry.}
\label{fig:SU2multiple}
\end{figure}

Ensembles of $U(1)$ surface operators can be introduced in several ways.
The first is to add multiple Lagrangian branes
at different locations in a toric geometry, as in Figure \ref{fig:SU2multiple}.
The most computationally tenable setup is that of Figure \ref{fig:SU2multiple}(b),
with at most one brane per gauge leg --- although in theory it should not be significant how the branes are distributed.

Each Lagrangian brane in such a geometry has its own disk-instanton parameter $z_i$.
In a dual brane construction, each D2-brane corresponding to these Lagrangians
would be attached to its \emph{own} regulating NS5'-brane.
It is in this sense that the surface operators are mutually non-interacting,
coupling to the 4d gauge theory but not to each other.
We would therefore expect that the equivariant vortex partition functions
for the worldvolume theory of this ensemble of surface operators will \emph{factorize},
at least in the usual limit of counting spin but not R-charge.
In other words, we should have
\be Z^{\rm vortex}_{p\,U(1){\rm 's}}(\{z_i\};q) = Z^{\rm vortex}_{U(1)}(z_1;q)\times\cdots\times Z^{\rm vortex}_{U(1)}(z_p;q)\,. \label{vortexfactored}\ee
Let us therefore test this on the BPS side.

For the geometry of Figure \ref{fig:SU2multiple}(b), coming from four-dimensional $U(2)$ theory with $N_f=2$, the refined topological vertex computes
\begin{align} &Z^{\rm open}_{\rm BPS}(z_1,z_2;q_1,q_2) \notag \\
 &\qquad = \sum_\mu {\textstyle\frac{s_{\mu_1}(Q_3z_1)s_{\mu_2}(z_2)C_{\lambda_1\circ\circ}(q_2,q_1)(-Q_1)^{|\lambda_1|}C_{\lambda_1^t\lambda_2\mu_1}(q_1,q_2)(-Q_2)^{|\lambda_2|}C_{\lambda_3\lambda_2^t\circ}(q_2,q_1)(-Q_3)^{|\lambda_3|}C_{\lambda_3^t\circ\mu_2}(q_1,q_2)}
 {C_{\lambda_1\circ\circ}(q_2,q_1)(-Q_1)^{|\lambda_1|}C_{\lambda_1^t\lambda_2\circ}(q_1,q_2)(-Q_2)^{|\lambda_2|}C_{\lambda_3\lambda_2^t\circ}(q_2,q_1)(-Q_3)^{|\lambda_3|}C_{\lambda_3^t\circ\circ}(q_1,q_2)}} \notag \\
 &\qquad = \sum_{\m_1,\m_2=0}^\infty \frac{\prod_{j=1}^{\m_1}(1-Q_{\tilde{\bar{m}}_1^1}q_2^{-\frac12}q_1^{j-\frac12})(1-Q_{\tilde{\bar{m}}_2^1}q_2^{-\frac12}q_1^{j-\frac12})}
 {\prod_{j=1}^{\m_1}(1-q_1^j)(1-Q_{\tilde{m}}q_1^{\m_1+1-j})}\, \label{SU2multipleBPSref} \\
 &\qquad\qquad\times \frac{\prod_{j=1}^{\m_2}(1-Q_{\tilde{\bar{m}}_1^2}q_2^{-\frac12}q_1^{j-\frac12})(1-Q_{\tilde{\bar{m}}_2^2}q_2^{-\frac12}q_1^{j-\frac12})}{\prod_{j=1}^{\m_2}(1-q_1^j)(1-Q_{\tilde{m}}q_2^{-1}q_1^{j-\m_2})}
 q_2^{\,\m_1+\frac{\m_2}{2}}q_1^{-\frac{\m_1^2}{2}}(-Q_{\tilde{m}})^{\m_1}z_1^{\m_1}z_2^{\m_2}\,, \notag
\end{align}
where we have used two-dimensional masses
\be Q_{\tilde{\bar{m}}_1^1} = Q_1\,,\quad Q_{\tilde{\bar{m}}_2^1} = Q_2^{-1}\,,\quad Q_{\tilde{\bar{m}}_1^2} =  Q_1Q_2Q_3\,,\quad Q_{\tilde{\bar{m}}_2^2}=Q_3\,,\quad Q_{\tilde{m}} = Q_2Q_3\,. \ee
Expression \eqref{SU2multipleBPSref} would factorize were it not for the $q_2^{-1}$ in the denominator factor $(1-Q_{\tilde{m}}q_2^{-1}q_1^{j-\m_2})$. In the limit $(q_1,q_2)\to (q,1)$, this is not a problem, and we indeed find
\begin{align} 
&Z^{\rm open}_{\rm BPS}(z_1,z_2;q,1) = Z^{\rm top}_{\rm BPS}(z_1;q,1)\,Z^{\rm bottom}_{\rm BPS}(z_2;q,1) \\
&\qquad =\sum_{\m_1,\m_2=0}^\infty \prod_{j=1}^{\m^1}
\frac{(1-Q_{\tilde{\bar{m}}_1^1}q^{j-\frac12})(1-Q_{\tilde{\bar{m}}_2^1}q^{j-\frac12})}{(1-q^j)(1-Q_{\tilde{m}}^{-1}q^{j-1})}
\prod_{j=1}^{\m_2}
\frac{(1-Q_{\tilde{\bar{m}}_1^2}q^{j-\frac12})(1-Q_{\tilde{\bar{m}}_2^2}q^{j-\frac12})}{(1-q^j)(1-Q_{\tilde{m}}q^{j})}q^{-\frac{\m_1}{2}}z_1^{\m_1}z_2^{\m_2}\,,
\notag \end{align}
where $Z^{\rm top}_{\rm BPS}$ and $Z^{\rm bottom}_{\rm BPS}$ are as in \eqref{Zbottom}--\eqref{Ztop}.

Our other way to engineer an ensemble of mutually noninteracting $U(1)$ surface operators is to place a stack of Lagrangian branes on top of each other.\footnote{Note that topological vertex computations as described here do \emph{not} encode interactions among these branes. Additional Ooguri-Vafa factors would need to be introduced (at least in the unrefined limit $q_1=q_2$) to account for interactions \cite{AMV-allloop}.} In the unrefined limit $q_1=q_2$, the resulting BPS partition function can be computed with the topological vertex by including a factor
\be s_{\mu^t}(z_1,z_2,...,z_p) \ee
for the stack, where the $z_i$ are the eigenvalues of the $U(p)$ holonomy of the gauge field on the $p$ Lagrangians. Note that this Schur function vanishes unless $\mu^t$ has length $\leq p$.

\begin{figure}[hbt]
\centering
\includegraphics[width=1.2in]{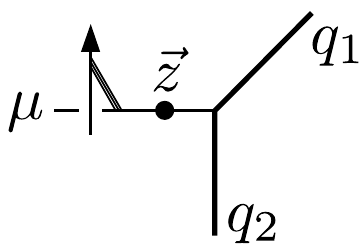}
\caption{A stack of $p$ Lagrangian branes in a decoupled $U(1)$ geometry.}
\label{fig:U1stack}
\end{figure}

For the refined topological vertex 
it would be natural to replace this Schur function with a MacDonald function
\be P_{\mu^t}^{\rm McD}(z_1,...,z_p;q_1,q_2)\,. \label{McD} \ee
We add a superscript ``McD'' here to signify that this is the function defined in \cite{McD}, which is slightly from that used for the refined topological vertex in \cite{IKV}.
Note that for a single brane $P_{\mu^t}^{\rm McD}(z;q_1,q_2)=s_{\mu^t}(z)$, so we recover our vortex-counting results. In the simplest case of a stack of $p$ Lagrangian branes on the (decoupled) conifold, as in Figure \ref{fig:U1stack}, we calculate
\begin{align} Z^{\rm open}_{\rm BPS}(\{z_i\};q_1,q_2) &= \sum_\mu P_{\mu^t}^{\rm McD}(z_1,...,z_p;q_1,q_2) C_{\circ\circ\mu}(q_1,q_2) \notag \\
&= \prod_{k=1}^p \prod_{r=0}^\infty \frac{1}{1-z_kq_2^{\,\frac12}q_1^r} \\
&= \prod_{k=1}^p\left(\sum_{\m_i=0}^\infty\frac{q_2^{\,{\m_i}/{2}}z_i^{\m_i}}{(1-q_1)\cdots(1-q_1^{\m_i})}\right)\,.
\end{align}
This is exactly the expected form of the partition function from first-principles BPS counting, as in \eqref{openBPS}.
In the 2d homological limit limit $q_1\to q=e^{-\beta\hbar}$, $q_2\to1$, $z\to \beta z$ and $\beta\to 0$, we find
\be Z^{\rm open}_{\rm BPS}(\{z_i\};q_1,q_2)\to \exp\left(\frac{z_1+ \ldots + z_p}{\hbar}\right) = \prod_{i=1}^p Z^{\rm vortex}_{U(1)}(z_i;q)\,.\ee
Again, this has the expected form of a product \eqref{vortexfactored}.

\subsection{Surface operators from knots and links}

A large class of half-BPS surface operators can be constructed from knots and links in a 3-sphere.
Specifically, given a knot (or link) $K$ in $\S^3$, one can construct \cite{Taubes}
a (special) Lagrangian submanifold in the conifold geometry \eqref{conifold}.
Therefore, the geometric engineering setup \eqref{surfeng}
leads to a family of half-BPS surafce operators in $U(1)$ gauge theory
naturally associated to knots and links:
\be
\boxed{\phantom{\int} \begin{array}{c@{\qquad}c@{\qquad}c}
\text{knot $K$} & \leadsto & \text{$\frac{1}{2}$-BPS surface operator} \,
\label{knotsurf}
\end{array}\phantom{\int}}
\ee

Moreover, in this case the normalized partition function of the refined BPS invariants
turns out to be closely related to homological knot invariants \cite{GSV}.
In particular, the term with monopole number $\m=1$ gives
the superpolynomial \cite{Superpolynomial} of the knot $K$,
\be
Z^{{\rm vortex}}_{\m=1} (Q_{\Lambda},q_1,q_2) = \bar \CP ({\bf a}, q,t)
\ee
with the following identification of variables:
\begin{eqnarray}
\sqrt{q_{1}} & = & q \nonumber \\
\sqrt{q_{2}} & = & -t q \label{qqqviaaqt} \\
Q_{\Lambda} & = & - t {\bf a}^{-2}
\nonumber
\end{eqnarray}
For example, the superpolynomial for the figure-eight knot ${\bf 4_1}$ is
\be
\bar \CP ({\bf 4_1}) = \frac{{\bf a} - {\bf a}^{-1}
+ (q^{-1} + {\bf a}^2q^{-1}t)({\bf a} q^{-1} - {\bf a}^{-1} q) ({\bf a}^{-2} t^{-2} + q^2 t)}{q-q^{-1}}
\ee

Notice, the ``decoupling limit'' $Q_{\Lambda} \to 0$
that describes the contribution of the two-dimensional sector due to
a surface operator has a very simple interpretation in terms of knot homologies.
Namely, according to \eqref{qqqviaaqt} this limit is encoded in the bottom row
of the superpolynomial $\bar \CP$, {\it i.e.} the terms with the lowest power of ${\bf a}$.
For example, after a suitable regularization, for the torus knots $T_{2,2p+1}$ we obtain
\be
Z^{{\rm vortex}}_{k=0, \m=1} = \sum_{i=0}^p \frac{q_1^{i-p+\frac{1}{2}} q_2^i }{1-q_1}
\ee
where we used \eqref{qqqviaaqt}.
In other words, given a knot (or link) $K$ one can compute
the ``$sl(\infty)$ knot homology'' $\CP \vert_{{\bf a}=0}$ via vortex counting
in the two-dimensional $\CN=(2,2)$ theory that describes the surface operator \eqref{knotsurf}.
In this expression, the universal factor $(1 - q_1)$ in the denominator of $Z^{{\rm vortex}}$
corresponds to the center-of-mass position of a $\m=1$ vortex on the plane $D = \R^2$.


\subsection{Closed BPS invariants}
\label{sec:GT}

Geometric transitions \cite{GV-largeN} relate open and closed BPS invariants and offer a different, interesting perspective of instanton counting in the presence of surface operators. In particular, in the simplest case, the geometric transition in a toric geometry with Lagrangian branes predicts the equivalence of instanton partition functions in the following geometrically engineered gauge theories:
\be \boxed{\begin{array}{c}
 \text{4d $\CN=2$ theory w/ group $G\times G$} \\
 \text{+ ``degenerate'' bifundamental} \\
 \text{matter of mass $m_b=\hbar/2$}
  \end{array}}
\quad\longleftrightarrow\quad
\boxed{\begin{array}{c}
 \text{4d $\CN=2$ theory w/ group $G$} \\
 \text{and a $U(1)$ surface operator\,.}
 \end{array}}
\label{GTgauge}
\ee
The scales $\Lambda_1,\,\Lambda_2$ of the $G\times G$ theory on the left are related to the scale $\Lambda$ and the surface operator FI parameter $z$ of the theory on the right as $\Lambda = \Lambda_1\Lambda_2$ and $z = \Lambda_1$. Similarly, in the decoupling limit $\Lambda\to 0$ that has been investigated in much of this paper, we find a predicted equivalence
\be \boxed{\begin{array}{c}
 \text{4d instanton counting for} \\
 \text{$\CN=2$ theory w/ group $G$\,,} \\
 \text{scale $\Lambda$}
  \end{array}}
\quad\longleftrightarrow\quad
\boxed{\begin{array}{c}
 \text{2d vortex counting for} \\
 \text{$\CN=(2,2)$ theory w/ group $U(1)$\,,} \\
 \text{parameter $z=\Lambda$\,.}
  \end{array}}
\label{GTvortex}
\ee
This time, if the 4d gauge group is $G=U(N)$, there should be $N_f=N$ flavors of matter. All but one bare mass parameters $m_i$ are equal to Coulomb vevs $a_i$, while one differs by $\hbar/2$.

\begin{figure}[htb] \centering
\includegraphics[width=4.0in]{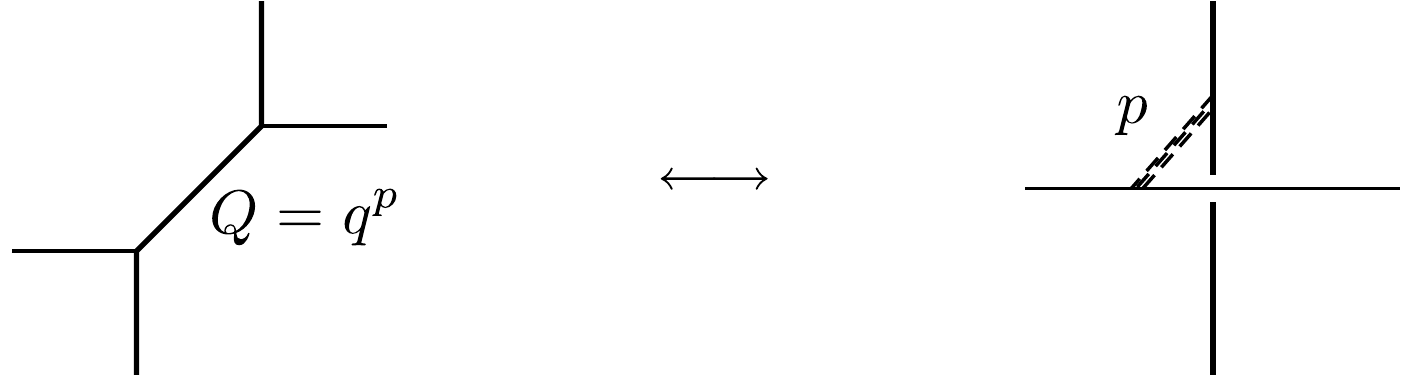}
\caption{The original geometric transition: $p$ Lagrangian branes on $T^*\S^3$ are replaced with the closed geometry $\CO(-1,-1)\to\cp^1$.}
\label{fig:GTconifold}
\end{figure}

To understand how these dualities come about, let's review a few facts about geometric transitions. Strictly speaking, the geometric transition is only known to hold only for BPS counting in the unrefined limit%
\footnote{Note that this is \emph{not} the same as the 2d vortex limit $q_1=q$, $q_2=1$ that we usually take. The physical interpretations of $q_1,q_2$ at the end of Section \ref{sec:GE} show that in the $q_1=q_2$ limit we count (\eg) 2d spin together with R-charge.} %
$q_1=q_2=q=e^{-\beta\hbar}$, so we will restrict to this case for the moment.
In its original version (Figure \ref{fig:GTconifold}, the geometric transition provided a duality between \emph{open} BPS invariants for $p$ Lagrangian branes in the deformed conifold geometry $T^*\S^3$ and \emph{closed} BPS invariants in the resolved conifold geometry $\CO(-1,-1)\to\cp^1$, where the K\"ahler parameter of the $\cp^1$ takes a special discrete value $Q=q^p$.%
\footnote{In \cite{GV-largeN}, the duality was phrased in terms of open topological string theory rather than BPS states, but the latter is more relevant for our perspective.} %
The transition was soon extended, however, to more general framed (compactified) Lagrangian branes in toric geometries --- precisely the types of Lagrangian branes we have been using to construct surface operators \cite{AV-engineering, AMV-allloop}.

\begin{figure}[htb]
\centering
\includegraphics[width=5in]{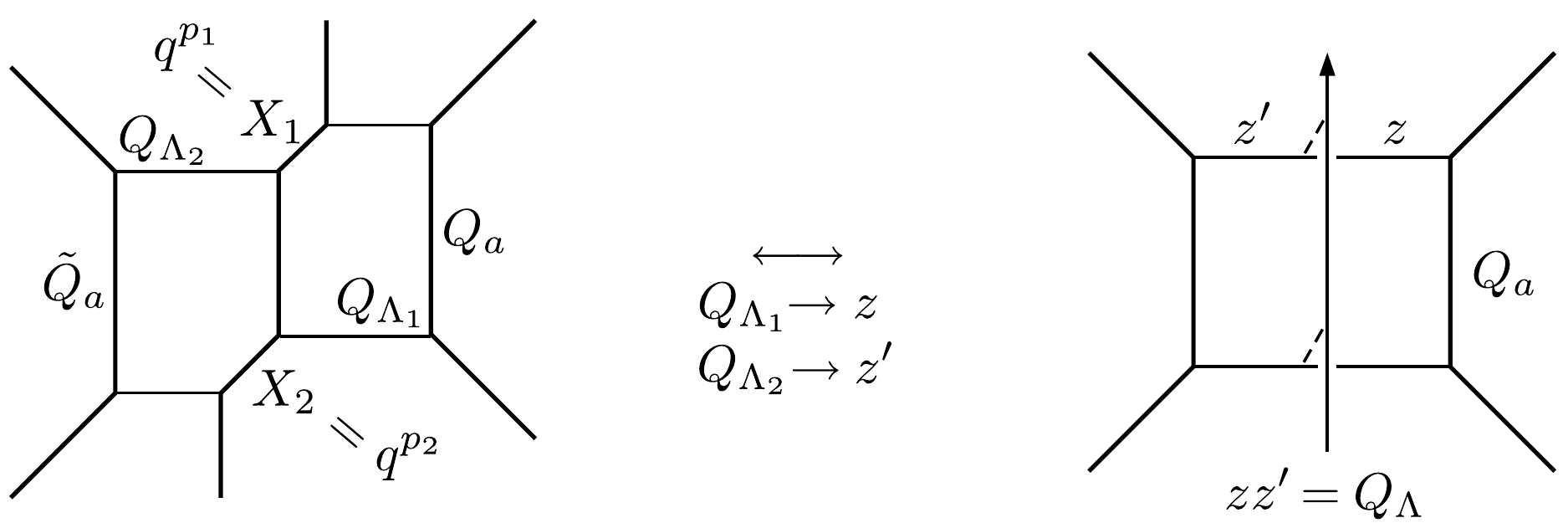}
\caption{An example of the geometric transition motivating \protect\eqref{GTgauge}. On the left, we have relations $\tilde{Q}_aX_1=Q_aX_2$, with $Q_a=e^{-\beta(a_2-a_1)},\;\tilde{Q}_a=e^{-\beta(\tilde{a}_2-\tilde{a}_1)}$\,. The bare bifundamental mass is $Q_{m_b}:=e^{-\beta m_b}=\sqrt{X_1X_2}$\,.}
\label{fig:GTgeneral}
\end{figure}

The typical situation that we are interested in is shown in Figure \ref{fig:GTgeneral}. The closed side, on the left, is a toric geometry that engineers $\CN=2$ $\tilde{G}\times G$ gauge theory (here for $G=\tilde{G}=U(2)$), with a bifundamental hypermultiplet. The bare mass of this hypermultiplet is related to Coulomb parameters as $-2m_b= a_1+a_2-\tilde{a}_1-\tilde{a}_2$. We could also insert additional fundamental matter for $G$ and antifundamental matter for $\tilde{G}$.

By setting the K\"ahler parameters $X_1,\,X_2$ to discrete values $X_1=q^{p_1}$ and $X_2=q^{p_2}$, one should reproduce BPS counting for the open geometry on the right hand side. This open geometry engineers $\CN=2$ gauge theory with gauge group $G$, and with a surface operator. The surface operator comes from two stacks of $p_1$ and $p_2$ Lagrangian branes, respectively, that are all framed by a single additional toric degeneration locus.

Note that in the $\tilde{G}\times G$ theory, the bifundamental mass parameter becomes
\be m_b=\frac{p_1+p_2}{2}\,\hbar\,, \label{p1p2mass} \ee
and the Coulomb parameters are related by $\tilde{Q}_a = q^{p_2-p_1}Q_a$. In terms of the parameters $2\tilde{a}=\tilde{a}_1-\tilde{a}_2$ and $2a=a_1-a_2$ of the $SU(2)\times SU(2)$ part of the theory, we have
\be \tilde{a}-a=\frac{p_1-p_2}{2}\,\hbar\,.\ee
Classically, $Q_a$ and $\tilde{Q}_a$ are equal, and they become identically equal if we place our branes symmetrically, with $p_1=p_2$.

Here we have not been too careful about shifts of $Q_{\Lambda_1},\,Q_{\Lambda_2},\,Q_\Lambda,\,z$ by powers of $q$. These do enter actual calculations --- see our examples further below --- but are not relevant in (\eg) the homological limit of equivariant instanton counting.

Note that in a geometric transition such as this, the BPS partition function on the closed $\tilde{G}\times G$ side corresponds to a BPS partition function on the open side where interactions between different Lagrangian branes are included \cite{AMV-allloop}. In other words, one counts BPS D2-branes stretching between Lagrangians, or, alternatively, worldsheet instantons connecting multiple Lagrangian branes.%
\footnote{Interactions of this type can by reproduced in topological vertex computations by inserting additional Ooguri-Vafa factors.} %
This is \emph{unlike} the multiple-surface-operator examples in Section \ref{sec:multiple}, where the Lagrangian branes and the surface operators were mutually noninteracting. For $p_1+p_2>1$ branes, we obtain a higher-rank two-dimensional gauge theory supported on a single, nonelementary surface operator.

For a single Lagrangian brane $(p_1+p_2=1)$ the geometric transition of Figure \ref{fig:GTgeneral} reproduces the duality of gauge theories in \eqref{GTgauge} in the case $G=U(2)$. It is clear that the duality should extend to general $U(N)$ or $SU(N)$ gauge groups, via geometric transitions in the corresponding toric diagrams. Moreover, given the discussion of vortex counting in Sections \ref{sec:GE}--\ref{sec:GEexamples}, sending $\Lambda_2\to 0$ in such geometries immediately leads to the 4d-2d equivalence \eqref{GTvortex}.

Although the geometric transition holds strictly only in the unrefined limit $q_1=q_2$, one might wonder whether it could be extended to provide a duality for refined BPS invariants, with $q_1\neq q_2$. At least in simple cases, the answer appears to be affirmative. Such simple cases include single elementary surface operators in the decoupled geometries that correspond to vortex counting. In the remainder of this section, we provide several explicit examples of this refined correspondence.

\subsubsection{Refined geometric transition}
\label{sec:refGT}

\begin{figure}[htb]
\centering
\vspace{.1in}
\includegraphics[width=4.7in]{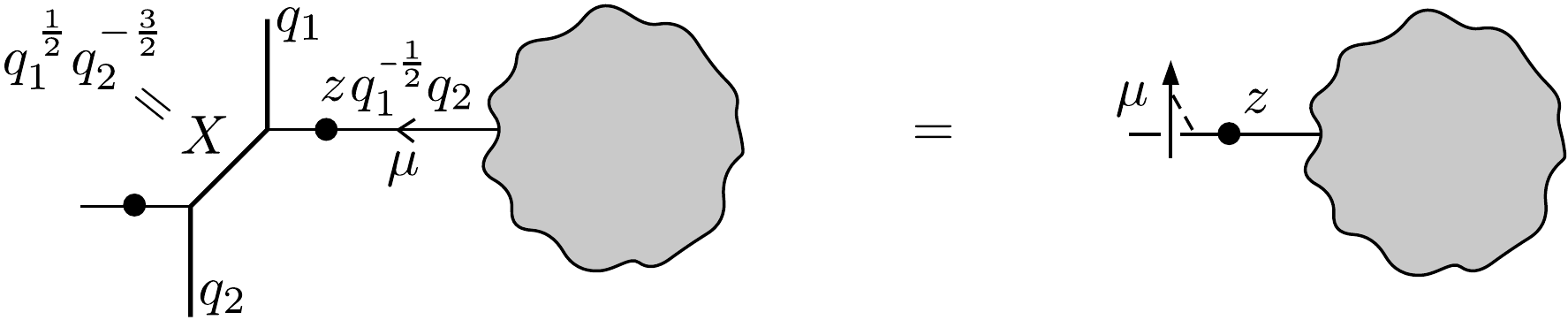}
\caption{A refined geometric transition on an external leg. The K\"ahler parameter $X=q_1^{\frac12}q_2^{-\frac32}$ for the $\cp^1$ on the left is equivalent to one Lagrangian brane on the right.}
\label{fig:GTblock}
\end{figure}

Let's consider surface operators in theories whose four-dimensional dynamics have been decoupled by sending $\Lambda\to0$, or $Q_\Lambda\to 0$. In the corresponding engineering geometries, all Lagrangian branes are attached to external legs. We can then understand the basic refined geometric transition in terms of the building block shown in Figure \ref{fig:GTblock}.

The shaded region in this figure represents the remaining part of an arbitrary toric geometry.%
\footnote{Strictly speaking, the remainder of this toric geometry should not intersect the framing locus of the Lagrangian branes, when this framing locus is infinitely extended. A potential intersection is not a serious concern for unrefined amplitudes (at $q_1=q_2$), but it may cause problems in refined partition functions --- see for example our $U(2)$ example in Section \ref{sec:GTU2}.} %
Denoting the part of the diagram on the external leg as $Z_{\mu^t}(z)$, it is straightforward to check the explicit algebraic relation
\begin{align} Z^{\rm closed}_{\mu^t}(z) &:= \sum_\mu\frac{\sum_\lambda C_{\lambda\circ\circ}(q_1,q_2)(-X)^{|\lambda|}C_{\lambda^t\circ\mu^t}(q_2,q_1)}{\sum_\lambda C_{\lambda\circ\circ}(q_1,q_2)(-X)^{|\lambda|}C_{\lambda^t\circ\circ}(q_2,q_1)}(-zq_1^{-\frac12}q_2)^{|\mu|} \label{Xpreprod} \\
 &= s_{\mu^t}(z) =: Z^{\rm open}_{\mu^t}(z)\,. \label{Xpreprodopen}
\end{align}
Note that the closed partition function here has been normalized so that $Z_{\mu^t}^{\rm closed}(z=0)=1$, the same normalization we would want for the open partition function. The relation in Figure \ref{fig:GTblock} therefore equates refined open BPS counting in the presence of a single Lagrangian brane to refined closed BPS counting with an extra $\cp^1$.

(A careful reader may notice that the unrefined version of the closed K\"ahler parameter $X$ is $X=q^{-1}$, rather than $X=q$ as in the preceding discussion. This is a rather trivial distinction, which can be understood in terms of placing single-row rather than single-column partitions on the brane, or a surface operator in the $\epsilon_1$ plane rather than the $\epsilon_2$ plane.)

More generally, let us try to keep $X$ arbitrary, and look at the closed side of a putative refined geometric transition. The closed partition function in \eqref{Xpreprod} takes the form:
\be Z^{\rm closed}_{\mu^t} = \left(\frac{q_1}{q_2}\right)^{\frac{|\!|\mu^t|\!|^2}{2}}P_{\mu}(q^{-\rho};t,q) \prod_{i,j=1}^\infty \frac{1-Xq_1^{i-\frac12-\mu^t_j}q_2^{j-\frac12}}{1-Xq_1^{i-\frac12}q_2^{j-\frac12}}\,. \label{Xprod} \ee
If we set
\be X = q_1^{r-\frac12}q_2^{\frac12-s}\, \ee
for integers $r,s\geq 1$, then the product in \eqref{Xprod} vanishes unless
\be \mu^t_s \leq r-1\,. \ee
In particular, taking $(r,s)=(1,1)$ forces $\mu=\circ$, and corresponds to an ``open'' partition function with no branes at all.

\begin{figure}[htb]
\centering
\includegraphics[width=1.8in]{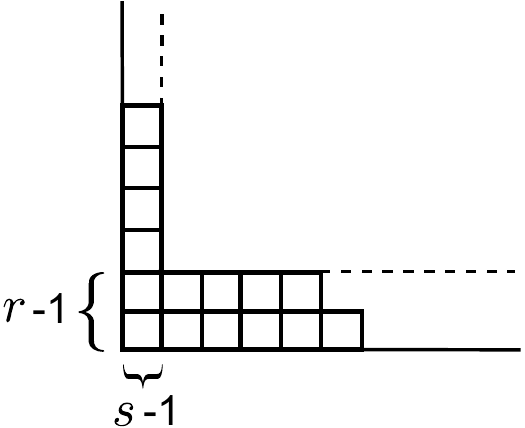}
\caption{A ``hook-shaped'' partition with $r-1=2$ rows and $s-1=1$ column.}
\label{fig:hookpartition}
\end{figure}

A general choice of $r$ and $s$ restricts $\mu^t$ to be ``hook-shaped,'' with at most $r-1$ rows and $s-1$ columns, as in Figure \ref{fig:hookpartition}. We optimistically expect that via open-closed duality this choice of $X$ would engineer a nonelementary surface operator supported on the surface
\be
\label{rssupp}
w_1^{r-1} w_2^{s-1}=0 \quad \subset \R^4\,,
\ee
where $w_1$ and $w_2$ are complex coordinates on $\R^2_{\epsilon_1}$ and $\R^2_{\epsilon_2}$, respectively.
Equivalently, it can be described in terms of a two-dimensional $\CN=(2,2)$ gauge theory
with gauge group $U(r-1) \times U(s-1)$, where the enhanced non-abelian gauge symmetry
is a consequence of the multiplicity $(r-1)$ resp. $(s-1)$.
Note, that in the brane construction of surface operators one arrives at the same conclusion.

Although we will not consider complete, non-decoupled theories in our examples (largely due to the complicated nature of the refined geometric transition and the refined vertex in non-decoupled geometries), it is interesting to imagine a putative refined version of a transition such as in Figure \ref{fig:GTgeneral}. Working in $U(N)\times U(N)$ four-dimensional theory, we can set $N$ K\"ahler parameters $X_i$ equal to $q_1^{\,r_i-\frac12}q_2^{\,\frac12-s_i}$. Presumably, this would transition to a $(r,s)$-type surface operator in $U(N)$ theory, with $r-1=\sum_{i=1}^N(r_i-1)$ and $s-1=\sum_{i=1}^N(s_i-1)$. The corresponding formula for the bare bifundamental mass in the $U(N)\times U(N)$ theory, which generalizes \eqref{p1p2mass}, is
\be
-m_b = \frac{\epsilon_1+\epsilon_2}{2}+\frac{r-1}{N}\epsilon_1+\frac{s-1}{N}\epsilon_2\,.
\label{GTrefmass}
\ee

\subsubsection{$U(1)$ theory}
\label{sec:GTU1}

The building block of Figure \ref{fig:GTblock} can be applied directly to a $U(1)$ theory in the 2d-4d decoupling limit $Q_\Lambda\to 0$. Let us take the simple decoupled geometry in Figure \ref{fig:U1brane}(b) of Section \ref{sec:U1surf}, corresponding to a surface operator in pure $\CN=2$ Maxwell theory. The closed geometry obtained via geometric transition is shown below in Figure \ref{fig:U1GT}. A specialization of relation \eqref{Xpreprod}--\eqref{Xpreprodopen} assures us that with parameters as given we must have
\be Z^{\rm closed}(z;q_1,q_2) = Z^{\rm open}_{\rm BPS}(z;q_1,q_2) = \prod_{i=1}^\infty\frac{1}{1-q_1^{i-1}q_2^{\,\frac12}z}\,, \ee
reproducing the open result in \eqref{U1vxopenprod}. In the limit $q_1\to q=e^{-\beta\hbar}$, $q_2\to 0$, we know that there is a relation $Z^{\rm open}_{\rm BPS}(z;q,1)= Z^{\rm vortex}_{\rm K-theory}(z;q)$. However, on the closed side of the transition, $Z^{\rm closed}(z;q,1)$ can also be interpreted as the 5d instanton partition function for a $U(1)$ gauge theory with one antifundamental hypermultiplet of mass $-m_b=\frac{\epsilon_1}{2}+\frac{3\epsilon_2}{2}$, in the limit $\epsilon_2\to 0$. Thus there is an equivalence of instanton/vortex counting in
\be \boxed{\begin{array}{c}
 \text{4d $U(1)$ theory w/} \\
 \text{antifundamental $m_b=-\hbar/2$} \end{array}}
 \quad\longleftrightarrow\quad
 \boxed{\text{2d $U(1)$ theory w/ massless chiral\,.}}
\ee
We could keep both $\epsilon_1$ and $\epsilon_2$ as parameters if we counted vortices with respect to R-charge as well as spin.

\begin{figure}[htb]
\centering
\includegraphics[width=3.8in]{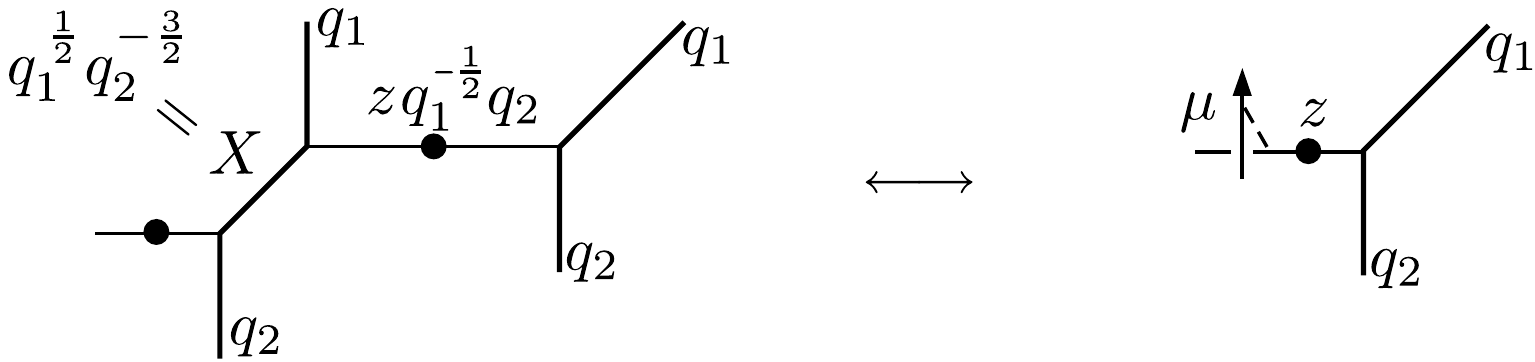}
\caption{Refined geometric transition for a surface operator in $U(1)$ theory.}
\label{fig:U1GT}
\end{figure}

\begin{figure}[htb]
\centering
\includegraphics[width=4in]{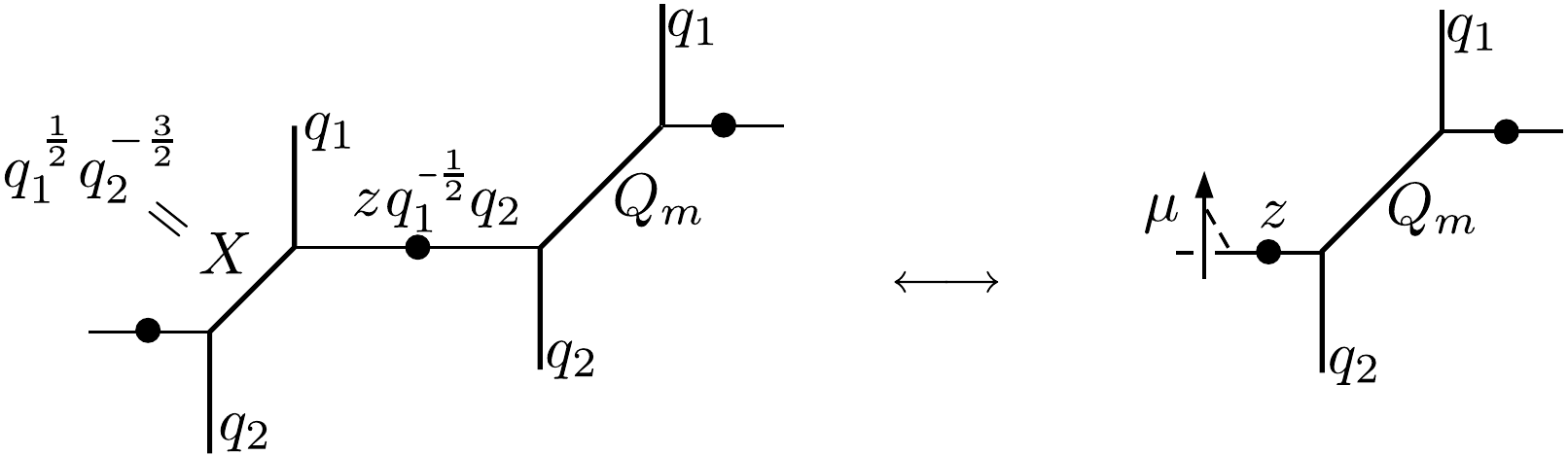}
\caption{Refined geometric transition for $U(1)$ theory with matter.}
\label{fig:U1mGT}
\end{figure}

Similarly, we may add a fundamental flavor to the four-dimensional theory, as in Figure \ref{fig:U1mGT}. Relation \eqref{Xpreprod}--\eqref{Xpreprodopen} assures us that
\be Z^{\rm closed}(z,Q_m;q_1,q_2) = Z^{\rm open}_{\rm BPS}(z,Q_m;q_1,q_2) = \prod_{i=1}^\infty \frac{1-q_1^{i-1/2}Q_mz}{1-q_1^{i-1}q_2^{1/2}z} \overset{\overset{q_2\to1}{\beta\to0}}{\longrightarrow} Z^{\rm vortex}(z,\tilde{m};\hbar)\,,\ee
matching the open formula of Section \ref{sec:U1mbrane}. Now the precise equivalence at $\epsilon_2\to 0$ is
\be \boxed{\begin{array}{c}
 \text{4d $U(1)$ theory w/ fundamental ($m$),} \\
 \text{antifundamental ($m_b=-\hbar/2$) } \end{array}}
 \quad\longleftrightarrow\quad
 \boxed{\begin{array}{c}
 \text{2d $U(1)$ theory w/ massless fund.} \\
 \text{+ antifundamental  $(\tilde{\bar{m}}\simeq m$)\,.}
 \end{array}}
\ee

\subsubsection{$U(2)$ theory}
\label{sec:GTU2}

Let us finally consider a surface operator in the decoupled limit of four-dimensional $U(2)$ theory. For later comparison with CFT results, we will add two flavors of fundamental matter. (Superconformal $SU(2)$ theory would have two flavors of antifundamental matter as well, but these decouple in the $Q_\Lambda\to 0$ limit.) The open BPS partition function in this case was given in \eqref{Zbottom}--\eqref{Ztop} of Section \ref{sec:generalsurf}, for two different choices of Lagrangian brane placement.

\begin{figure}[htb]
\centering
\includegraphics[width=4.5in]{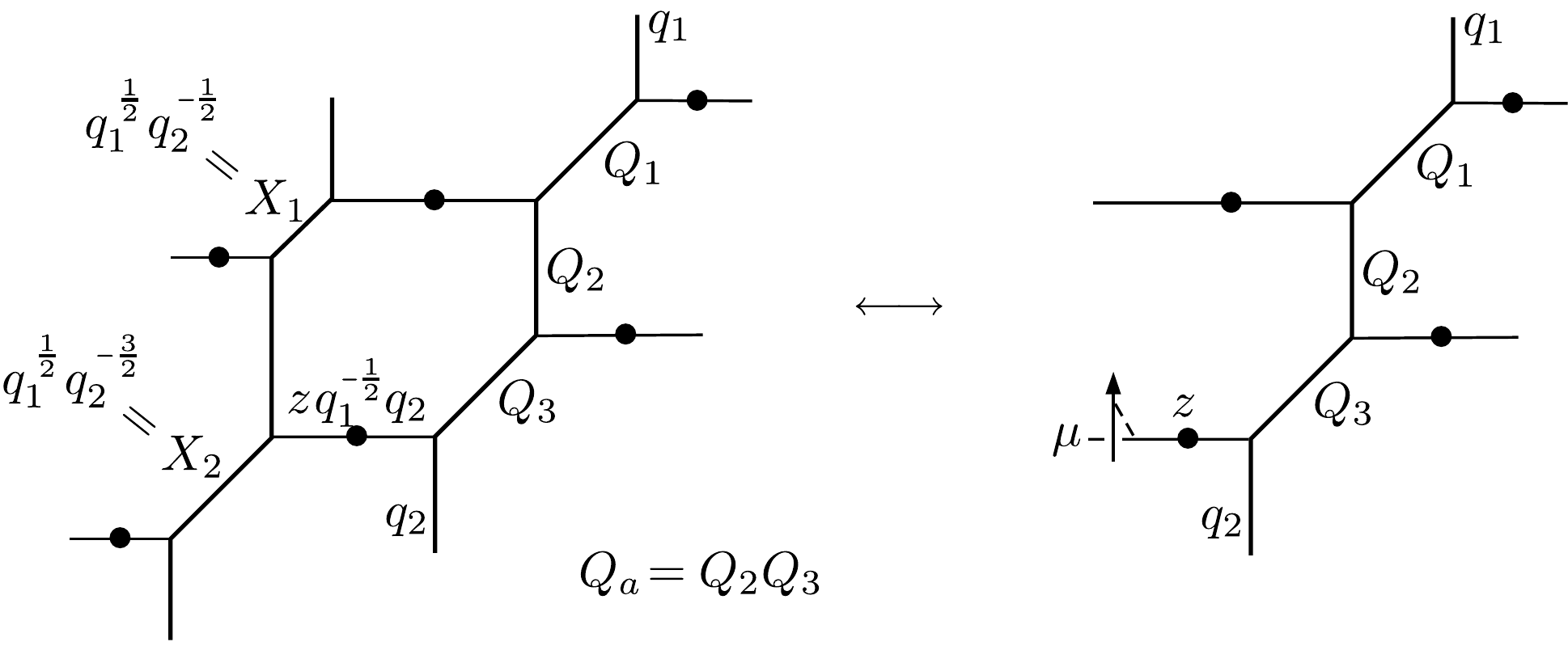}
\caption{Refined geometric transition for a brane on the bottom leg in decoupled $U(2)$ theory.}
\label{fig:GTbottom}
\end{figure}

For $U(2)$ theory, the building block of Figure \ref{fig:GTblock} cannot be applied directly. The reason, as shown in Figure \ref{fig:GTbottom}, is that the framing locus of a Lagrangian brane on one of the gauge legs always passes across the other gauge leg. In the case of the unrefined geometric transition, this would not have been a problem: one would simply assign trivial K\"ahler parameter $X_1=1$ to the potential coming from the resolution of this second crossing. But we saw in Section \ref{sec:refGT} that there is no such thing as a trivial parameter in a refined transition: zero branes corresponds to $X_1 = q_1^{\,\frac12}q_2^{-\frac12} \neq 1$.

\begin{figure}[htb]
\centering
\includegraphics[width=4.5in]{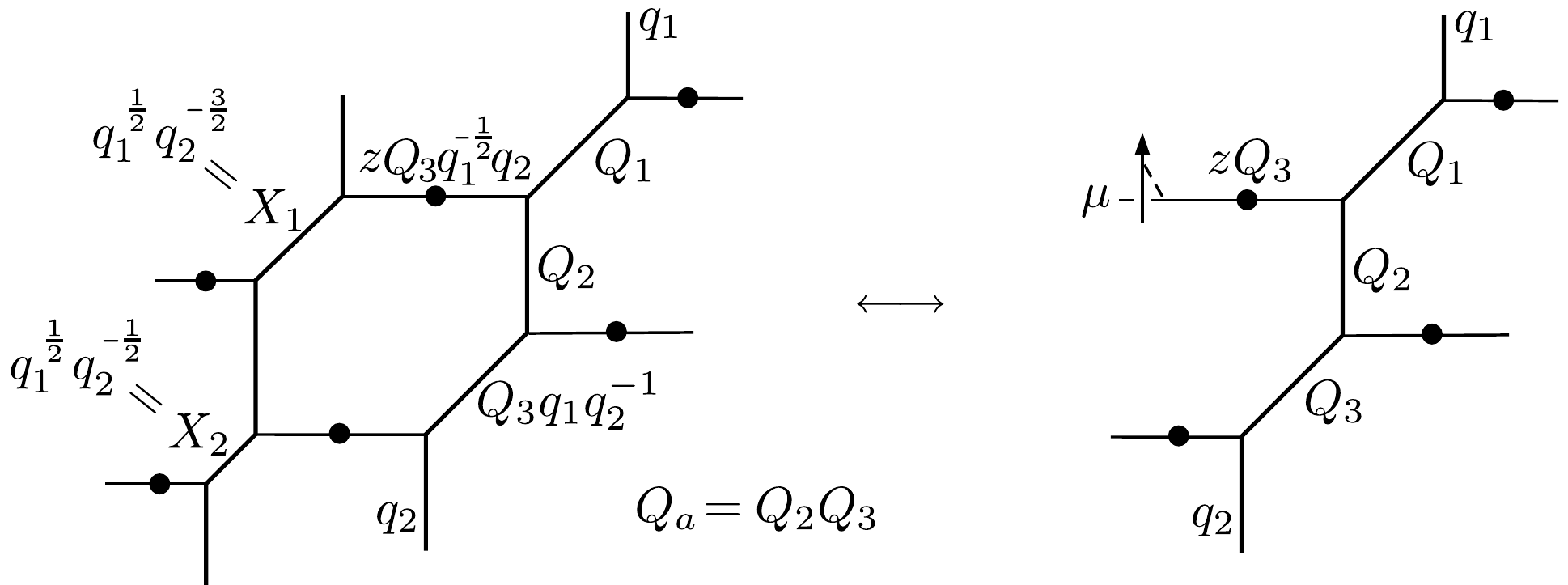}
\caption{Alternative brane placement (top leg) in the geometric transition.}
\label{fig:GTtop}
\end{figure}

Despite this difficulty, the refined geometric transitions for single branes in the $U(2)$ geometry still turn out to work. For a brane on the bottom leg (Figure \ref{fig:GTbottom}), we find
\begin{align} Z^{\rm closed}(z,Q_1,Q_2,Q_3) &= Z^{\rm open,\,bottom}_{\rm BPS}(z,Q_1,Q_2,Q_3) \notag \\
&= \sum_{\m=0}^\infty \prod_{j=1}^{\m}\frac{(1-Q_3q_2^{-\frac12}q_1^{j-\frac12})(q-Q_aQ_1 q_2^{-\frac12}q_1^{j-\frac12})}{
(1-q_1^j)(1-Q_aq_2^{-1}q_1^{j})}q_2^{\m/2}z^\m\,.
\label{GTbottom}
\end{align}
For a brane on the top leg, it is necessary to shift $Q_3\to Q_3 q_1q_2^{-1}$ while keeping $zq_1^{-\frac12}q_2Q_3\to zq_1^{-\frac12}q_2Q_3$ constant, as in Figure \ref{fig:GTtop}, in order to obtain the same partition function as \eqref{Ztop}. This shift can be traced to the extra interaction between the framing locus and the bottom gauge leg, as just discussed, and the new nontrivial instantons thereby created from the resolved $\cp^1$ in the closed geometry. Then we find the exact equivalence
\begin{align} Z^{\rm closed}_{\rm BPS}(z,Q_1,Q_2,Q_3) &= Z^{\rm open,\,top}_{\rm BPS}(z,Q_1,Q_2,Q_3) \notag \\
 &= \sum_{\m=0}^\infty\prod_{j=1}^\m \frac{(1-Q_1q_2^{-\frac12}q_1^{j-\frac12})(1-Q_2^{-1}q_2^{-\frac12}q_1^{j-\frac12}))}{(1-q_1^j)(1-Q_a^{-1}q_1^{j-1})}q_1^{-\frac{\m}{2}}q_2^{\,\m}z^\m\,.
\label{GTtop}
\end{align}

With either placement of the branes, an appropriate identification of $Q_{1,2,3}$ with 2d fundamental and antifundamental masses as discussed in Section \ref{sec:generalsurf} leads to $Z^{\rm open}_{\rm BPS}(z;q_1,q_2) \to Z^{\rm vortex}(z;\hbar)$ as $q_1\to q=e^{-\beta\hbar}$, $q_2\to 1$, and $\beta\to 0$. On the other hand, the closed geometries in Figures \ref{fig:GTbottom}--\ref{fig:GTtop} engineer four-dimensional $U(2)$ theories with antifundamental matter of (bare) masses $-\bar{m}_1=a_1+\frac{\epsilon_1+\epsilon_2}{2}$ and $-\bar{m}_2=a_2+\frac{\epsilon_1+3\epsilon_2}{2}$ (or $-\bar{m}_1=a_1+\frac{\epsilon_1+3\epsilon_2}{2}$, $-\bar{m}_2=a_2+\frac{\epsilon_1+\epsilon_2}{2}$). Sending $\epsilon_1\to\hbar$ and $\epsilon_2\to 0$, we find the equivalence of instanton/vortex counting in
\be
\boxed{\begin{array}{c}
\text{4d $U(2)$ theory w/} \\
\text{2 fund. + 2 antifund. hypers,} \\
\text{antifund masses $\bar{m}_i+a_i=-\hbar/2$} \end{array}}
\quad\longleftrightarrow\quad
\boxed{\begin{array}{c}
\text{2d $U(1)$ theory w/} \\
\text{1 massless + 1 massive fund.,} \\
\text{2 massive antifund. chirals\,.}
\end{array}}
\ee


\subsubsection{Nonelementary surface operators} 
\label{sec:GTp}

As a final example, we consider a geometric transition involving multiple branes. Since the refined geometric transition is not fully well understood in such cases,%
\footnote{In particular, the proper way to include multiple branes on the open side, with potential Ooguri-Vafa factors, has not yet been worked out. It is likely that the answer will involve MacDonald functions associated to the branes, as in \eqref{McD}.} %
we will restric to the unrefined limit $q_1=q_2=q=e^{-\beta\hbar}$. Thus, in two-dimensional surface-operator theories, we will count a combination of spin and R-charge.

\begin{figure}[htb]
\centering
\includegraphics[width=4.5in]{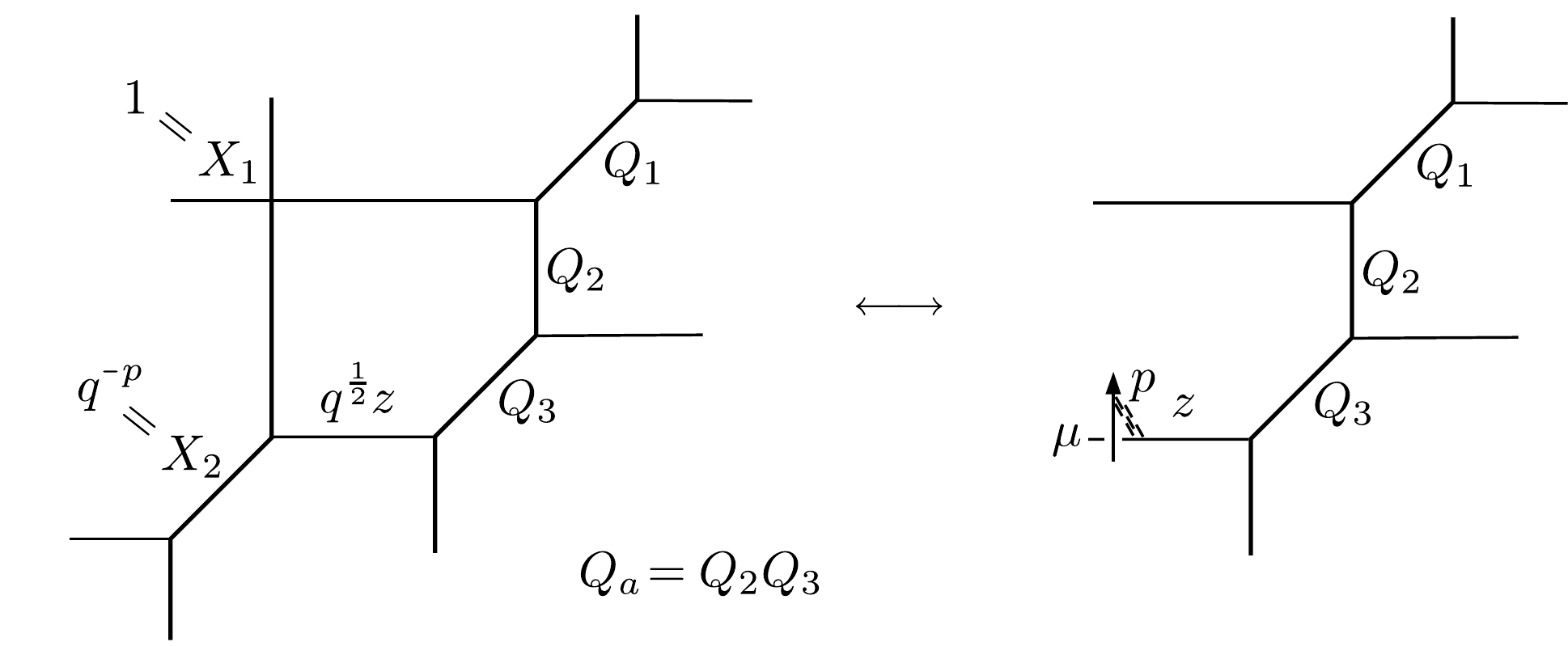}
\caption{Unrefined geometric transition for $p$ branes on the bottom leg of a decoupled $U(2)$ geometry.}
\label{fig:GTpbranes}
\end{figure}

Let's take the simplest case of $p$ branes on the bottom leg of an $SU(2)$ geometry, in the decoupled limit $Q_\Lambda\to 0$. We place the branes as if they had engineered a surface operator in the $\epsilon_1$-plane before the unrefined limit was taken --- \ie\ we choose parameters so that the partition $\mu$ on the branes is restricted to have $p$ \emph{rows}. On the closed side, the appropriate geometry is shown in Figure \ref{fig:GTpbranes}. Now that $q_1=q_2$, we can simply take closed K\"ahler parameters to be $X_2=q^{-p}$, and $X_1=1$ (completely eliminating interactions between the two legs). The normalized closed partition function is completely equivalent to the open partition function that corresponds to the right of Figure \ref{fig:GTpbranes} --- where the branes are now associated with a factor
\be s_{\mu^t}(z,q^{-1}z,...,q^{-p+1}z) = z^{|\mu|}q^{-n(\mu^t)}\prod_{(i,j)\in\mu^t}\frac{(1-q^{p+j-i})}{(1-q^{-h(i,j)})}\,. \label{Schurp} \ee
(As usual, $n(\mu)=\frac{|\!|\mu^t|\!|}{2}-\frac{|\mu|}{2}$ and $h(i,j)=\mu_i+\mu^t_j-i-j+1$ is the hook length.) In other words,
\begin{align} &Z^{\rm closed}_{\rm BPS}(z,Q_1,Q_2,Q_3) =
 Z^{\rm open}_{\rm BPS}(z,Q_1,Q_2,Q_3) \\
  &\qquad= \sum_\mu z^{|\mu|} s_{\mu^t}(1,q^{-1},...,q^{-p+1}) \prod_{(i,j) \in\mu}\frac{(1-Q_3q^{i-j})(1-Q_1Q_2Q_3q^{i-j})}{(1-q^{h(i,j)})(1-Q_2Q_3q^{i-j})}\,. \label{GTpUR}
\end{align}
Note that the Schur function of $p$ variables  \eqref{Schurp} restricts $\mu$ to have at most $p$ rows.

We have engineered a surface operator supported on the surface (scheme) $w_2^p=0$ in $\R^4$.
Physically, we should think of this operator as containing a $U(p)$ worldvolume theory in a fully un-Higgsed phase.
Interactions between the $p$ branes that compose the surface operator are completely absorbed
in the shifts $z\to q^{-i}z$ in the Schur function eigenvalues, as in \eqref{Schurp}.
For a two-dimensional $U(p)$ theory,
there is a single exponentiated FI parameter $z$ that functions as the vortex-counting parameter.
In terms of a Levi classification, we \emph{still} have $\LL = U(2)$ (for a 4d $G=U(2)$ theory) and $\Lambda_\LL\simeq \Z$.
However, instead of a basic singularity \eqref{asing} for the gauge field in four dimensions,
nonelementary surface operators with $p>1$ produce higher order singularities
(corresponding to poles of order $p$), in the mathematical literature known as {\it wild ramification}.
With $p$ branes, the duality between instanton counting in 4d and vortex counting in 2d now looks like
\be
\boxed{\begin{array}{c}
\text{4d $U(2)$ theory w/} \\
\text{2 fund. + 2 antifund. hypers,} \\
\text{antifund masses $-\bar{m}_1=a_1$,}\\\text{$-\bar{m}_2=a_2+p\hbar$} \end{array}}
\quad\longleftrightarrow\quad
\boxed{\begin{array}{c}
\text{2d $U(p)$ theory w/} \\
\text{1 massless + 1 massive fund.,} \\
\text{2 massive antifund. chirals\,.}
\end{array}}
\ee
The homological limit $\beta\to0$ of \eqref{GTpUR} gives a prediction for the vortex partition function of $\CN=(2,2)$ $U(p)$ theory, now counting vortices with respect to spin plus R-charge.


\section{Comparison with conformal field theory}
\label{sec:cft}

It has been suspected for some time that four-dimensional (homological) instanton counting in $\CN=2$ gauge theories should be related to conformal field theories. The simplest such examples involved abelian $U(1)$ theory and a free-fermion or free-boson CFT \cite{LMN, NO}.
%
%
In the past year, great progress was made in extending such $U(1)$ results to $\CN=2$ theories with gauge group $SU(N)$ or a product of $SU(N)$ factors \cite{AGT}, and Wilson loops and surface operators were added to the gauge theory/CFT correspondence \cite{AGGTV, DrukkerTeschner, AT}. In this section, we will make some brief connections between vortex partition functions and conformal field theory. Indeed, the results of \cite{AGGTV} can be used to reformulate as CFT conformal blocks many of the vortex/BPS partition functions previously described, and we will give examples in the case of $U(1)$ and $SU(2)$ 4d theories.

The relation between $\CN=2$ gauge theories with group $G\times G$ and theories with group $G$ and a surface operator, which was motivated in Section \ref{sec:GT} via geometric transitions, may be also much more familiar in the context of conformal field theory. In \cite{AGGTV}, degenerate vertex operator insertions were used to engineer surface operators for $\CN=2$ gauge theories. However, conformal blocks with degenerate insertions can also be interpreted as special limits of ordinary conformal blocks, which in turn count instantons in four-dimensional theories with product gauge groups.
We find that such degenerate limits in CFT map perfectly to the picture of geometric transitions.

\subsection{Degenerations and decouplings}
\label{sec:CFTdd}

In \cite{AGT}, homological instanton partition functions of four-dimensional gauge theory are expressed as CFT conformal blocks. Recall that arbitrary $n$-point functions in a CFT can be computed from conformal blocks and 3-point-function coefficients. The basic idea (\cf\ \cite{DiFrancesco}) is to reduce an $n$-point function (say of primary fields) to 3-point functions by inserting complete bases of states:
\begin{align} &\langle V_\infty(\infty)V_{n-2}(1)\cdots V_1(z_1)V_0(0)\rangle \notag \\
 &\leadsto \int d^{n-3}\alpha \sum_{\{\mb{k}_i,\mb{k}_i'\}} \langle V_\alpha(\infty)V_{n-2}(1)|\alpha_{n-3},\mb{k}_{n-3}\rangle H_{\mb{k}_{n-3},\mb{k}'_{n-3}}^{-1}\langle \alpha_{n-3},\mb{k}_{n-3}'|\cdots \label{Fdef}\\
 &\hspace{2.8in} \cdots |\alpha_1,N_1\rangle H_{\mb{k}_1,\mb{k}_1'}^{-1} \langle \alpha_1,\mb{k}_1'|V_1(z_1)V_0(0)\rangle\,, \notag
\end{align}
where the $\alpha_i$ label the primary states and the $\mb{k}_i,\mb{k}_i'$ label descendants. The inverse Gram matrix $H_{\mb{k},\mb{k}'}^{-1}$ provides a proper normalization for these complete bases. The conformal block $\CF_{V_\infty\;\;\;\alpha_{n-3}\;\;\cdots\;\;\alpha_1\;\;\;V_0}^{\;\;\;V_{n-2}\;\;\;\cdots\;\;\; V_2\;\;\;V_1}(z_{n-3},...,z_1)$ (here, for a sphere $n$-point function) is the holomorphic part of the quotient of the integrand in \eqref{Fdef} by itself with all descendants $N_i,N_i'$ set to zero --- \ie\ the integrand normalized by 3-point functions of primaries. It can be computed order-by-order in the $z_i$ using only the operator algebra of the CFT. This particular block is associated to the diagram
%
\be \raisebox{-.4in}{\includegraphics[height=.8in]{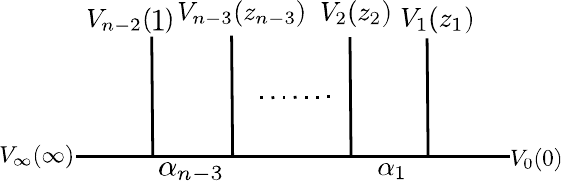}\,\,.}
 \label{nptblock} \ee
Just like the instanton partition function encodes information in a topological sector of a gauge theory, the conformal blocks encode the robust, ``topological'' part of a CFT, \emph{i.e} the part that is not dependent on the particular conformal model that is being considered.

A conformal block diagram like \eqref{nptblock} can be thought of as the skeleton of the ``Gaiotto curve'' \cite{Gaiotto-dualities}, a quotient of a gauge theory's Seiberg-Witten curve. External vertex operators $V_i$ correspond to punctures on the curve, and their momenta are related to bare masses of matter in the gauge theory, while internal momenta $\alpha_i$ are related to Coulomb parameters. The positions $z_i$ of insertions determine the complex structure of the curve, hence (for conformal theories) they determine the UV couplings $e^{2\pi i\tau}\sim$``$\Lambda$'' that function as instanton-counting parameters.

For example, instanton counting in an $SU(N)$ superconformal theory, with $N$ fundamental and $N$ antifundamental flavors, is given by the 4-point block
\be
\raisebox{-.4in}{\includegraphics[height=.75in]{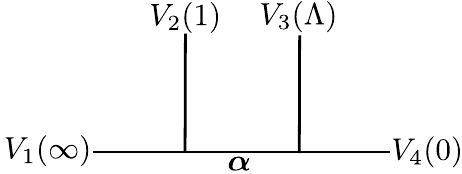}}
 \label{4block}
\ee
in $A_{N-1}$ Toda theory. The internal momentum $\bm{\alpha}$, a vector of length $N$-1, encodes the Coulomb parameters of the theory, while the external momenta of the operators $V_i$ are linear combinations of the fundamental and antifundamental bare mass parameters. As explained in \cite{Wyllard-AN, Tachikawa-Toda}, one should choose $V_2$ and $V_3$ to be level-1 degenerate for all higher $W$-algebras. Then $V_2,\,V_3$ are each associated with a single independent momentum, while $V_1,\,V_4$ each have a full complement of $N-1$ momenta, so that the total number of independent external momenta is exactly $2N$. Similarly, the 5-point block
\be \raisebox{-.4in}{\includegraphics[height=.8in]{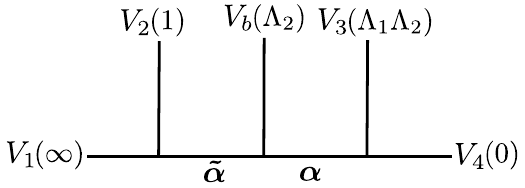}}\label{5block} \ee
corresponds to $SU(N)\times SU(N)$ superconformal theory, with $N$ fundamental (or antifundamental) hypermultiplets for each factor and one bifundamental. The operators $V_2$, $V_b$, $V_3$ are all level-1 degenerate for higher $W$-algebras. In particular, the single unconstrained momentum of $V_b$ encodes the bifundamental bare mass.

In \cite{AGGTV}, instanton counting in the presence of surface operators was conjectured to be reproduced by conformal blocks with fully degenerate insertions. For example, in Liouville theory, a 5-point conformal block with a center vertex operator that is $(r,s)$-degenerate for the Virasoro algebra,
\be \raisebox{-.4in}{\includegraphics[height=.8in]{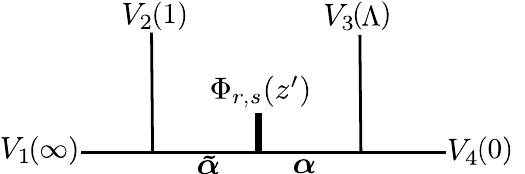}\,\,,}  \label{5blockd} \ee
should correspond to \emph{some} surface operator in superconformal $SU(2)$ theory. A similar construction exists for $A_{N-1}$ Toda theory, where the middle insertion can roughly be thought of as level-1 degenerate for all higher $W$-algebras and $(r,s)$-degenerate for Virasoro. Such diagrams should be compared to open toric engineering geometries for $SU(N)$ theory with surface operators, as in Figure \ref{fig:branechoices2}(b) or the right side of Figure~\ref{fig:GTgeneral}.

Now, in principle, we could ignore the fact that the momentum of $\Phi_{r,s}$ in \eqref{5blockd} is Virasoro-degenerate, and think of the conformal block as reproducing instead a $SU(N)\times SU(N)$ theory with bifundamental matter, as in \eqref{5block}. Then \eqref{5blockd} is just a limit of \eqref{5block} where the bifundamental mass takes a special small, discrete value. But this is exactly the setup we encountered on the closed side of geometric transitions for toric geometries, as in the left side of Figure \ref{fig:GTgeneral}. And now we have a good idea of what \emph{kind} of surface operator the insertion $\Phi_{r,s}$ should create: a non-elementary operator supported on the surface
\be w_1^{r-1}w_2^{s-1}=0\quad\subset\;\R^4\,,\ee
as in \eqref{rssupp}. The relation to geometric engineering also clarifies the meaning of approximate momentum conservation in the CFT. In the presence of a degenerate operator, the degenerate OPE forces the difference $\tilde{\bm{\alpha}}-\bm{\alpha}$ in a block like \eqref{5blockd} to take one of a discrete, finite set of values. These values translate directly to the choices of arrangements of $r+s-2$ Lagrangian branes on toric gauge legs, on the open side of a geometric transition.%
\footnote{Such a correspondence was also suggested in \cite{AGGTV}, in the language of D2-D4-NS5 brane-engineering models for surface operators. There, placement of toric Lagrangian branes translated to placement of D2-branes.}

Throughout most of this paper, we have focused on 2d vortex counting in the decoupling limit $\Lambda\to 0$. Let us simply note that it is very easy to pass to this limit in CFT as well. For a conformal block like \eqref{5blockd}, taking $\Lambda\to 0$ corresponds to ``cutting'' on either the $\bm{\tilde{\alpha}}$ or $\bm{\alpha}$ legs --- \cf\ Figure \ref{fig:branechoices2} --- and replacing the cut momentum with an external vertex operator. For example, if we choose to keep $z$ as a vortex counting parameter, we should cut on $\bm{\tilde{\alpha}}$, obtaining
\be  \raisebox{-.4in}{\includegraphics[height=.8in]{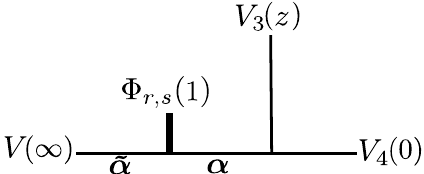}\,\,.} \ee
This conformal block \emph{directly} reproduces vortex counting in the 2d $\CN=(2,2)$ theory on the surface operator itself. Just as in Section \ref{sec:GT}, we could alternatively interpret this block as instanton counting in a 4d $SU(N)$ theory with $N$ fundamental and $N$ antifundamental chirals, where the bare masses of the antifundamentals (encoded in $\bm{\tilde{\alpha}}$) have small, finite shifts away from the Coulomb parameters (encoded in $\bm{\alpha}$).

We proceed to illustrate these ideas explicitly for $U(1)$ and $SU(2)$ gauge theories, and their corresponding free-boson (or $U(1)$ current) and Liouville CFT's.


\subsection{$U(1)$ CFT}
\label{sec:CFTU1}

Instanton partition
functions of $U(1)$ gauge theories can be reformulated
as correlation functions of a free boson/free fermion theory
\cite{LMN,NO,C-O}. In this theory a bifundamental hypermultiplet of mass $m$ is associated with a vertex operator  
\begin{align}
V_m(z) =  \exp\left( \frac{- m + \frac{\e_1 + \e_2}{2}}{\sqrt{\e_1 \e_2}} \phi_-(z)  \right)  \exp\left(  \frac{ m + \frac{\e_1 + \e_2}{2}}{\sqrt{\e_1 \e_2}} \phi_+(z)  \right)
\end{align}
that acts on the free-boson Fock space.
Here, $\phi(z) = \phi_-(z) + \phi_+(z)$ is the decomposition of the free boson field in positive
and negative modes. An elementary surface operator in the $\e_1$-plane
should correspond to a vertex operator $\Phi_{1,2}(z)$
%
%
with degenerate mass $-m_b = \frac{1}{2} \e_1 + \frac{3}{2} \e_2$. 

In the unrefined limit $\e_2 = - \e_1$ the degenerate vertex operator
$\Phi_{1,2}(z)$ is simply a free fermion field $\Psi(z)$. 
This is expected from the description of a surface operator as a
Lagrangian brane in the open BPS theory, since such a Lagrangian brane 
is well-known to have a dual description as a free fermion field  \cite{adkmv,remodelingBmodel,DHSV}.

Let us consider the superconformal four-dimensional $U(1)$ theory, coupled to a fundamental hypermultiplet of mass $m$
and an antifundamental hypermultiplet of mass~$\overline{m}$\,.
The instanton partition function of this theory has a dual interpretation
as a 4-point block (for $U(1)$ current algebra) of the free boson/fermion theory on a sphere \cite{LMN},
\be \raisebox{-.4in}{\includegraphics[height=.8in]{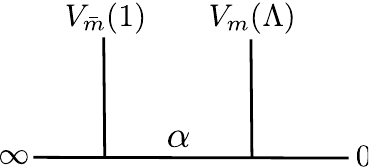}\;\;.} \ee
The two vertex operators $V_m$ and $V_{\overline{m}}$ correspond to the two hypermultiplets,
and the internal momentum $\alpha$ encodes the Coulomb parameter $a$ of the gauge theory.

\begin{figure}[t] \centering
\includegraphics[width=4.5in]{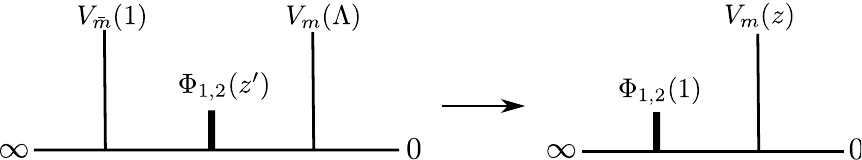}
\caption{Five-point conformal block corresponding to a surface operator in the
four-dimensional $U(1)$ gauge theory coupled to a fund. hyper of
mass $m$ and an antifund. hyper of mass $\bar{m}$. In the decoupling
limit $\Lambda \to 0$ this conformal block is mapped to the four-point block on the right.  \label{CFTSOU1}}
\end{figure}

Adding an elementary surface operator to the
$U(1)$ gauge theory corresponds to inserting a degenerate vertex
operator $\Phi_{1,2}$ in the 4-point block, as in Figure~\ref{CFTSOU1}.
In the decoupling limit $\Lambda \to 0$ the surface operator partition
function obtained from the free boson/fermion correlator is
%
\begin{align}
Z^{\rm CFT} = \langle \emptyset | \, \Phi_{1,2}(z) \, V_{m}(1) \, | \emptyset \rangle \,, 
\end{align}
where the state $| \emptyset \rangle$ is the Dirac vacuum in the free fermion theory.
Inserting a complete basis of states gives a power series in $z$
\begin{align}
Z^{\rm CFT} = \sum_R z^{|R|} \langle \emptyset | \, \Phi_{1,2}(1)  |R \rangle \, \langle R | V_{m}(1) \, | \emptyset \rangle \,,
\end{align}
where $R$ is a Young tableau and $|R \rangle$ represents the corresponding state in the Fock space.\footnote{A definition of $|R \rangle$ in terms of the free fermion theory can be found (\eg) in Section 5~of~\cite{NO}.}
This series expansion is equal to the homological vortex partition function
\begin{align}
Z^{\rm vortex} = \sum
z^{\m} \prod_{j=1}^{\m} \frac{-m + (j- \frac{1}{2}) \e_1 + \frac{1}{2} \e_2}{j \e_1} \,.
\end{align}
For $(\epsilon_1,\epsilon_2)\to (\hbar,0)$ (our usual vortex limit) and a redefinition $m\to -m+\hbar/2$, this reproduces vortex-counting \eqref{U1massvortex} for the two-dimensional abelian Higgs model with an extra antifundamental chiral of mass $m$.


\subsection{$SU(2)$ CFT}

Instanton counting for gauge theories whose gauge groups are a product of $SU(2)$ factors is reproduced by conformal blocks of Liouville theory \cite{AGT}.%
\footnote{Here we just discuss surface operators in \emph{superconformal} $SU(2)$ theory. However, we should emphasize that all constructions and relations can be extended to non-conformal theories, with different matter contents, by using (\eg) the formalism of \cite{Gaiotto-AF}.} %
The equivariant parameters $\e_1,\,\e_2$ are then related to the Liouville parameters $b^{\pm1}$ as
\be \e_1= b \eta\,,\qquad \e_2=\frac{\eta}{b}\,,\ee
where $\eta$ is a fixed overall mass scale (called $\hbar$ in \cite{AGT,AGGTV}). The central charge is $c=1+6Q^2$, with $Q=b+b^{-1}$. The Virasoro algebra is sufficient to classify all states; therefore, primary fields have a single momentum $\alpha$, in terms of which their conformal weight is $h = \alpha(Q-\alpha)$.

In a conformal block such as
\be  \raisebox{-.4in}{\includegraphics[height=.8in]{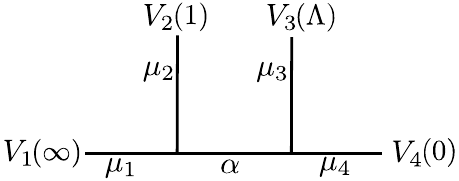}}   \ee
for superconformal $SU(2)$ theory with two fundamentals and two antifundamentals, the internal momentum is related to the Coulomb parameter $a$ as
\be \alpha = \frac{Q}{2}+\frac{a}{\eta}\,. \ee
The momenta $\mu_i$ of the four insertions $V_i$ are related to bare fundamental masses $m_1,m_2$ and antifundamental masses $\bar{m}_1,\bar{m}_2$ as%
\footnote{We use conventions --- as described in Appendix \ref{app:GE} --- where bare masses are all shifted by $(\e_1+\e_2)/2$ from the standard values appearing in Nekrasov partition functions. This differs slightly~from~\cite{AGT}.}
\begin{subequations} \label{Lmasses}
\begin{align} \mu_1 = \frac{Q}{2}+\frac{\bar{m}_1-\bar{m}_2}{2\eta}\,,&\qquad \mu_2 = \frac{Q}{2}+\frac{\bar{m}_1+\bar{m}_2}{2\eta}\,, \\
\mu_3 = \frac{Q}{2}+\frac{{m}_1+{m}_2}{2\eta}\,,&\qquad \mu_4 = \frac{Q}{2}+\frac{{m}_1-{m}_2}{2\eta}\,.
\end{align}
\end{subequations}
The one insertion coordinate ``$\Lambda$'' that is not fixed by conformal symmetry is the UV coupling of the superconformal theory.

In a similar way, superconformal $SU(2)\times SU(2)$ theory with a bifundamental hypermultiplet of mass $m_b$ corresponds to the block
\be \raisebox{-.4in}{\includegraphics[height=.8in]{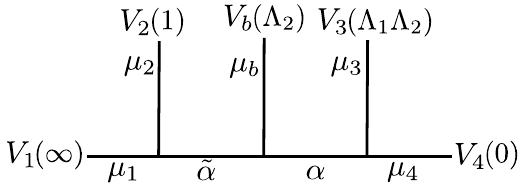}\;\;.}  \label{5ptSU2} \ee
The two $SU(2)$ factors have couplings $\Lambda_1,\Lambda_2$, Coulomb parameters are given by $\tilde{\alpha}=\frac{Q}{2}+\frac{\tilde{a}}{\eta}$, $\alpha = \frac{Q}{2}+\frac{a}{\eta}$, fundamental and antifundamental masses are related to external momenta $\mu_i$ as in \eqref{Lmasses}, and the bifundamental bare mass is related to the momentum $\mu_b$ of $V_b$ as
\be \mu_b = \frac{Q}{2}+\frac{m_b}{\eta}\,. \ee

Degenerate primary fields $\Phi_{r,s}$ of the Virasoro algebra have conformal weights that correspond to momenta
\be \alpha_{r,s} := \frac{Q}{2}-\frac{rb}{2}-\frac{s}{2b} = \frac{1-r}{2}b+\frac{1-s}{2b}\,,\qquad r,s\in\Z_{>0}\,.\ee
Following \cite{AGGTV}, an insertion of $\Phi_{r,s}$ in a conformal block such as
\be \raisebox{-.4in}{\includegraphics[height=.8in]{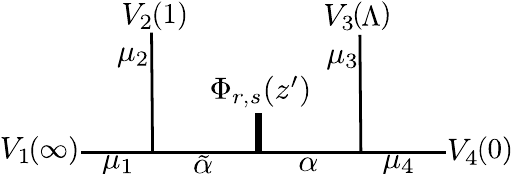}}  \label{5ptdSU2} \ee
introduces a surface operator in the $SU(2)$ gauge theory. As discussed in Section \ref{sec:CFTdd}, we can also view this as a conformal block \eqref{5ptSU2} for $SU(2)\times SU(2)$ theory, in the special limit
\be \mu_b\to \alpha_{r,s}\,,\qquad\text{or}\qquad m_b\to -\frac{r\e_1}{2}-\frac{s\e_2}{2}\,. \ee
We find precisely the degenerate mass described in \eqref{GTrefmass} of Section \ref{sec:GT}, in the context of refined geometric transitions. Therefore, we may expect $\Phi_{r,s}$ to realize a surface operator supported on $w_1^{r-1}w_2^{s-1}=0\;\subset\R^4.$

To compare with vortex counting, consider the decoupling limit of \eqref{5ptdSU2}, \ie\ the conformal block
\be  \raisebox{-.4in}{\includegraphics[height=.8in]{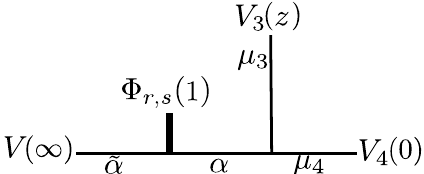}\;\;.}  \label{CFTdecoupSU2} \ee
This conformal block should directly give homological vortex counting for the 2d $\CN=(2,2)$ theory on the $(r,s)$ surface operator.


For example, a $\Phi_{1,2}$ insertion in \eqref{CFTdecoupSU2} should correspond to vortex counting in an abelian 2d theory with two fundamental chirals (one massless) and two antifundamental chirals.
The degenerate OPE with $\Phi_{1,2}$ imposes one of the two ``momentum-conservation'' relations
\be (+)\quad\tilde{\alpha}-\alpha=\frac{1}{2b}\,,\qquad\text{or}\qquad (-)\quad\tilde{\alpha}-\alpha=-\frac{1}{2b}\,.
\label{SU2momcons} \ee
Comparing to Section \ref{sec:GTU2} (especially the diagrams~\ref{GTbottom}--\ref{GTtop}) we see that these choices correspond to geometric engineering of surface operators with a Lagrangian brane on (+) the ``top'' gauge leg, or (--) the ``bottom'' gauge leg.

The conformal block \eqref{CFTdecoupSU2} can be computed order-by-order in $z$ using operator algebra. To this end, define the Gram matrix $H_{\mb{k'},\mb{k}}=\langle \alpha,\mb{k}'|\alpha,\mb{k}\rangle$, and the ``vertex''
\be S_{\mb{k},\mb{k}'}(\alpha_1,\alpha_2,\alpha_3;x)
 := \frac{\langle \alpha_1,\mb{k}|V_{\alpha_2}(x)|\alpha_3,\mb{k}'\rangle}{\langle \alpha_1|V_{\alpha_2}(x)|\alpha_3\rangle}\,,
\ee
where $|\alpha,\mb{k}\rangle = L_{-\mb{k}}|\alpha\rangle = L_{-k_1,-k_2,...,-k_K}|\alpha\rangle$ denotes the Virasoro descendant of a primary state with momentum $\alpha$. We impose $k_1\geq k_2\geq...\geq k_{K}$, so that $\mb{k}$ is a partition. Conformal symmetry shows that $S_{\mb{k},\mb{k'}}(\alpha_1,\alpha_2,\alpha_3;x) = x^{|\mb{k}|-|\mb{k}'|}S_{\mb{k},\mb{k'}}(\alpha_1,\alpha_2,\alpha_3;1)$. Then the conformal block \eqref{CFTdecoupSU2} is
\be \CF(z) = \sum_{\mb{k},\mb{k}'} S_{\circ,\mb{k}}(\tilde{\alpha},\alpha_{1,2},\alpha;1) H^{-1}_{\mb{k},\mb{k}'}  S_{\mb{k}',\circ}(\alpha,\mu_3,\mu_4;1)\,z^{|\mb{k}'|}\,. \label{Fcomp} \ee
Setting $\tilde{\alpha}=\alpha\pm\frac{1}{2b}$, we obtain
\begin{align} \CF^{(+)}(z) &=
(1-z)^{-\frac{\epsilon_2}{\eta^2}\left(m_1+m_2-\frac{\e_+}{2}\right)} \notag \\
&\quad \times\sum_{\m=0}^\infty \prod_{j=1}^{\m} \frac{(a-m_1+(j+\frac12)\e_1+\frac{\e_2}2)(a-m_2+(j+\frac12)\e_1+\frac{\e_2}2)}{j\e_1(2a+j\e_1+\e_2)}z^\m\,, \label{CBtop} \\
\CF^{(-)}(z) &=
 (1-z)^{-\frac{\epsilon_2}{\eta^2}\left(m_1+m_2-\frac{\e_+}{2}\right)}\notag \\
&\quad \times\sum_{\m=0}^\infty \prod_{j=1}^{\m} \frac{(-a-m_1+(j+\frac12)\e_1+\frac{\e_2}2)(-a-m_2+(j+\frac12)\e_1+\frac{\e_2}2)}{j\e_1(-2a+j\e_1+\e_2)}z^\m \,, \label{CBbottom}
\end{align}
with $\epsilon_+=\epsilon_1+\epsilon_2$.

Alternatively, we could have used the degenerate equation for $\Phi_{1,2}$\,,
\be \left(L_{-2}-\frac{3}{2(2h_{1,2}+1)}L_{-1}^2\right)|\Phi_{1,2}\rangle = 0 \ee
(where $h_{1,2} = \alpha_{1,2}(Q-\alpha_{1,2})$ is the conformal weight of the primary field $\Phi_{1,2}$) to derive a differential equation for the conformal block $\CF(z)$. After removing the prefactor $(1-z)^{-\frac{\epsilon_2}{\eta^2}\left(m_1+m_2-\frac{\e_+}{2}\right)}$, this reduces to the hypergeometric differential equation and directly gives (\cf\ Appendix B of \cite{AGGTV})
\begin{align}
\CF^{(\pm)}(z)  &=
 (1-z)^{-\frac{\epsilon_2}{\eta^2}\left(m_1+m_2-\frac{\e_+}{2}\right)}\notag \\ &\quad\times {}_2F_1\left(\frac{\pm a-m_1}{\epsilon_1}+\frac{\epsilon_2}{\epsilon_1}+\frac32,\,\frac{\pm a-m_2}{\epsilon_1}+\frac{\epsilon_2}{\epsilon_1}+\frac32;\,\frac{\pm2a}{\epsilon_1}+\epsilon_2+1;\,z\right)\,.
\end{align}

The inverse of the prefactor $(1-z)^{-(...)}$ in these expressions is a standard $U(1)$ contribution, which should be added (or in this case removed) from the conformal blocks in order to obtain a precise correspondence with vortex counting --- because our surface operator naturally lives in a $U(2)$ 4d theory rather than an $SU(2)$ theory. The remaining hypergeometric functions are precisely the homological ($\beta\to 0$) limits of $Z^{\rm open, top}_{\rm BPS}(z)$ and $Z^{\rm open, bottom}_{\rm BPS}(z)$ from \eqref{GTbottom} and \eqref{GTtop}, assuming the standard identification of four-dimensional mass and Coulomb parameters with K\"ahler parameters in Appendix \ref{app:GE}.
Sending $(\epsilon_1,\epsilon_2)\to (\hbar,0)$ then reproduces the expected vortex-counting partition functions for two-dimensional $\CN=(2,2)$ theory. In case (+), the (classical) 2d fundamental chiral masses are properly identified as $\tilde{m}_{1,2}=(0,2a)$ and antifundamental chiral masses as $\tilde{\bar{m}}_{1,2}=(a-m_1,a-m_2)$ (\cf\ the end of Section \ref{sec:generalsurf}, with $a=a_1=-a_2$). In case (--) the identification is $\tilde{m}_{1,2}=(0,-2a)$ and $\tilde{\bar{m}}_{1,2}=(-a-m_1,-a-m_2)$.

\subsubsection{Higher degenerations}

As a final example, let's explicitly consider a higher degenerate insertion of the form $\Phi_{1,s}$\,, with $r=1$. By our arguments, this should produce a $U(s-1)$ surface operator in $SU(2)$ theory, classically supported on the surface $w_2^{s-1}=0$ in $\R^4$. Again, we look at the decoupled limit $\Lambda\to 0$. In the conformal block diagram \eqref{CFTdecoupSU2}, a $\Phi_{1,s}$ insertion imposed momentum conservation of the form
\be \tilde{\alpha}-\alpha \;\in\;\left\{ -\frac{s-1}{2b},\ldots,\frac{s-3}{2b},\frac{s-1}{2b}\right\}\,.\ee
The choice $\tilde{\alpha}-\alpha=\frac{p_1-p_2}{2b}$, with $p_1,p_2\geq0$ and $p_1+p_2=s-1$, corresponds to placing $p_1$ branes on the ``top'' gauge leg and $p_2$ branes on the ``bottom'' gauge leg in a toric diagram that geometrically engineers the surface operator.

The computation of the conformal block \eqref{CFTdecoupSU2} with $\Phi_{1,s}$ inserted is almost identical to the computation \eqref{Fcomp}; only the degenerate momentum $\alpha_{1,2}\to \alpha_{1,s}$ and the $\tilde{\alpha}-\alpha$ relation changes. Now we have
\be \CF(z) = \sum_{\mb{k},\mb{k}'} S_{\circ,\mb{k}}(\tilde{\alpha},\alpha_{1,s},\alpha;1) H^{-1}_{\mb{k},\mb{k}'}  S_{\mb{k}',\circ}(\alpha,\mu_3,\mu_4;1)\,z^{|\mb{k}'|}\,. \ee
For simplicity, let us choose $\tilde{\alpha}-\alpha=-\frac{s-1}{2b}$, corresponding to $s-1$ branes on a ``bottom'' toric gauge leg, as in Figure \ref{fig:GTpbranes}. Then, for example, we find
\begin{align} \CF(x) &= (1-z)^{-\frac{(s-1)\epsilon_2}{\eta^2}\left(m_1+m_2-\frac{\e_+}{2}\right)} \notag \\
 &\quad\times\left[1+(s-1)\frac{(-a-m_1+\frac{\epsilon_1+\epsilon_2}{2})(-a-m_2+
\frac{\epsilon_1+\epsilon_2}{2})}{\epsilon_1(-2a+\epsilon_1+\epsilon_2)}x+\CO(x^2)\right]\,. \label{Fsblock}
\end{align}
Alternatively, since the insertion $\Phi_{1,s}$ satisfies a degenerate equation $(L_{-s}+\ldots)\Phi_{1,s}=0$ at level $s$, it is possible to find the block $\CF(x)$ as a solution to an order-$s$ differential equation.

In the limit $\epsilon_1,\epsilon_2\to\hbar$, we expect this conformal block to reproduce vortex counting (equivariant with respect to spin plus R-charge) in a $U(s-1)$ 2d theory with two fundamental chirals (of masses zero and $-2a$) and two antifundamental chirals (of masses $-a-m_1$, $-a-m_2$). This  computation has been done in Section \ref{sec:GTp}, via geometric engineering and BPS counting. After removing the prefactor $(1-z)^{-(\cdots)}$, again a $U(1)$ (anti)contribution, it is easy to check order-by-order, for fixed $s$, that the $\epsilon_1,\epsilon_2\to\hbar$ limit of \eqref{Fsblock} exactly reproduces the homological limit $\beta\to0$ limit of \eqref{GTpUR}. The identification of parameters is $Q_1=e^{\beta(a-m_1)}$, $Q_2=e^{-\beta(m_2-a)}$, and $Q_3=e^{\beta(a+m_2)}$ as usual, having set $a_1=-a_2=a$ in the geometrically engineered $U(2)$ theory.


\acknowledgments

We would like to thank M.~Aganagic, C.~Beem, A.~Borodin, A.~Braverman, A.~Gorsky, C.~Keller, H.~Nakajima, J.~Song,
and E.~Witten for very useful discussions,
and C.~Vafa for collaboration at an earlier stage of this project.
The work of SG and LH is supported in part by NSF grant PHY-0757647.
The work of SG is also supported in part by DOE grant DE-FG03-92-ER40701
and in part by the Alfred P. Sloan Foundation.
Opinions and conclusions expressed here are those of the authors
and do not necessarily reflect the views of funding agencies.


\appendix

\section{Two-dimensional $\Omega$-background}
\label{appendix:Omegabackground}

In this appendix we spell out details of the $\Omega$-background $\R^2_{\hbar}$ in two dimensions. Most of the analysis is similar to that in four dimensions \cite{Nekrasov}, but we will encounter a few important differences in the resulting expressions, in particular in the contribution of gauge and matter fields to the two-dimensional contour integrals.\footnote{Part of this analysis has been performed in \cite{Shadchin-2d}. We also found \cite{Shadchin-thesis} very helpful.}

The two-dimensional $\Omega$-background $\R^2_{\hbar}$ simply refers to a two-dimensional plane $\R^2$ endowed with a $U(1)$ rotation symmetry that leaves the origin fixed. We choose the $U(1)$ symmetry to be generated by the two-dimensional vector field
\begin{align*}
 V^{\rho} = \Omega^{\rho}_{\hspace{1mm} \sigma} x^{\sigma},
\end{align*}
where
\begin{align*}
\Omega = \left( \begin{array}{cc} 0 & \hbar \\ - \hbar & 0 \end{array} \right)
\end{align*}
and $\hbar \in \C$. The complex vector field $V^{\rho} = V^{\rho}_1 + i V^{\rho}_2$ is the sum of two real vector fields  that both generate a Lorentz rotation in $\R^2$.

The Lagrangian of a $\CN=(2,2)$ gauge multiplet $\Sigma = \{ A_{\rho}, \sigma,  \la_{\pm}, \bar{\la}_{\pm},D \}$ in the $\Omega$-deformed background $\R^2_{\hbar}$ is given by
\begin{align}\label{eqn:Omegadeformedgaugeaction}
\CL^{2d,\Omega}_{\mathrm{gauge}} = & - \frac{1}{4 g^2} F_{\rho \sigma} F^{\rho \sigma} - \frac{1}{g^2} \left( D_{\rho} \sigma + V^{\sigma} F_{\sigma \rho}  \right) \wedge *   \left( D_{\rho} \bar{\sigma} + \bar{V}^{\sigma} F_{\sigma \rho}  \right) \\
&  +  \frac{1}{2 g^2}   [\sigma + V^{\sigma} D_{\sigma},~ \bar{\sigma} +  \bar{V}^{\rho} D_{\rho}]^2  + \frac{1}{2} D^2  \notag \\ & + \frac{i}{g^2}  \bar{\la}_+ D_{z} \la_+ + \frac{i}{g^2} \bar{\la}_- D_{\bar{z}} \la_- \notag \\
& - \frac{i \sqrt{2}}{g^2} \bar{\la}_+ [ \sigma +  V^{\rho} D_{\rho}, \la_-]  - \frac{i \sqrt{2}}{g^2} \bar{\la}_- [ \bar{\sigma} +  \bar{V}^{\rho} D_{\rho},  \la_+], \notag
\end{align}
which reveals that the $\Omega$-deformation can be purely described by a shift $\sigma \mapsto \sigma + V^{\rho} D_{\rho}$ that turns the adjoint scalar into a differential operator.

The above Lagrangian may be obtained by reducing the four-dimensional $\CN=1$ gauge multiplet in the background specified by the metric
\begin{align*}
ds_4^2 = G_{\mu \nu} dx^{\mu} dx^{\nu} = g_{\rho \sigma} \left( dx^{\rho} + V^{\rho}_a dx^a \right)  \left( dx^{\sigma} + V^{\sigma}_b dx^b \right) + (dx^1)^2 + (dx^2)^2,
\end{align*}
where $\rho, \sigma \in \{0,3\}$ and $a, b \in \{1,2\}$. Lagrangians for other two-dimensional interactions can be derived similarly. For instance, the Lagrangian for a chiral multiplet $Q = \{ \phi, \psi_{\pm}, F \}$ reads
\begin{align}\label{eqn:Omegadeformedchiralaction}
\CL^{2d, \Omega}_{\mathrm{chiral}} =
&  - D_{\rho} \bar{\phi} D^{\rho} \phi
-  \left( [ \bar{\sigma}, \phi]  + \bar{V}^{\rho} D_{\rho} \phi \right)  \left( [\sigma,\bar{\phi}] + V^{\rho} D_{\rho} \bar{\phi} \right)
\\
& -  \left( [ \sigma, \phi]  + V^{\rho} D_{\rho} \phi \right)  \left( [\bar{\sigma},\bar{\phi}] + \bar{V}^{\rho} D_{\rho} \bar{\phi} \right) + D \bar{\phi} T \phi
 + |F|^2 \notag \\
&  + i \bar{\psi}_+ D_z \psi_+ + i \bar{\psi}_- D_{\bar{z}} \psi_-
 \notag \\
& + i \sqrt{2} \bar{\psi}_+ [ \sigma T + V^{\rho} D_{\rho} , \psi_-]  + i \sqrt{2} \bar{\psi}_-  [ \bar{\sigma} T + \bar{V}^{\rho} D_{\rho},  \psi_+] \notag \\
 & - i \sqrt{2} \bar{\phi} \left( \psi_- T \la_+ - \psi_+ T \la_-\right) - i \sqrt{2} \phi \left( \bar{\la}_- T \bar{\psi}_+ - \bar{\la}_+ T \bar{\psi}_- \right), \notag
\end{align}
where $T$ represents the Hermitean generators of the
gauge group in the representation defined by the chiral multiplet. Furthermore, the FI-term plus theta-term
\begin{align}\label{eqn:OmegadeformedFIaction}
\CL^{2d,\Omega}_{FI} = -r D + \frac{\theta}{2 \pi} F_{03},
\end{align}
stays invariant under the $\Omega$-deformation.

Not all two-dimensional $\CN=(2,2)$ couplings can be obtained by a dimensional reduction from four dimensions. However, it is a simple matter to also write down the $\Omega$-transformations for such terms.  As an example, let us consider the twisted mass term for a chiral multiplet $Q$
\begin{align}\label{eqn:Omegadeformedtwistedmassaction}
\CL^{2d}_{\widetilde{\mathrm{mass}}} = \int d^4 \theta ~ Q^{\dag} e^{2V_1} Q  =& ~ - \partial_{\rho} \bar{\phi} \partial^{\rho} \phi - 2 \widetilde{m} \bar{\widetilde{m}}  \phi \bar{\phi}
 + |F|^2 \\
&~  + i \bar{\psi}_+ \partial_z \psi_+ + i \bar{\psi}_- \partial_{\bar{z}} \psi_- - \sqrt{2} \widetilde{m} \bar{\psi}_+ \psi_-  -  \sqrt{2} \bar{\widetilde{m}} \bar{\psi}_-  \psi_+, \notag
\end{align}
where $V_1 = \theta^+ \bar{\theta}^- \widetilde{m} + \mathrm{h.c.}$. Since this interaction term equals the kinetic term (\ref{eqn:Omegadeformedchiralaction}) for the chiral multiplet, when we set $A=\la=D=0$ and exchange $\sigma$ with the twisted mass $\widetilde{m}$, it is clear that the twisted mass needs to be shifted similarly as the adjoint scalar $\sigma$,
\begin{align*}
\widetilde{m} \mapsto \widetilde{m} + V^{\rho} \partial_{\rho},
\end{align*}
for supersymmetry to be preserved in the $\Omega$-deformed background. Twisted masses thus enter the two-dimensional $\Omega$-deformed Lagrangian in the same way as the adjoint scalar $\sigma$.


\section{Closed BPS invariants and instanton counting}
\label{app:GE}

We recall here several details of the geometric engineering construction for four-dimensional gauge theories, in the context of BPS/instanton counting.
We summarize a set of conventions that leads to a consistent identification of $Z^{\rm BPS}$ with $Z^{\rm inst}_{\rm K-theory}$. (See, \eg, \cite{HIV} for further information.)

\begin{figure}[htb]
\centering
\includegraphics[width=6.5in]{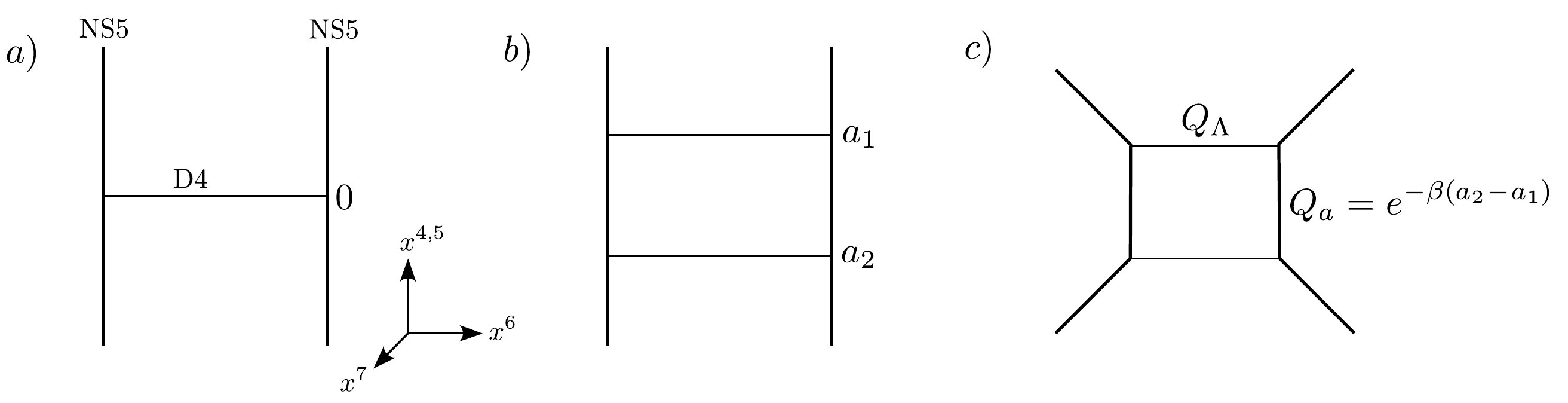}
\caption{a) A stack of D4-branes realizing $U(N)$ gauge theory with unbroken gauge symmetry. b) The $U(N)$ theory at a generic point on the Coulomb branch (for $N=2$). c) The corresponding toric geometry.}
\label{fig:GE1}
\end{figure}

At the origin of the Coulomb branch, an $\CN=2$ gauge theory with gauge group $U(N)$ can be realized on a stack of D4-branes stretched between parallel NS5-branes, as in Figure \ref{fig:GE1}(a). (Also \cf\ Figure \ref{branefig}.) The D4-branes sit at the origin of the $x^4+ix^5$ coordinate; and the distance between NS5-branes in the $x^6$ direction (complexified to $x^6+ix^{10}$ in an M-theory lift), measured at various positions in $x^4+ix^5$, is the running gauge coupling $\tau$. At a generic point on the Coulomb branch, the D4-branes move apart, as in Figure \ref{fig:GE1}(b), with $x^4+ix^5$ positions corresponding to the eigenvalues $a_i$ of the adjoint scalar. The resulting theory is then engineered in a toric geometry as in \ref{fig:GE1}(c). Differences in Coulomb eigenvalues become fiber K\"ahler parameters (here $Q_a=e^{-\beta(a_1-a_2)}$), and the scale $\Lambda$ derived from the running of the gauge coupling becomes a base K\"ahler parameter (here $Q_\Lambda = \beta^4\Lambda^4$). \\

\begin{figure}[htb]
\centering
\includegraphics[width=6.2in]{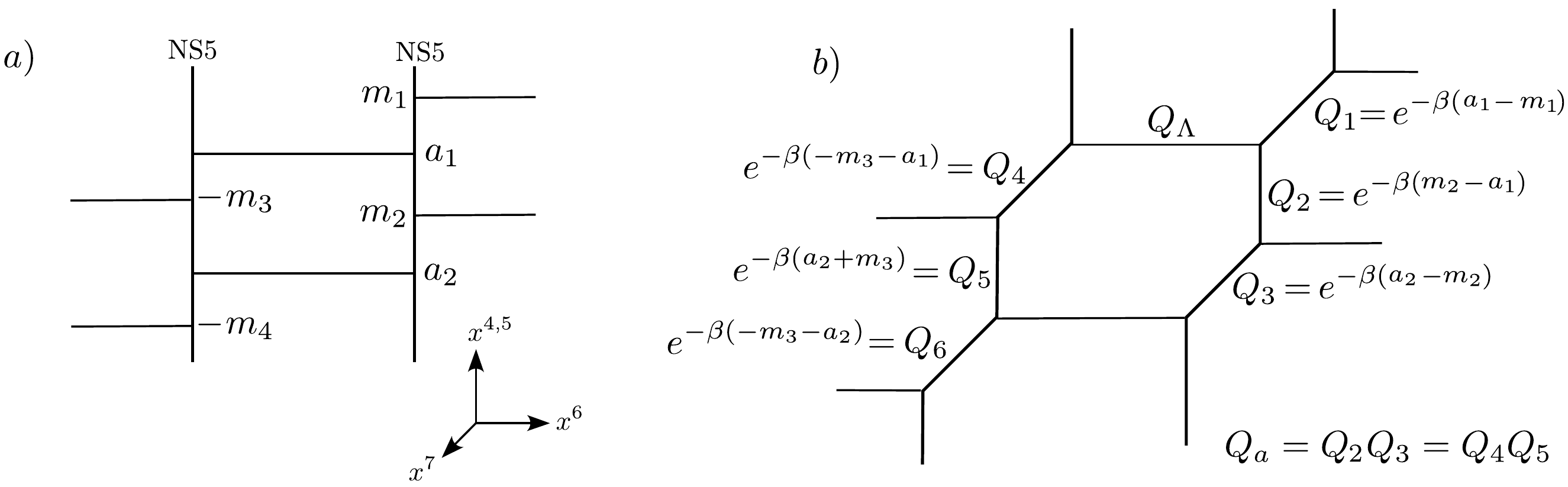}
\caption{Adding flavors and choosing K\"ahler parameters, in (a) the brane construction and (b) the corresponding toric geometry. Here the gauge group is $(S)U(2)$.}
\label{fig:GE2}
\end{figure}

It is possible to add fundamental and antifundamental hypermultiplet flavors to the $U(N)$ theory --- up to $N_f=2N$ of them --- by attaching semi-infinite D4-branes outside the NS5's. In Figure \ref{fig:GE2}(a), the D4's stretching to the right are fundamental hypermultiplets, and their $x^4+ix^5$ positions label their bare masses. The D4's stretching to the left are antifundamental, and their $x^4+ix^5$ positions correspond to the \emph{negative} of their bare mass. Although fundamental and antifundamental hypermultiplets are completely equivalent in $\CN=2$ gauge theory, they behave slightly differently with respect to the $\Omega$-deformation. In accord with conventions in (\eg) \cite{AGGTV}, we shift the bare gauge theory masses occurring in instanton expressions by $m\to m+\frac{\epsilon_1+\epsilon_2}{2}$. Then a fundamental with mass $m$ is equivalent an antifundamental with mass $-m$.

On the Coulomb branch of $U(N)$ theory, after the gauge group has been broken to $U(1)^N$, the physical BPS masses of hypermultiplets are calculated as differences between bare masses and adjoint-scalar vevs. Correspondingly, these gauge-theory BPS states are realized by open string stretching between gauge and flavor D4 branes. A toric geometry that engineers $U(N)$ theory with flavors, and can be used for instanton counting is shown in Figure \ref{fig:GE2}(b).\footnote{In fact, several different toric geometries can correspond to the same brane construction, with the horizontal external legs describing degeneration loci placed between different inner horizontal legs. They are all related by flop transitions.}

\vspace{.2cm}

\begin{figure}[htb]
\centering
\includegraphics[width=5.5in]{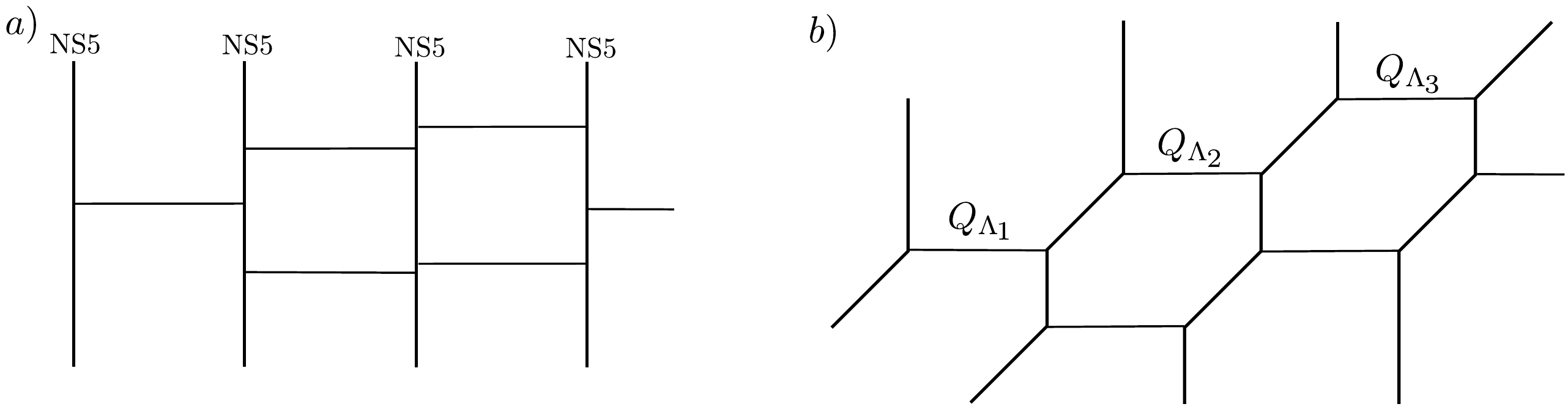}
\caption{A brane construction and toric geometry for a product of gauge groups, here $U(1)_1\times U(2)_2\times U(2)_3$, with one fundamental flavor for $U(2)_3$.}
\label{fig:GE3}
\end{figure}

Finally, one can consider an $\CN=2$ theory with a product gauge group $U(N_1)\times U(N_2)\times\cdots \times U(N_r)$. (This will become quite important later, in Section \ref{sec:GT}.) A typical D4-NS5 setup looks like Figure \ref{fig:GE3}(a). A bifundamental hypermultiplet for each pair of consecutive gauge groups $U(N_n)\times U(N_{n+1})$ is automatically present, and its bare mass is the difference of the \emph{average} of Coulomb parameters ($a_i^{(n)}$ and $a_j^{(n+1)}$) for the two groups. The corresponding toric geometry appears in Figure \ref{fig:GE3}(b).

%

\bibliographystyle{JHEP_TD}
\bibliography{duality}

\end{document}